# Superconductivity in particle accelerators

*Lecture*

C.Z. ANTOINE

*Université Paris-Saclay, CEA, Département des Accélérateurs, de la Cryogénie et du Magnétisme, 91191, Gif-sur-Yvette, France.* claire.antoine@cea.fr

## Content

# Preamble

This text is a full transcription of a lecture was given in the Joint Universities Accelerator School (JUAS) in 2023. A much shorter version is to be published in the JUAS book in 2024, along with all other lectures from the JUAS school.

The aim of this lecture is to provide to physicists non-expert in materials science a glimpse on how superconducting materials are tailored (and must be chosen) for their specific application in accelerators.

## Accelerators

In a simplified view, in an accelerator, one needs a source of charged particles, electrical field to accelerate them and magnetic field to deviate them. For decades, copper has been used for both acceleration and deviation, to fabricate either radiofrequency (RF) cavities or electromagnets. It is possible to replace Copper by a superconductor (SC[1]), but then, the specifications of the superconductors differ a lot wherever magnetic field or electric field generation is concerned.

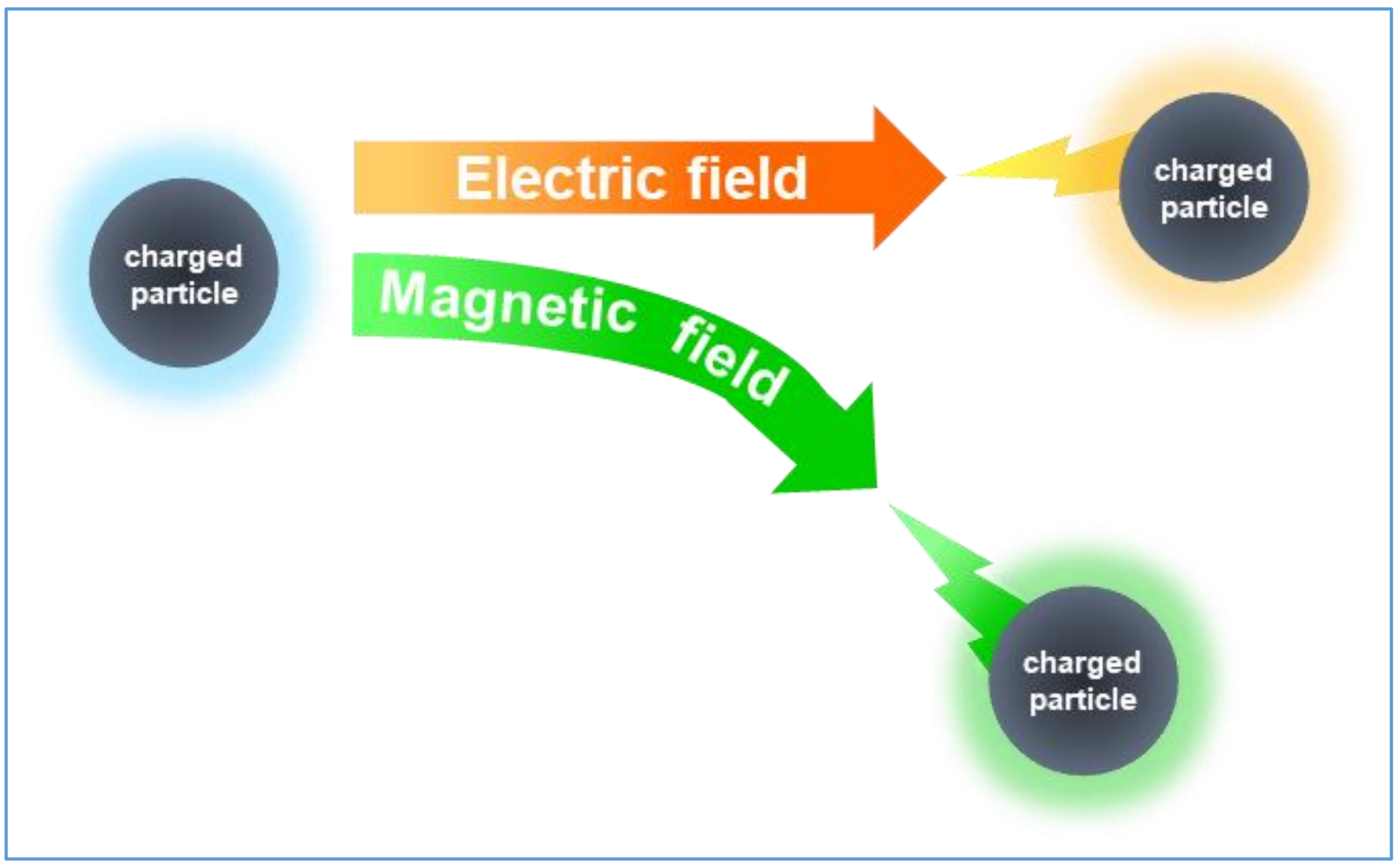

*Figure 1. Charged particles are accelerated by electric field and deviated by magnetic field.*

[1] In the following "SC" means either "superconducting" or "superconductor" depending on the context.

# 1 Introduction

Superconductors can be used for electromagnets, for curvature and focalization of the beam while other superconductors can be used for the fabrication of RF cavities. There are closely related field of developments worth knowing (see Figure 2):

- Magnets for medical imaging, which are in fact at the origin of the development of superconducting magnets
- Fusion magnets for plasma confinement
- Electrical engineering, with the fabrication of high $T_C$ wires/ tapes that can be cooled with (inexpensive) liquid nitrogen
- Josephson Junctions, which are junction of two superconductors with a very thin insulating layer in-between. They present original electronic properties.
    - SC electronics (RSFQ logic): the electron in a superconductor are ballistic, it can render this kind of electronic very fast.
    - SQUIDS
        - Magnetic field detection
        - Bolometers
- Nanodetectors (constrictions, wires…)

*Electrical engineering: transport, storage, energy conversion. There exist already SC current limiters, transmission lines, and coming soon: transformers and motors. Should allow enormous reductions in losses (Electricity transmission now generates significant energy losses by Joule effect.) For the electricity transmission network in France, for example, these losses are estimated on average at 2.5% of the Consumption, i.e. 11.5 TWh per year) Main difficulty: manufacture of HTC SC cables (liquid nitrogen cooled) in long length (because HTC SC are ceramic)*

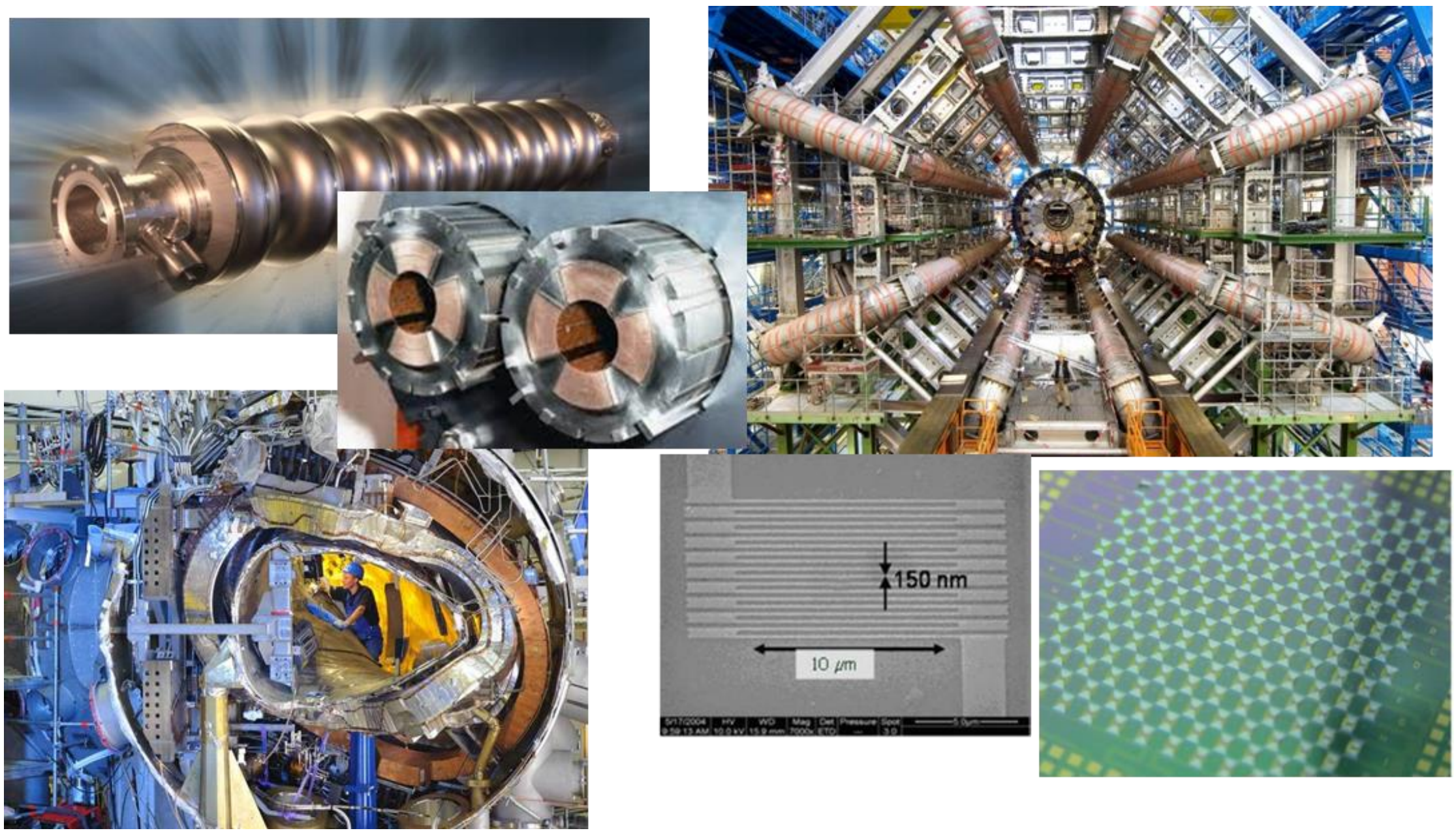

*Figure 2. Images focusing quadrupoles LHC in NbTi, NbTi Atlas detector magnets, 9-cell cavity for Nb electrons (ILC, XFEL), NbTi W7X stellator coils, NbSi bolometer networks (SQUIDS, Application to spatial IR detection), single photon NbN detectors.*

## 1.1 Superconductors in accelerators: why?

### 1.1.1 DC applications, electromagnets.

In DC, in specific conditions, the resistance of a superconductor is strictly equal to 0[1]. Superconductivity allows phenomenal savings in weight, volume and energy consumption. For instance, SC cables about 1 mmx10 mm section, as shown on Figure 3 can drive several 10 000 A where a copper cable would require a 10 cm x 10 cm section.

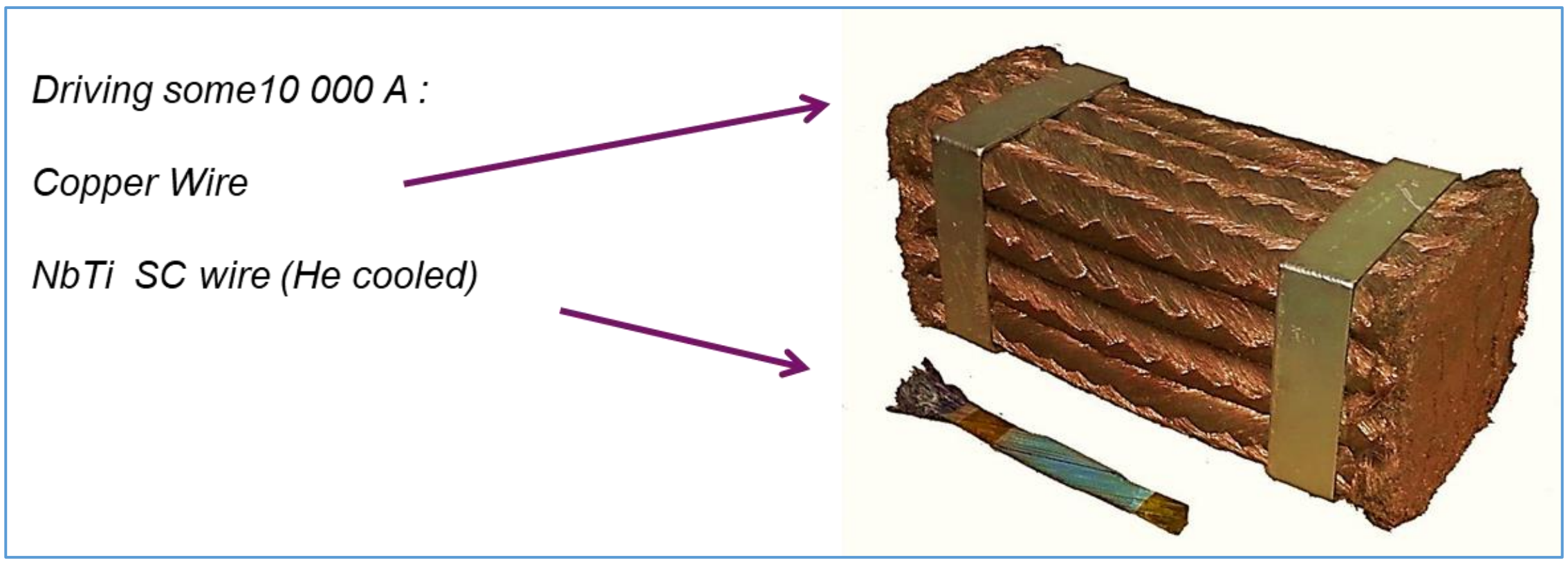

*Figure 3. Section comparison for copper vs superconducting cable (curtesy of B. Harvieu)*

All the curvature magnets from the LHC accelerator at CERN have been fabricated with this kind of cable resulting in a 27 km circumference machine, which needs the equivalent of one power plant, mostly to supply the liquid helium cryogenic system (mostly during summertime, when the said plant is not used to heat the Geneva city ☺).

If the same magnets would have been built out of copper, even with the help of Iron cores to concentrate the field, the accelerator would have been ~100 km in circumference, and would have required ~4 power plants. Indeed, although Copper is a very good electrical conductor, it still has a (feeble) Joule effect, so most of the 4 power plant would be busy producing a lot of heat.

Magnetic field is often required in detectors to curve the trajectory of charged particles and help identifying them, but no detecting element can be placed at the location of the magnet. Here again, even with the presence of cumbersome cryostats, superconductors allow a large advantage in "transparency", when a SC magnet occupies roughly one third of the volume (and weight) of copper magnet.

### 1.1.2 RF cavities.

For RF cavities, the situation is more complex, it depends a lot on the duty cycle

Superconducting cavities exhibit very quality factors at high duty cycle, but are limited around a few 10s of MV/m, the exact figure depending on the cavity geometry. Copper cavities can go to very high fields (some 100s of MV/m), but only with very low duty cycle. Duty cycle of 1 (beam power 100% of the time, also called "continuous wave": CW) can only be reasonably achieved with superconductors.

[1] Be careful, it is not always true!

Table 1 gives the comparison between copper and niobium for a 700 MHz CW Proton linac dedicated to trigger a neutron source.

*Table 1. Figure of interest in the comparison Cu vs Nb to gain 1GeV proton linac (Reproduced with permission from J.L. Biarotte, CNRS)*

| | **Copper** | **Niobium (SC)** |
|---|---|---|
| **Surface resistance: $R_S$** | 7 mΩ | 10 nΩ |
| **Foreseeable accelerating field: $E_{acc}$** | 1,6 MV/m | 10 MV/m |
| **RF efficiency** <br> **$h_{RF}=P_{beam}/(P_{beam}+P_{cav})$** | 15% | 100% |
| **Cryogenic efficiency** <br> **($h_{cryo}=h_{Carnot}$ x $h_{thermo}$ avec** <br> **$h_{Carnot}=T_{froid}/(T_{chaud}-T_{froid})$)** | 100% | 0,2% |
| **>Global efficiency** <br> **$P_{fournie\ au\ faisceau}/P_{à\ la\ prise}$** | 7.5% | 49% |
| **Actual length to gain 1 GeV** | **833** m | 286 m |

For copper cavities, power dissipation is a huge constraint (most of the power transforms into heat instead of increasing beam energy); cavity design is driven by this fact. Small duty cycle often imply higher frequencies operation to compensate the luminosity.

For SC cavities, power dissipation is minimal; as most of the power goes to the beam, the cavity design decoupled from the dynamic losses and one gains the freedom to adapt design to specific application. Lower frequencies are available. From the design point of view, the cavities openings are larger and impedance is lower, which leads to reduced wake field and better emittance. In addition, alignment is easier with large openings, which is an advantage for several (10 s) of km long. Unfortunately, one loses on the cryogenic efficiency (Carnot yield ~ 0,15%, to be multiplied by a technical yield). So although $Q_0$ are ~ $10^5$ higher than for copper, the plug power gain is only between 2 and 500 times better with SC according to the applications. Even reducing costs a factor 2 is desirable, when dealing with research accelerators where a cryogenic plant cost several € and the annual power consumption is several 100s Millions of Watts for its ~30 years of existence.

In fact, it is not always desirable to use superconducting cavities. Figure 4 represents a coarse cost estimation in arbitrary unit. Only the area covered by the blue ellipsoid present some advantages in using superconductivity.

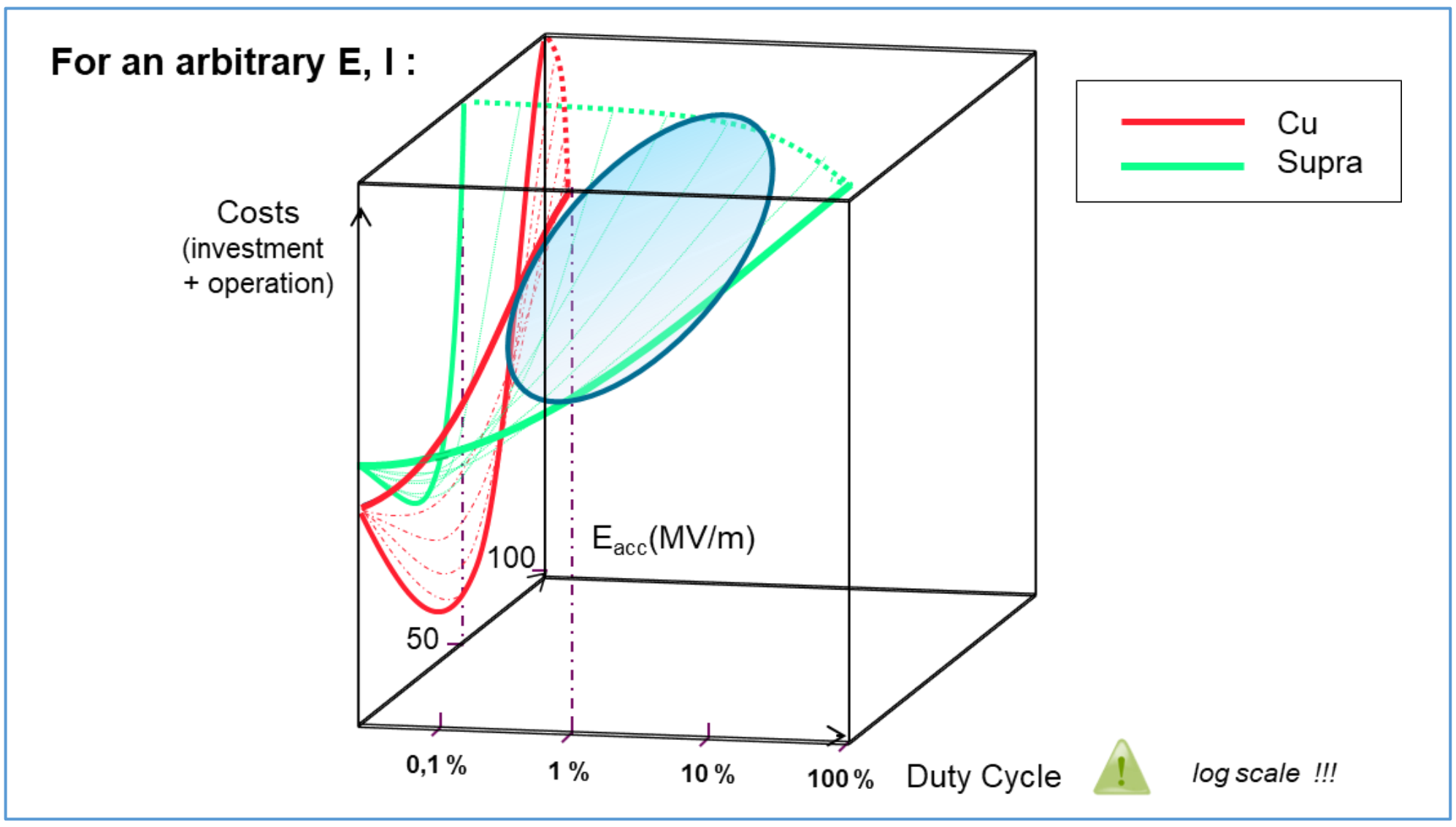

*Figure 4. Coarse estimation of costs for a superconducting vs copper accelerator.*

Typically, Linac costs include ~ 1/3 for the tunnel and construction, ~ 1/3 for the niobium and cryogenic material and ~ 1/3 for RF power supply and beam control. There is always an optimum accelerating field: if low field is fixed one needs a longer machine (longer tunnel, more cavities and cryostat…) and the cost increases. At high field, RF and cooling costs increase for Cu as cryogenic costs for SC.

Costs also increase with Duty Cycle.

> *Duty Cycle and luminosity: CLIC vs ILC (both are e+/e- collider projects, Higgs factory application)*
>
> *ILC is superconducting:  D.C. = 0.5% @ 1,3 GHz*
>
> *CLIC is normal-conducting: D.C. = 0,001% @ 12 GHz*
>
> *Frequency has been enhanced to compensate for low D.C.*

## 2 Superconductivity

In the following, we will describe general features for conventional superconductors. Describing superconductivity requires using thermodynamic notions on phase transition. For those not familiar with this approach, brief descriptions can be found in boxed text or annexes.

### 2.1 Simplified (and of course physically incorrect) mechanism:

When an electron travels between two atoms nucleus (small positive charge), they experience a slight attraction. If another electron arrives in between, it "feels" a slight increase of + charge due to the closed-up atoms and is attracted too, forming a Cooper pair with the first electron. This is observable only at low temperature, because at high temperature thermal agitation counterbalance that kind of movements.

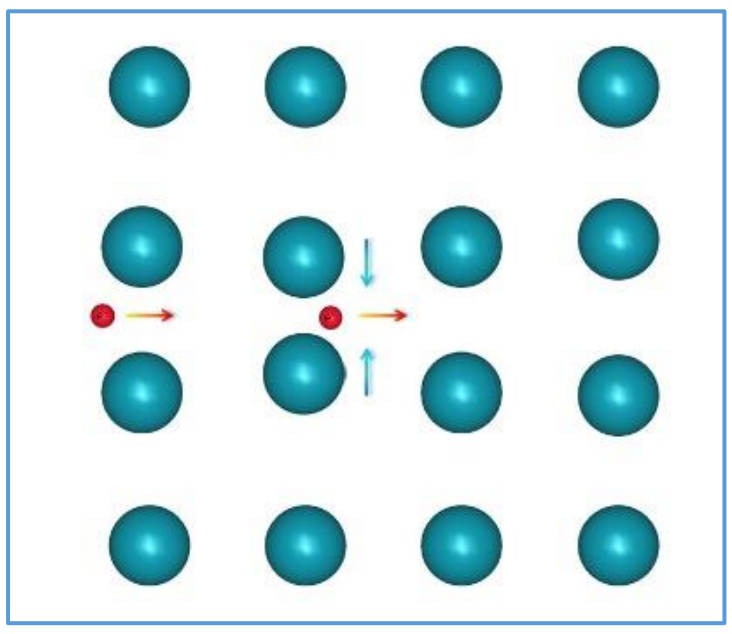

Figure 5. Simplified view of the Cooper pair coupling

In fact, the persistence of a superconducting phase is not only limited in temperature, but it is also limited by current density and or magnetic field (it will be detailed in § 2.1.2 below).

However, superconductivity is a quantum physics phenomenon. Since an electron pair is a Boson, one observes a Bose-Einstein type condensation and a gap opening in the conduction band. The gap opening is detailed in a boxed text below, which provides quick recalls on band theory.

#### 2.1.1 Meissner state

When a superconductor is held at low temperature, low current density and is submitted to an external magnetic field, supercurrent develop at its surface (over a few 10s nm) and generate an opposite magnetic field so that inside the material one gets:

$$\vec{B} = \mu_0(\vec{H} + \vec{M}) = 0 \quad (1)$$

All magnetic field lines are expulsed from the superconductor. This state is called "Meissner state", and is common to all superconductors. The depth over which the field is damped is called, the penetration depth and will be detailed in § 2.3.1

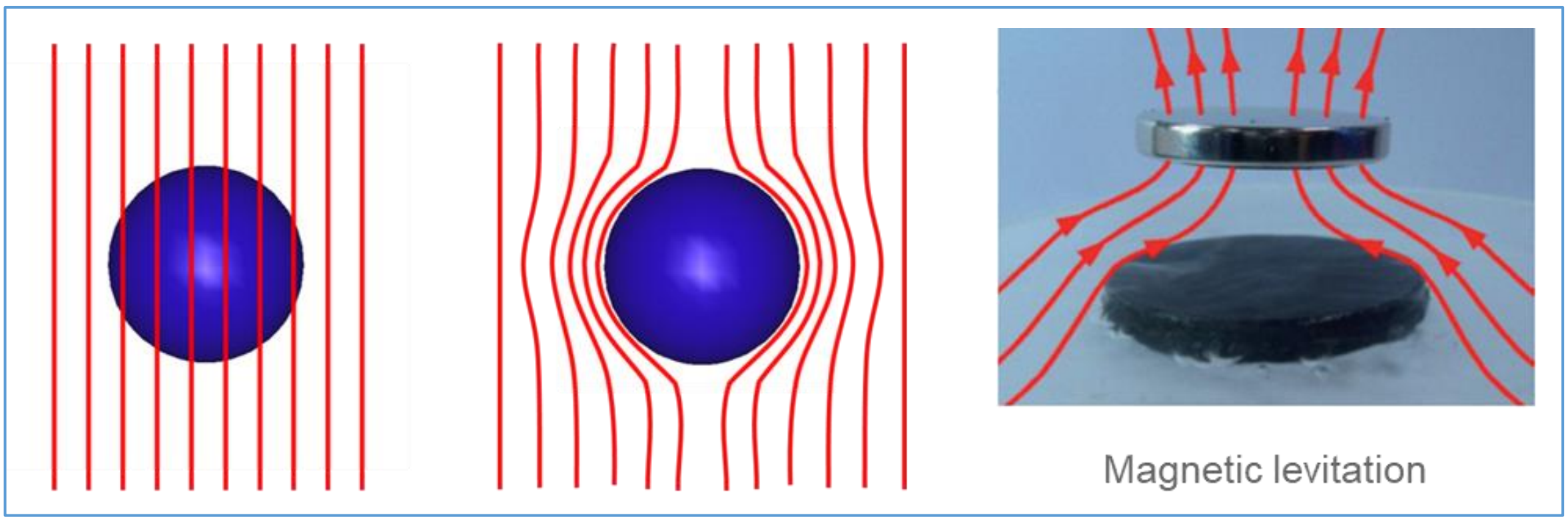

Figure 6. Left: in the normal state flux line fully passe through the materail. At the SC transition flux lines are expelled from the material. This effect is at the origin of the SC magntic levitation which in fact is sligthly different from purely magnetic repulsion as observed e.g. with permanent magnets.

**_Tesla, Gauss, Oersted…so many units ???_**

*Most of the time, in accelerator physics the magnetic field is expressed in Tesla (or its division), which is an abuse of language, since Teslas apply to Magnetic Induction (or Flux Density)* ***B****.* ***B*** *can also be expressed in Gauss. The magnetic field (in fact the magnetic field strength)* ***H*** *should be expressed in Oersted or A/m. So when the "applied field" is expressed in Tesla, it is in fact =* $\mu_0 H$*.*

*In the air, B=* $\mu_0 H$ *and μ is usually about 1, except in special cases, so 1 Gauss = 1 Oersted. The habit of using one or the other units depends if one is immerged in a field (e.g. geoscience where Gauss is used) or producing it (magnetic measurements where Oersted are used).*

*If in addition, one is in presence of a material with magnetic properties,* ***B*** *includes the magnetic response (magnetic moment* ***M****) generated by the applied field:* $\vec{B} = \mu_0 (\vec{H} + \vec{M})$

**_I Oe ~ 1 G ~ 0.1 mT ~ 0.07977 kA/m_**

### 2.1.2 Type I –Type II superconductors

For type I superconductors, when the magnetic field exceeds a critical value $H_C$, the material exhibit a transition to the normal conducting. $H_C$ is the thermodynamic transition field for type I SC.

Type II SC has a different behavior: they are in the Meissner state until the field reaches the first critical field, $H_{C1}$, a field weaker than $H_C$. Then some flux lines start to enter the material, forming a vortex (see Figure 7). A vortex is a single flux line (a quantum of flux) in a normal zone surrounded by screening currents. Around the vortices[1], the superconductivity persists until the field reaches the second critical field ($H_{C2}$), where the material finally becomes normal conducting. The phase between $H_{C1}$ and $H_{C2}$ is called the mixed state or the vortex state.

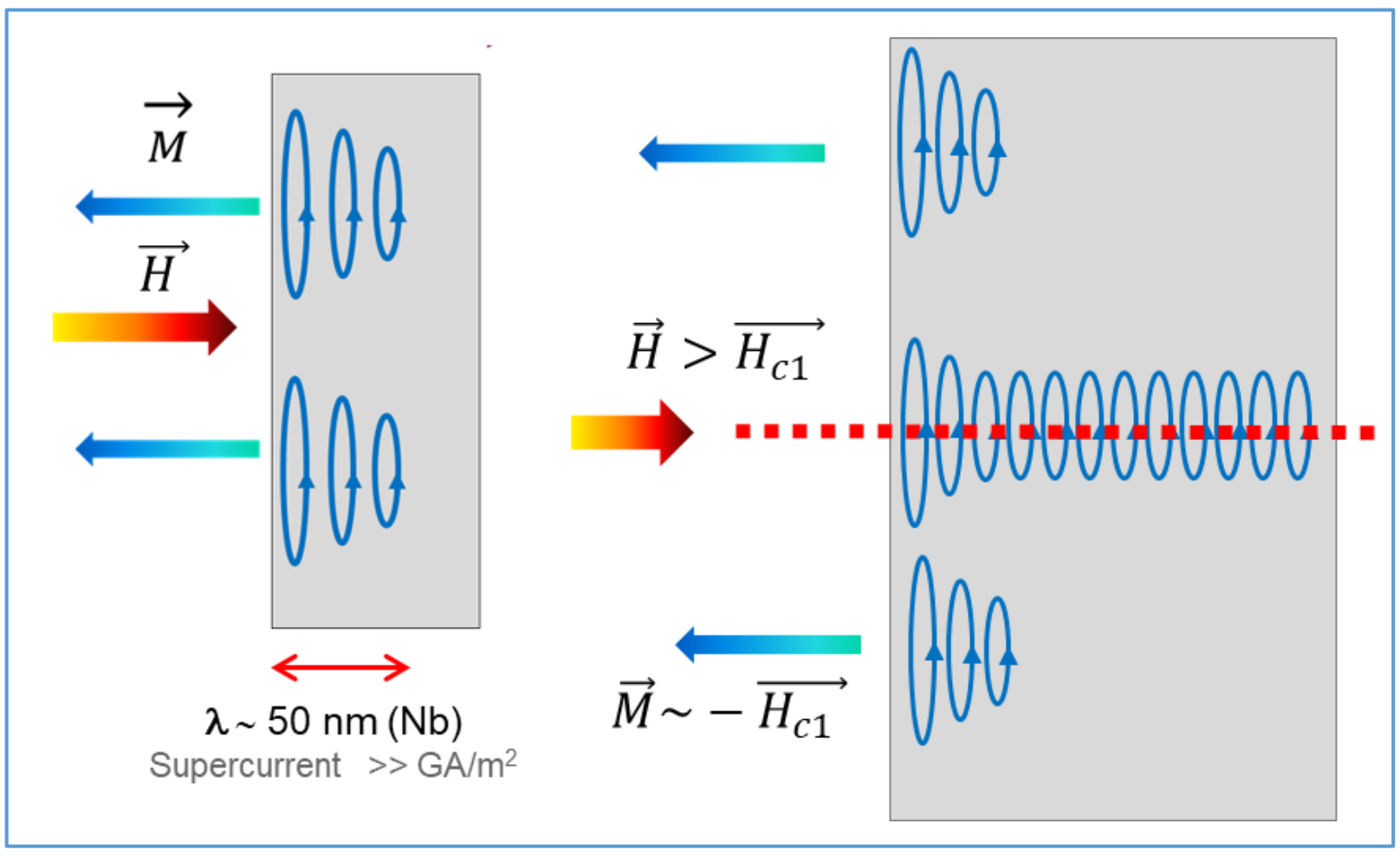

*Figure 7. Meissner state and mixed state. When submitted to an external field, surface supercurrent generate an opposite moment. For type II SC, above $H_{C1}$, some flux lines manage to penetrate the material, cancelling superconductivity in that region (see text below). Around flux lines, screening currents help to maintain the rest of the material in the superconducting state. This state is called "mixed state", since it exhibit a mixture of normal- and superconducting regions.*

In the mixed state, the external field is not fully screened anymore. This can be verified by magnetometry, by measuring the magnetic moments of superconducting samples as depicted in

[1] 1 vortex in singular, several vortices in plural!

Figure 8. Contrary to $H_C$, $H_{C1}$ and $H_{C2}$ are not thermodynamic, intrinsic figures, but depend on the purity and the mechanical state of the SC.

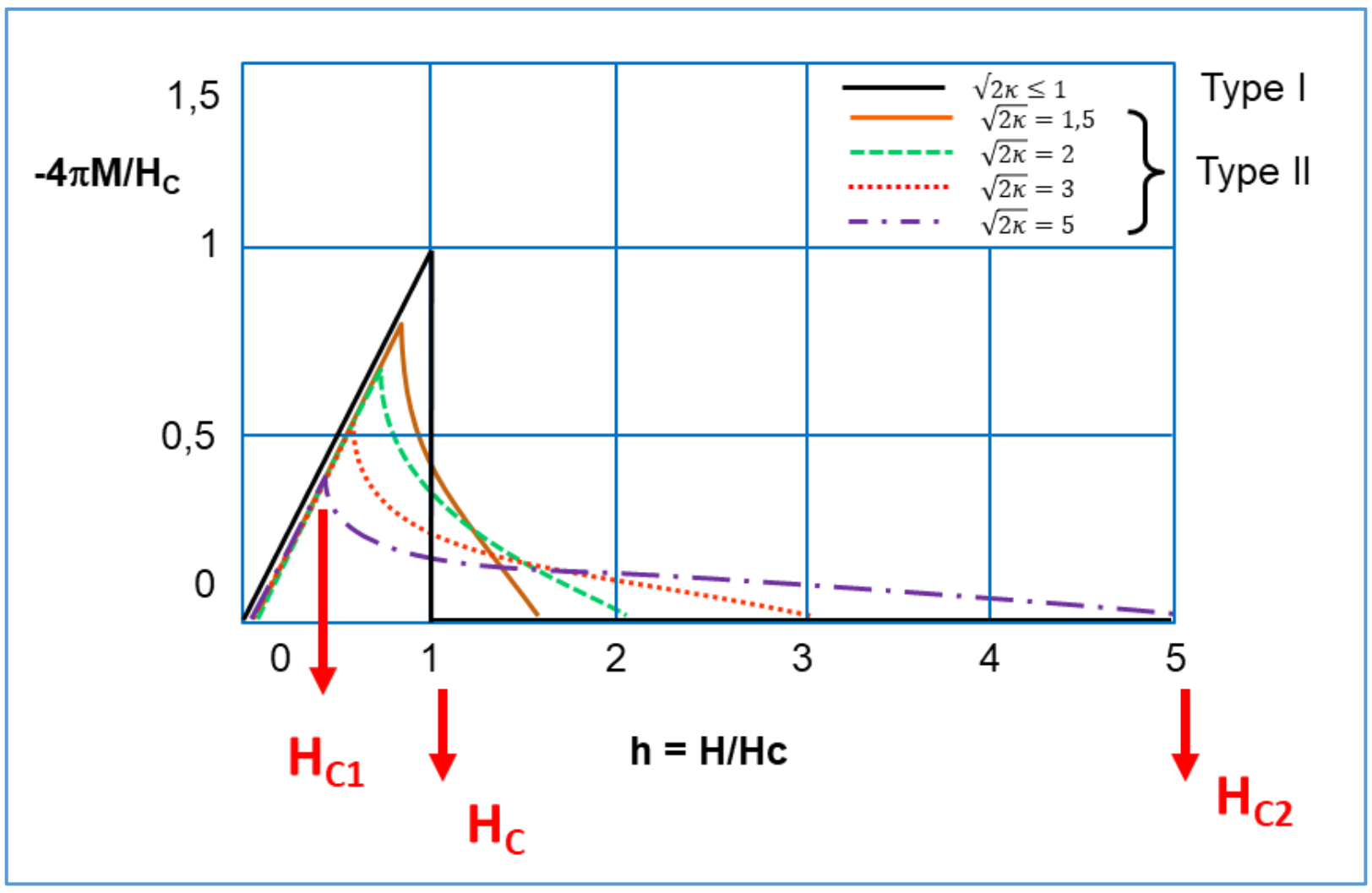

*Figure 8. Reduced magnetic moment of different types of defect free superconductors. According to Eq (1), the maximum moment is equal to $-\mu_0 H_C$ for type I superconductors. For type II superconducting, the screening is not total beyond $H_{C1}$. K is a figure of merit for a given superconductor and will be described below.*

There exist about 10 000s different superconductors. Conventional SC are metals or alloys, and their behavior can be described with BCS theory and its developments (see §2.2.5). In addition, there are also high Tc materials (Cuprates, Pnictides, most of them ceramics), organic superconductors (e.g. Twisted Bilayer Graphene, boron doped diamond, Beechgard salts), and many other exotic material (Heavy Fermion SC, 2-D materials, Bose Einstein condensate in atoms…) where BCS does not apply, and that are sources of active research in solid state physics [1].

Only a dozen superconductors led to effective applications. Most of them are conventional type II. In accelerators, for now, only conventional materials are in use. They all exhibit a low $H_{C1}$ and a high $H_{C2}$, and they operate in the mixed state… **EXCEPT Nb** (for RF applications)**.** Figure 9 represents the SC phase diagram for type II superconductors in the 3D (T, H, J) space.

$T_C$, $H_C$, $H_{C1}$, and $H_{C2}$, have been introduced previously. On this graph, one can observe also two other parameters: $J_C$ and $J_D$

$J_C$, the so-called "critical current" is in fact a technical limit devised by the magnet community and has no thermodynamic or physical meaning outside its field of application (it will be discussed in § 3.3.2). $J_D$, the depairing current is in fact the thermodynamic "critical" value: when the current density reaches this value, it becomes high enough to break the Cooper pair.

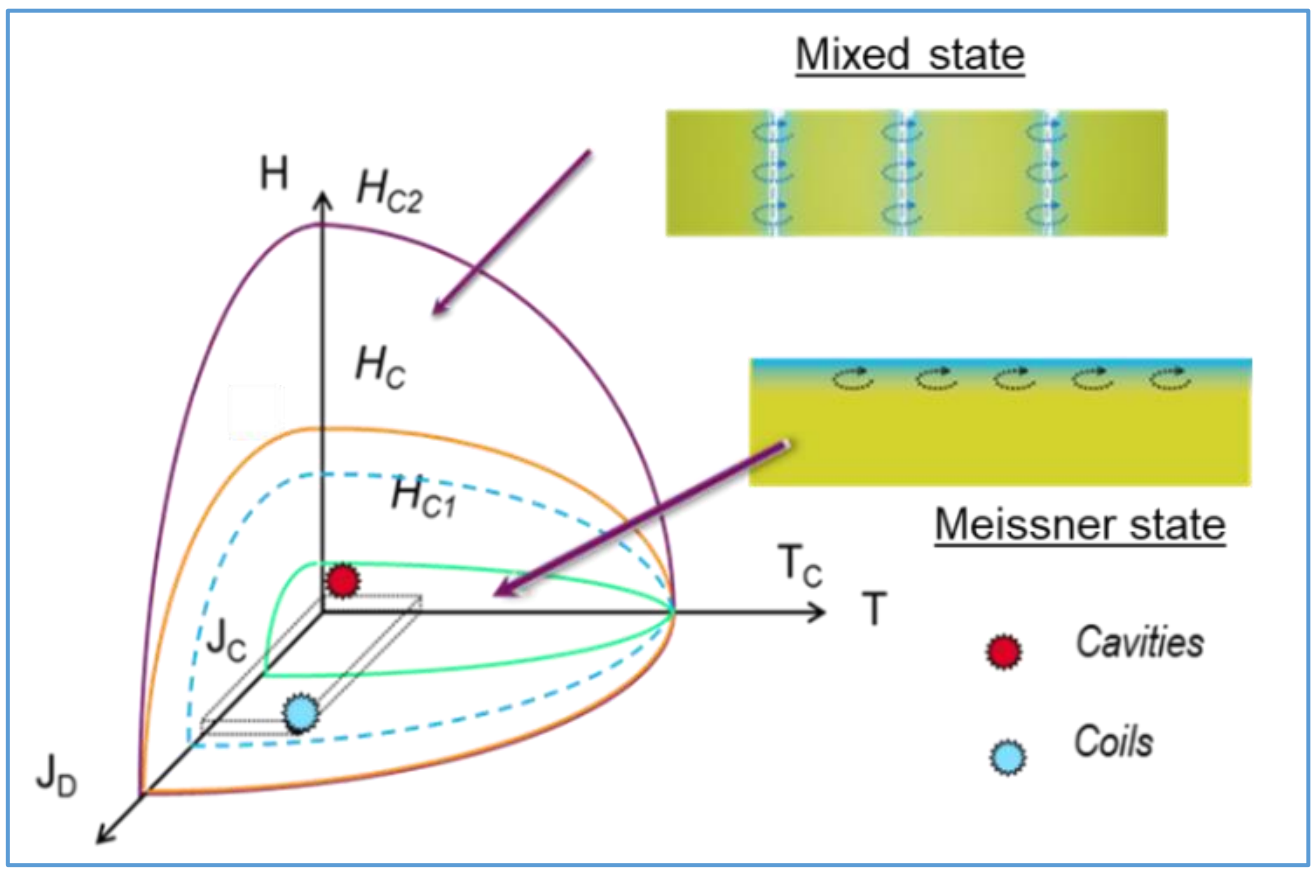

*Figure 9. Phase diagram of type II superconductors in temperature, magnetic field and current density. The red star indicates the operation range of SC cavities and the blue star indicates the range of operation of SC electromagnets.*

We have indicated that RF cavities, **contrary to EVERY OTHER SC applications**, operate in the Meissner state. Indeed, if one puts a vortex line in an high frequency oscillating field, it would start to oscillate, along with its normal conducting core, and that would give rise to huge thermal dissipations. That is why RF application is not based on the same material as magnet application. Up to today, only Niobium, the SC material with the highest known $H_{C1}$, has good performances in RF. This will be further detailed in § 4.2

### 2.1.3 Vortex and flux quantization, penetration depth, coherence length

In a defect-less material, each vortex contains one quantum of flux $\Phi_0 = h/(2e) = 2{,}06783376 \times 10^{-15}$ Weber (or Tesla/m$^2$); with *h the Planck constant and e is the electron charge*. The screening currents decrease over $\lambda$, the penetration depth, while the radius of the normal conducting zone is $\xi$, also called Cooper pair coherence length. $\lambda$ and $\xi$ are critical parameters that characterize each superconductors; they depend on its purity and mechanical state, and they diverge close to Tc. Note that the parameter $\kappa$, introduced in Figure 8, **is called the Ginsburg-Landau parameter and is equal to $\lambda/\xi$.** It also depends on the material state, but is Temperature independent. One can also define an **order parameter $\Psi$,** which is equal to 0 in the normal state and reaches positive values in the superconducting state[1]. Note that the gap is suppressed in the core of the vortex because the screening current density reaches $J_D$, which explains why the vortex core is normal conducting.

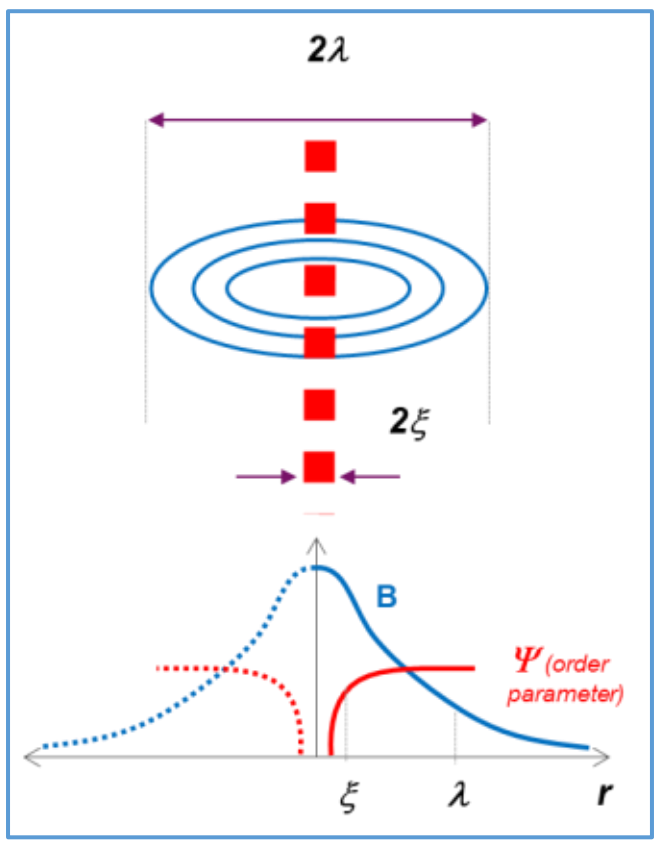

*Figure 10. Vortex description*

Vortices behavior is what conditions all applications. It will be further detailed in the next part of this lecture (§3).

### 2.1.4 Ginsburg-Landau parameter and applications

Figure 11 present another type of phase diagram: the reduced field $H/H_C$ in function of the Ginsburg-Landau parameter $\kappa$, allowing to better understand the difference between type I and type II SC. One can observe that most of the practical SC have a high $\kappa$ except Nb that is very close to the typeI/typeII frontier. The dotted blue curve shows in addition the existence of a "superheating field" $H_{SH}$. $H_{SH}$ is a metastable[2] field below which Meissner state can be observed above $H_C$ (type I) or $H_{C1}$ (type II). It is mentioned here because it is believed that when the magnetic field is parallel to the surface, as is the case in RF cavities, this metastable state can be favored, and it is considered to be the ultimate limit in SRF cavities. This point will be discussed further in § 3.2.2 and § 4.2

---

[1] Physical quantities show singular behaviors at phase transitions and one can define universal critical exponents or parameters, including an order parameter to characterize a phase transition. For superconductivity the critical parameters are $\lambda$ and $\xi$ and they diverge –as expected for critical parameters- at T= $T_C$.

[2] For example, very pure water can still be liquid (super cooled state) below 0)C. the slightest choc will make it become solid very quickly. See e.g. [(73) Watch supercooled water freeze - YouTube](). By definition, a metastable state is not very stable…

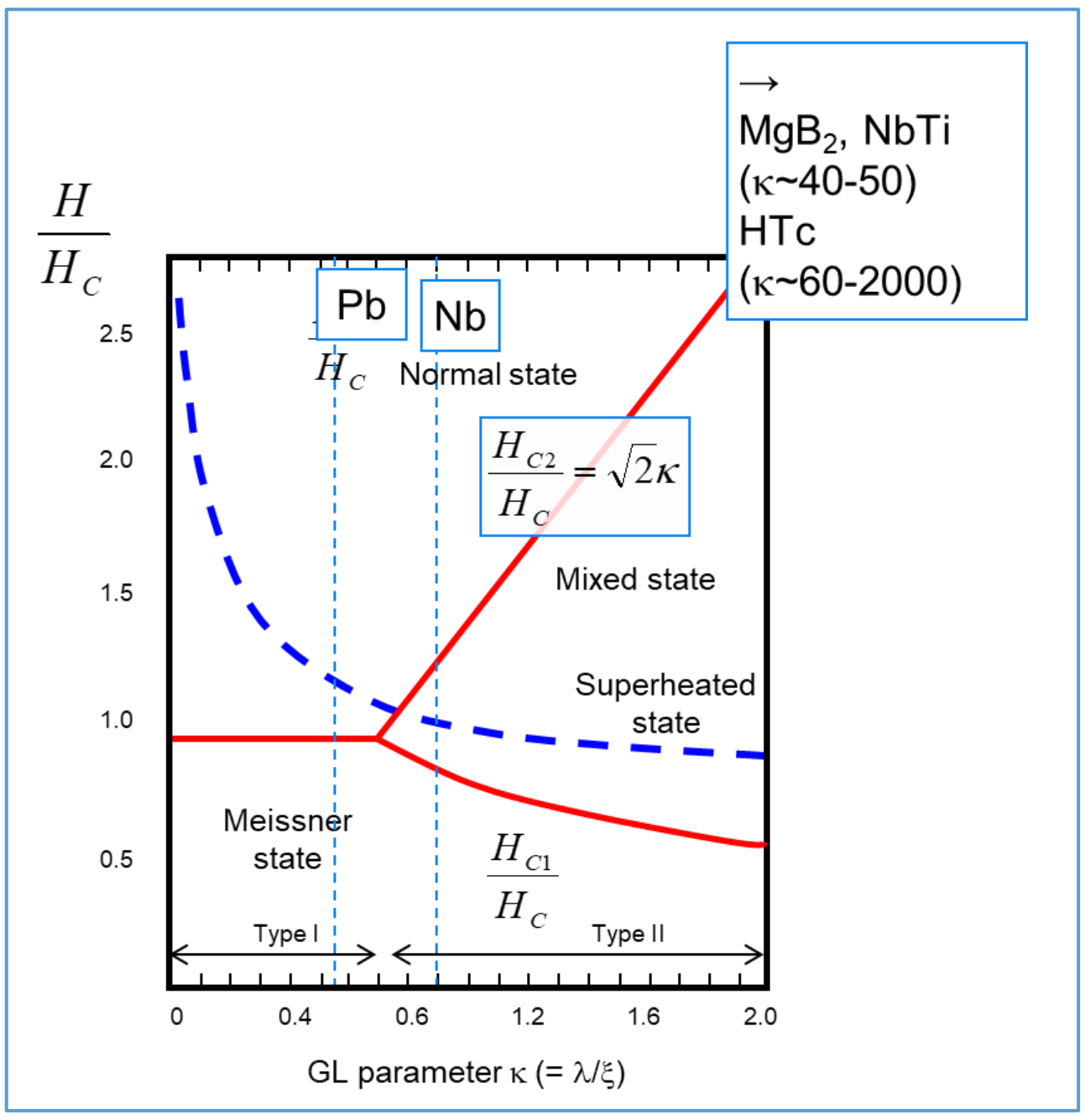

*Figure 11. Reduced field H/$H_C$ vs the Ginsburg-Landau Parameter $\kappa$. Superconductors with $\kappa > 1/\sqrt{2}$ are all type II. Nb is very close to the type I/type II frontier, and therefore exhibit a very high $H_{C1}$. Other practical type II exhibit much lower $H_{C1}$, which is more favorable for magnet applications. Blue dotted line figure superheating field, where Meissner state can be observed above its thermodynamic limits, as a metastable state.*

### 2.1.5 The importance of the electronic mean free path $\ell$

In the § 2.1.3, we have mentioned that λ and ξ depend on the purity and mechanical state of the superconductor. In fact, modifying $\ell$, one can tune the properties of the superconductor for a dedicated application, which is widely used for magnet wires. When $\ell$ decreases, ξ decreases while λ and κ increase, according to:

$$\frac{1}{\xi} = \frac{1}{\xi_0} + \frac{1}{\ell} \tag{2}$$

$$\lambda = \lambda_L.\left(\frac{\xi_0}{\xi}\right)^{\frac{1}{2}} = \lambda_L.\left(1 + \frac{\xi_0}{\ell}\right)^{\frac{1}{2}} \tag{3}$$

$\xi_0$, the Cooper pairs coherence length in the clean limit ($\ell$ >> ξ ), can be estimated with Ginzbourg-Landau model (§ 2.2.4) , calculated with BCS theory (§ 2.2.5):

$$\xi_0(T) = \frac{\hbar v_F}{\pi\Delta(T)} \text{ and } \lambda_0(T) = \sqrt{\frac{mc^2}{4\pi e^2 n_S(T)}} = \sqrt{\frac{m}{\mu_0 e^2 n_S(T)}} \tag{4}$$

In the dirty limit ($\ell$ ≤ ξ ), one has : $\xi \sim \sqrt{\xi_0 \ell}$ et $\lambda \sim \lambda_0 \sqrt{\frac{\xi_0}{\ell}}$

The size of a Cooper pair is assimilated to $\xi_0$ (at T=0) because it is a system coherence length. $\xi_0$ depends on T and diverge @ T~Tc (typical behavior of a critical phenomenon, $\lambda$ diverges as well - Figure 12). It also corresponds to diameter where order parameter is 0 inside a vortex: we will see that the $\xi$ parameter was initially introduced in the GL model to account for the variation of the order parameter at the vortex core.

NB. The size of a Cooper pair does not vary with T, only their density.

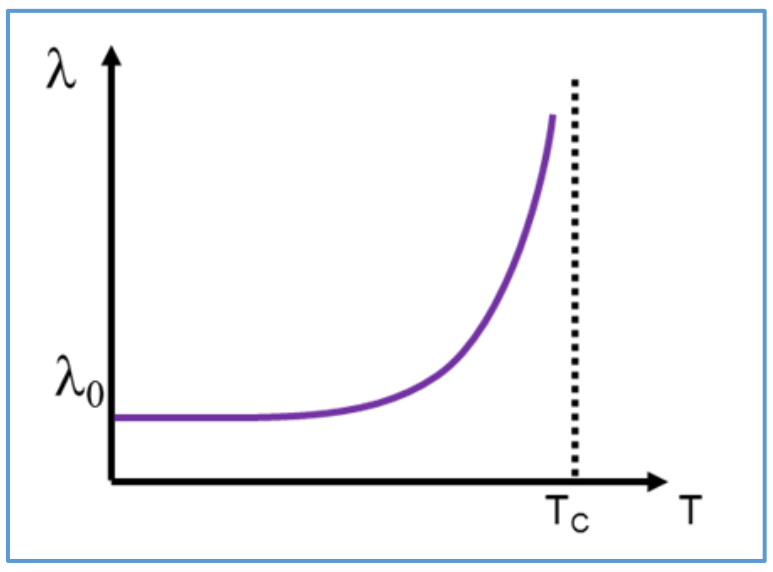

*Figure 12. Divergence of $\lambda$ close to $T_C$.*

## 2.2 Superconductivity: more details

Superconductivity remained a puzzle for decades for lack of finding a theory that could face all the physical aspects observed during the transition from the normal to the superconducting state. We present first the main experimental details that needed to be considered, and then an introduction to the various models or theories that have been proposed, along with their field of application. Indeed, there is no complete theory for superconductivity yet. In the literature, one can find many particular complex developments that apply to very specific conditions and have limited range of validity. It is the source of many apparent contradictions since various author do not use the same approximations. Let's admit that some of the theories are quite far for the real world, and have not yet applications.

### 2.2.1 Typical observations:

**1)** Jump on specific heat but not on latent heat in absences of field which means it is a $2^{nd}$ order transition[1]. The electronic contribution changes, but not the crystalline lattice one.

**2)** Isotopic Effect: the critical temperature of a superconductor behaves like Tc ~ $1/\sqrt{m}$, which tells there is a link with the atomic structure, but X-rays show that the crystalline lattice (atomic distances) is not affected.

**3)** There is a long-range order for electrons: there is an abrupt transition at Tc, which means that many e- are concerned. Another phenomenon, called proximity effect[2] is also resulting from this long-range order, but one will not detail it for now.

**4)** No change is observed in visible light absorption (which is related to normal conductivity $\rho$n), but an absorption band appears around $10^{11}$ -$10^{12}$ Hz (~ $10^{-3}$ -$10^{-4}$ eV, ~ some K) which traduce the opening of the superconducting gap, and is also related to Tc

**5)** The thermal conductivity drops since the e- that were involved in thermal conductivity are now grouped into Cooper pairs.

**6)** Mechanical state plays some role: isostatic pressure can play on Tc. Deformation, residual stress also plays on the electronic mean free path $\ell$ ($\ell$ has a huge role on SC properties).

**7)** No thermoelectric effects in type I SC because Cooper pairs do not transport entropy (but Vortices do ! see § 3)

---

[1] **1rst order Transitions:** heterogeneous systems, transition does not happen everywhere at the same time, e.g. boiling water= liquid + gas mixture. Involves latent heat, the free energy derivatives are discontinuous.
**$2^{nd}$ order transition:** homogeneous system, all parts transit at the same time, e.g. ferromagnetic transition, nematic transition in liquid Crystals. The free energy derivative are continuous, second derivative like Cp might diverge. Phases before and after transition have different symmetries. High energy phases are always uniform (s symmetry). Low temperature phase present some order (specific symmetry appears).

[2] Proximity effect is related to the coherence length of Cooper pairs. In short, a very thin normal conducting layer in contact with a superconductor can become superconducting. The full description of this phenomenon is beyond the scope of this lecture.

## *Gap opening: recalls on band theory*

*To form a molecule, one has to combine 2 atomic levels into 2 molecular levels: one "bounding" the other "antibonding". Bounding state energy is always lower than the initial atomic level energy, which is why a molecule is stable.*

*In a material many atomic levels have to combine, molecular states cannot occupy the same state of energy (Pauli exclusion principle) and come to lie close together, forming two bands (bounding and antibounding state are now called valence and conduction bands). In insulating and semiconducting material, the valance bound is full and the conduction band is empty, and electron can move only by jumping from the conduction band to the valance band across the "gap", which is several eV for an insulating material.*

*In metallic materials, the 2 bands overlap, leaving a conduction band filled only up to the fermi Level. Electron can move across the conduction band with a simple low energy thermal activation (~25 meV at room temperature).*

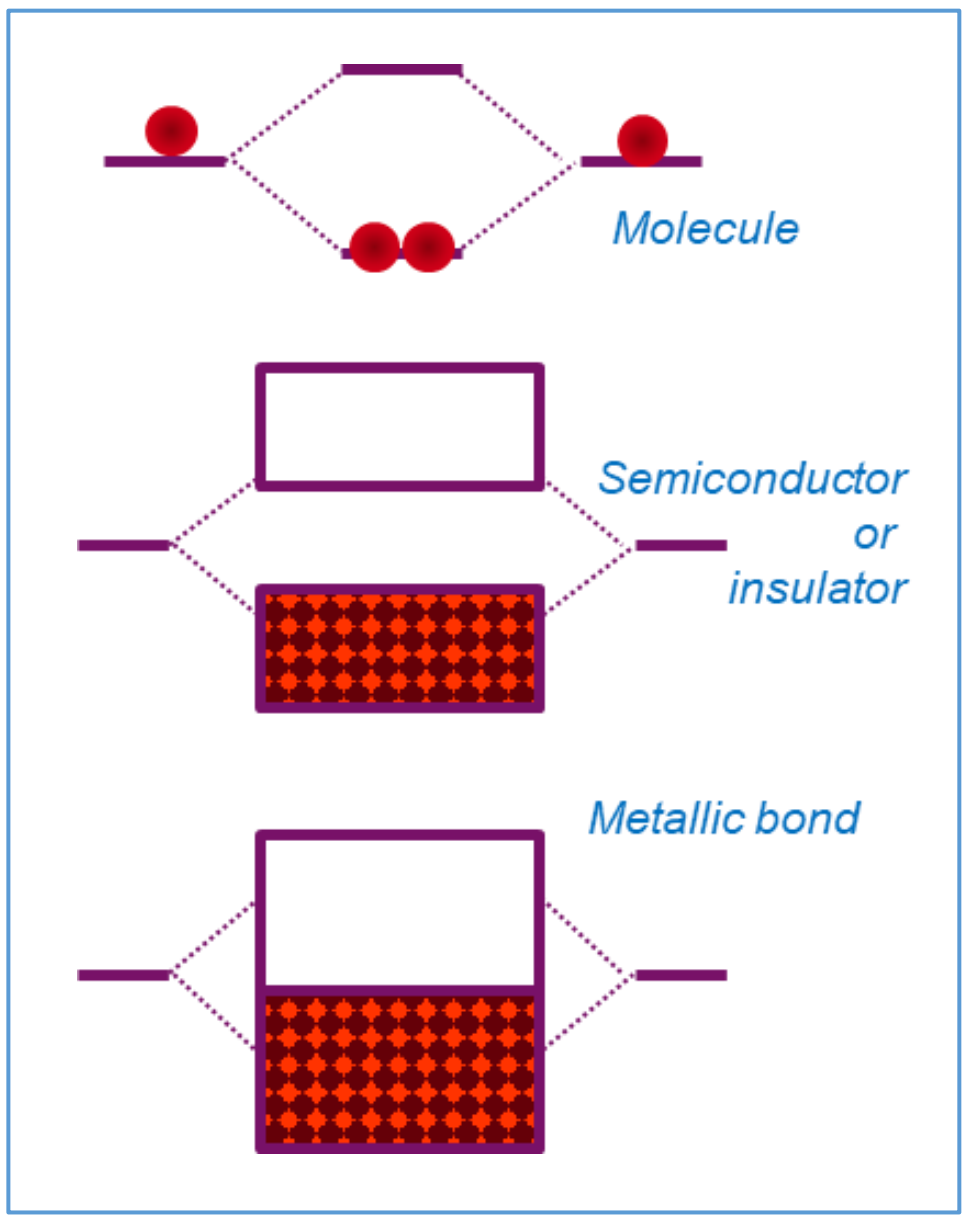

*Figure 13 : top: schematic combination of two atomic levels to form a molecule, Below, combination of many initial atomic level form energy bands. When they start to overlap, one obtains an unique band, filled only up to the Fermi Level EF (normal conducting metal).*

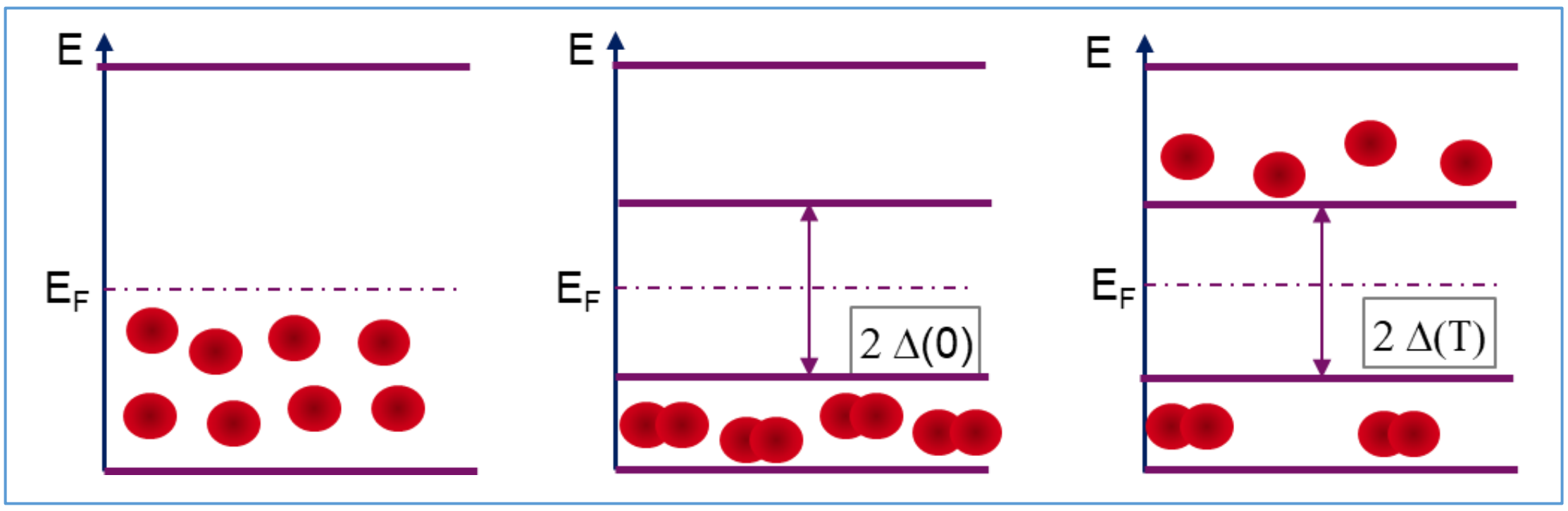

*Figure 14. Superconducting gap opening in the conduction band, close the Fermi Level*

*When reaching the superconducting state, this conduction band splits in two, opening a new gap (the superconducting gap, about 10-3-10-4 eV), forming a bounding state (the cooper pairs) and an antibounding state called quasiparticles that can be assimilated to normal electrons issued from broken Cooper pairs. At T=O K all the electrons from the initial conduction band are coupled into Cooper pairs. When T increases, some pairs dissociate by thermal activation. In DC those normal electrons are short-circuited, and the resistance of the superconductor is strictly 0; but in AC or RF, those normal electrons are at the origin of some dissipation. Note that the only electrons concerned by superconductivity are those close from the Fermi level. Core electron from lower energy orbitals do not participate in electrical conduction.*

### 2.2.2 Theories presentation

The first proposed models were macroscopic. They are the most used from the practical point. Then BCS theory was introduced, and several developments have been pursued to face its limitations. We will first make a general description, and the details each model further below.

**London: 2-fluids model (1935)**

The 2-fluids proposed by Gorter and Casimir (1934) to explain superfluidity in Helium. Then F. & H. London used it to explain the Meissner effect. It is a phenomenological classic theory. It explains Meissner effect and predicts λ (within a factor 2). It works well if **Js** is << $\mathbf{J_D}$ and $\boldsymbol{n_s}$ (superelectrons density) is uniform [1]. It is valid ∀ T, but low B since at high B: nonlinear behavior appears. Predictions are not very accurate for thin films.

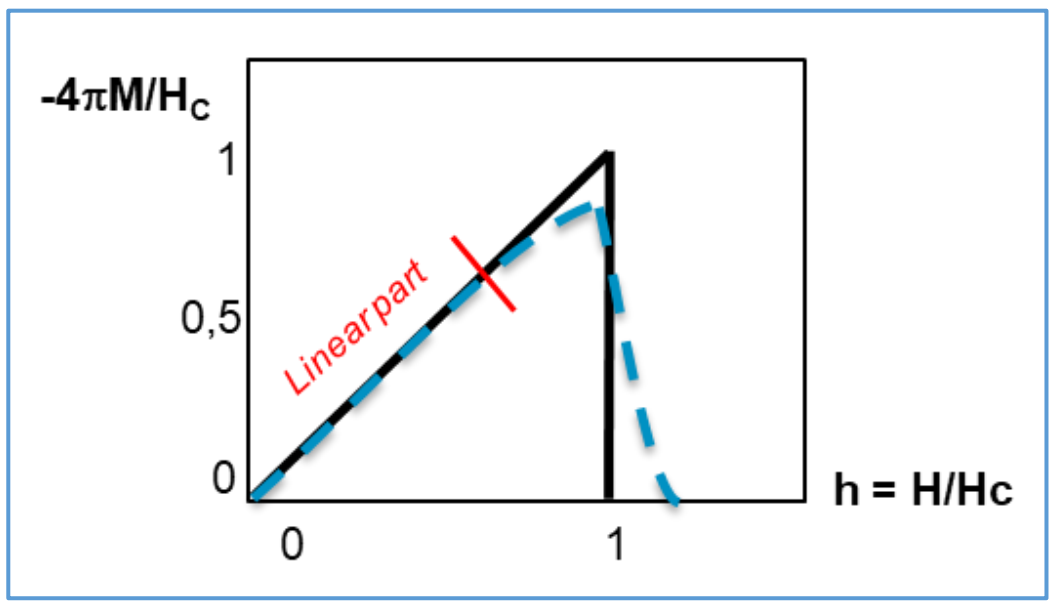

*Figure 15. Experimental magnetization curves compared to the theoretical model*

**Ginzburg landau (1950)**

This model is based on the combination of Electromagnetism + Thermodynamics, in particular second-order phase transitions. It introduces quantum aspects: one single Wave function for all Cooper pairs, the coherence length ξ, and the order parameter ψ called "wave function of condensed electrons". This order parameter measures the symmetry breaking energy U when the superconducting state appears. Condensation energy is $\frac{1}{2}\mu_0 Hc^2$. In the presence of H, $n_s$ can vary, the free energy can be written as a Taylor development. The solution is found by minimizing ψ (the wave function) and A (the vector potential) over all space, but developing order 1 or 2 is valid only near transition (T~$T_C$ or if T~ 0 K, H~$H_{C2}$). It also introduces some non-linear behavior. Numerical predictions are similar to London, but the prediction of thin films behavior is better. When H<<$H_{C2}$, ψ is ~constant in space.

The GLAC (Ginzburg-Landau-Abrikosov- Gor'kov) model is an extension of G.L. for κ >> 1,2 where the mean free path ℓ is very small ("dirty" SC).

These London and G.L. models rather apply to type II superconductors. For type I, the developments were proposed by Pippard.

**BCS (Bardeen-Schrieffer-Cooper), (1957)**

BCS is the most complete theory. It is a local (microscopic), quantum theory. It explains the main aspects of SC (gap, Tc, ….) and works at arbitrary l, arbitrary T, arbitrary H. The Cooper pairs are electrons with opposed moments p and spins (at the instant t), there is one single wave function for all Cooper pairs, it is a highly correlated system with many exchanges between pairs. The total energy of the interacting pairs remains constant although their moments are constantly changing (quantic behavior, delocalization: a cooper pair does not concern two determined individual electrons but two electrons with opposed moments and spins at each instant t).

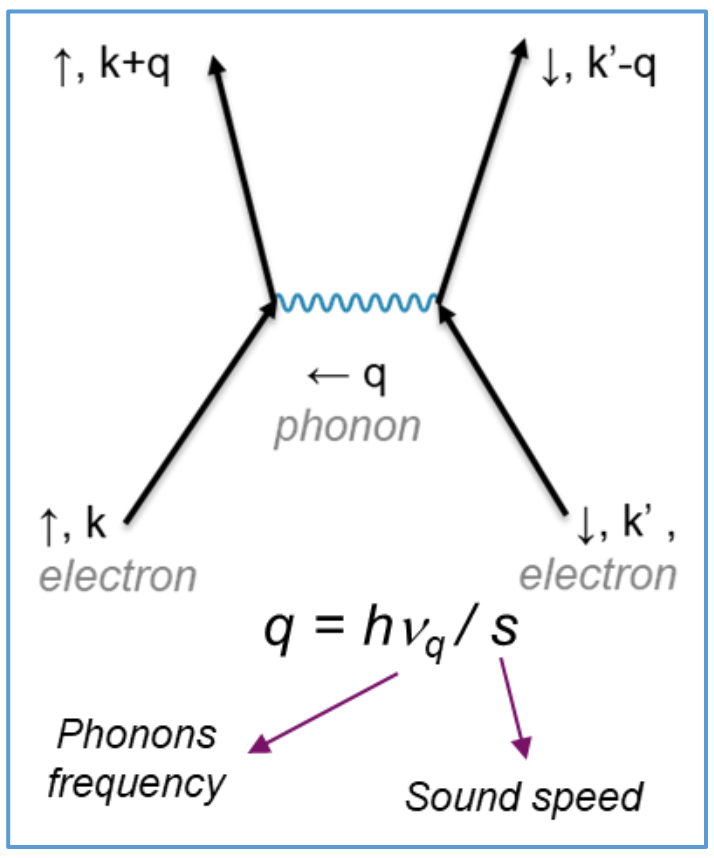

*Figure 16. Feynman electron-phonon interaction diagram.*

BCS theory is essentially used to make exact calculation and comfort GL hypothesis (same results at T~$T_C$). In general, it is not used for engineering predictions (too complex). Compared to G.L. model, BCS introduces the role of phonons, the existence Cooper pairs, and the BCS gap (slightly different from the one calculated in G.L). The e-/ phonon interaction is replaced with an effective potential and the use of second order perturbation theory. Nonetheless, BCS is not a complete theory; it works mainly for low coupling systems (low κ type I SC). High coupling systems like Nb and other type II need corrections.

[1] Not always true, see §3

**Extensions from BCS**

If there is a strong interaction between pairs, one can no longer make a perturbative treatment, it is necessary to make a variational approach. ***Eliashberg*** model directly takes into account the coupling e- / phonon.

Several other specific developments have been made, for instance:

**Bogolubov-De Gennes:** Arbitrary l, Arbitrary T, arbitrary H, weak coupling electron-phonon, order parameter depends on position. 1965

**Eilenberger:** Arbitrary l, Arbitrary T, Arbitrary H, type II. 1968 Semi-classical theory, still not valid at lower T, "clean " type II : takes into account some inhomogeneity

**Usadel:** Eilengerber with approx. dirty limit: λ >> ξ. 1970

**Radiofrequency**

Response to EM wave has been treated by ***Mattis-Bardeen*** in 1958 at weak field, in the Pippard (ξ>>λ) approximation, and using the 2-fluid approach ($\sigma = \sigma_N + i\sigma_S$) [2]. The extension for type II was proposed later on [3].

High current density ($J \sim J_D$): "non linear $R_{BCS}$" model has been developed by ***Gurevich*** in 2006, only for clean type II SC [4].

### 2.2.3 London: 2-fluids model (1935)

Conductivity has 2 components: $\vec{J} = \vec{J_n} + \vec{J_s}$, with normal electron density = $n_n/n$ and superfluid electrons density $n_s/n = 1 - n_n/n$ as shown in Figure 17. The normal conducting electrons are short circuited by superconducting ones.

$$\boldsymbol{J_s = n_s\, e\, v_s} \qquad \textbf{(5)}$$

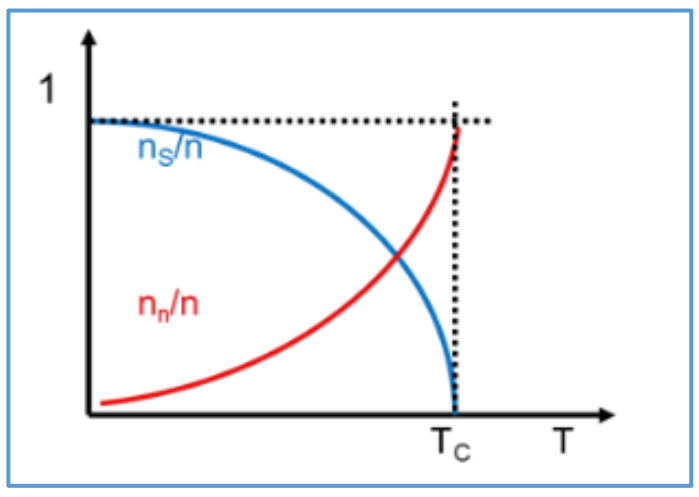

*Figure 17. Population of super-conducting and normal-conducting electrons vs temperature*

e is the electron charge, and $v_s$ is the Cooper pairs velocity.

The current is constant, electric field $\vec{E}$ cannot appear in the superconductor because otherwise, the super electrons would be continuously accelerated and the current would increase indefinitely. Only the super electrons carry the current in steady state, and they do not scatter on the impurities thanks to coherent state (all pairs are correlated, so to break one pair, it would be necessary to provide energy to destroy all pairs).

$$\overrightarrow{\mathbf{m\dot{v}_s}} = -\mathbf{e}\vec{\mathbf{E}} \qquad \textbf{(6)}$$

Equation (3) represents the electrodynamics acceleration of the electrons by $\vec{E}$.

By combining Maxwell equations with this 2-fluids approach one gets the 1rst London Equation (4):

$$\frac{\mathbf{d}\overrightarrow{\mathbf{J_S}}}{\mathbf{dt}} = \frac{\mathbf{e^2 n_S}}{\mathbf{m}}\vec{\mathbf{E}} = \boldsymbol{\lambda}_\mathbf{L}^2\vec{\mathbf{E}} \qquad \textbf{(7)}$$

With

$$\vec{J_S} = \frac{\vec{A}}{\mu_0 \lambda_L^2} \tag{8}$$

Thus

$$J(x) = \frac{H(0)}{\lambda_L} e^{\left(-x/\lambda_L\right)} \tag{9}$$

First London equation shows that the electric field differs from zero only when the current density varies over time. It describes the ballistic flow of superelectrons in place of Ohm's law (which describes the viscous flow of normal electrons in a metal) and gives Js as a function of the vector potential $\vec{A}$.

This equation suffers from the disadvantage that it is not gauge invariant, but is still true in the Coulomb gauge, where the divergence of A is zero. This equation holds for magnetic fields that vary slowly in space.

In the stationary case: one can neglect the displacement currents in Maxwell's 2nd equation, then one replaces rotE by its Maxwell equivalent in the second equation and one integrates according to time to get the second London equation (7):

Rotational of (4) $\nabla \times \frac{d\vec{J_S}}{dt} = \nabla \times \lambda_L^2 \vec{E}$ c ombined with Maxwell : $\nabla \times \vec{H} = \vec{J_S}$ ;

$$\nabla \times \vec{E} = -\mu_0 \frac{d\vec{H}}{dt} \tag{10}$$

London explains Meissner effect but underestimate λ. In this model, the magnetic field penetrates over $\lambda_L$:

$$B = B(0) e^{\left(-x/\lambda_L\right)} \text{ with } \lambda_L = \left(\frac{m}{e^2 n_s(T) \mu_0}\right) \tag{11}$$

Within this depth a complete screening of $H_0$ occurs.

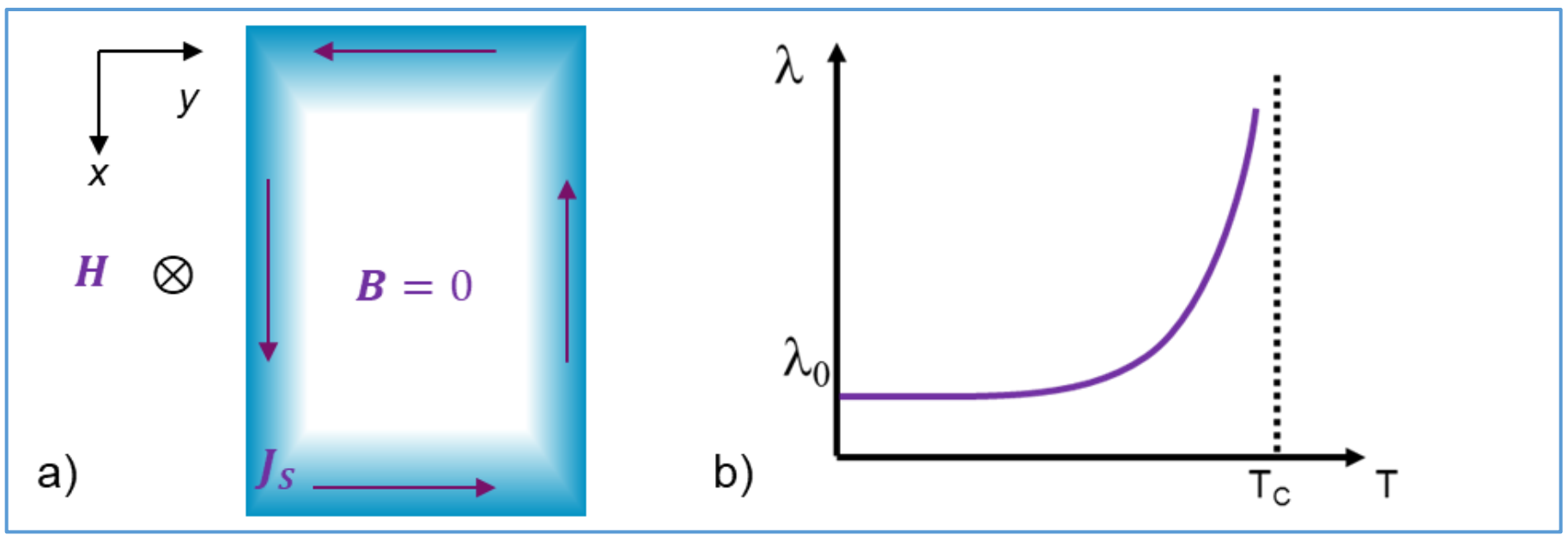

*Figure 18.a) Induction is equal to 0 in a superconductor except on a very thin surface layer where supercurrents occurs. b) Divergence of λ close to $T_C$. In the normal state, the penetration depth is replaced by δ and is worth several µm.*

The surface current density cannot exceed the depairing current density $J_D$. $J_D$ can be defined as the current density sufficient to brake Cooper pairs.

$$J_D(T) = \frac{H_C(T)}{\lambda_L(T)} \sim J_0 \left(1 - \frac{T^2}{{T_C}^2}\right)^{3/2} \tag{12}$$

With $H_C$: thermodynamic critical current.

*Notes*

- *In the coherent state, all Cooper pairs are correlated. So to break 1 pairs, it would be necessary to provide energy to destroy all pairs*
- *As the temperature increases from 0 to the critical temperature Tc, the density ns/n decreases from 1 (all e- from the conduction band are in form of Cooper pairs) to 0 (all Cooper pairs are broken) when the material becomes normal conducting (*Figure 17*).*
- *If the current is constant, electric field E cannot appear inside the superconductor because otherwise, the super electrons would be continuously accelerated and the current would increase infinitely.*
- *If there is no electric field, the normal electrons are at rest: only the super electrons carry the current in steady state, and they do not scatter on the impurities thanks to coherent state (see above)*
- *All Cooper pairs have the same phase in absence of current. A phase gradient appears only if a current appears*
- *Predicted $\lambda_L$ are smaller than experimentally measured $\lambda$ as London model does not take into account local effects such as pinning (see § 3)*

### 2.2.4 Ginzburg-Landau (1950) (GL model)

The GL is based on a mean field gauge theory, which is a universal approach for critical phenomena such as phase transitions. It consists into finding a macroscopic order of parameter $\psi$ that describes all freedom degrees of the system at the critical point. It is generally be guided by geometrical consideration (Gauge theory[1]). The second step consist into building up an effective free energy formula F[$\psi$] in the mean field approximation, i.e. where it is supposed that spatial and thermodynamic fluctuations of $\psi$ keep small compared to the system size. F[$\psi$] must respect microscopic symmetries of the system (gauge invariance).

Then the energy of the system U($\vec{r}$) can be replaced by g.$\delta(\vec{r})$ that can be treated as a weak perturbation (small d° in $\psi$ Taylor expansion). One chooses the effective potential g with same symmetry properties in the considered energy range, either with intuition, or either by minimizing the Free Energy. This approach allows providing the fundamental configuration $\psi_0$ and reconstructing the phase diagram.

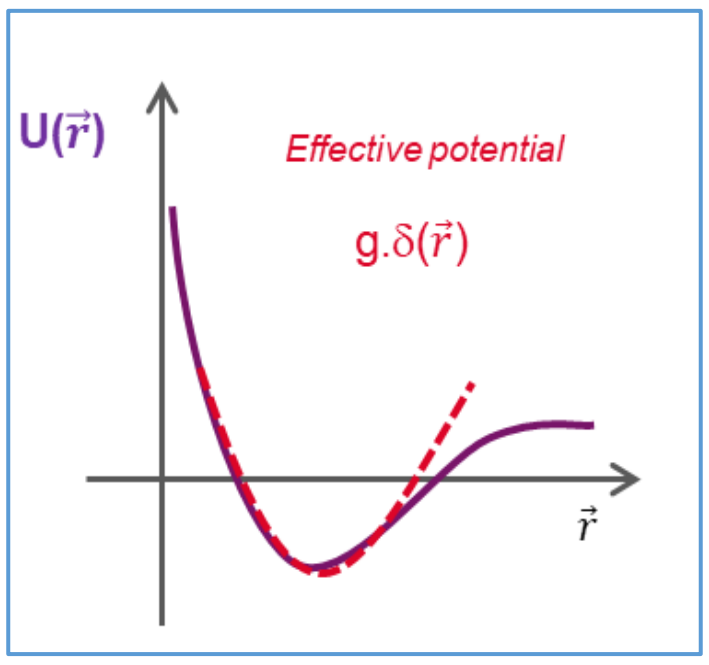

*Figure 19. Simplified effective potential compared to actual potential*

The resulting equation is equivalent to an effective Hamiltonian (so-called "Ginzburg-Landau" H)

The expression for the Free energy takes the form:

$$\mathcal{F}_0(T,\Psi,\vec{A}) = \mathcal{F}_0(T) + a(T-T_c)|\Psi|^2 + \frac{b}{2}|\Psi|^4 + \frac{1}{2m^*}\left|\left(\frac{\hbar}{i}\vec{\nabla} - e^*\vec{A}\right)\Psi\right|^2 + \frac{B^2}{2\mu_0} - \vec{H_0}\cdot\vec{B} \qquad (13)$$

[1] Gauge theory : field theory based on a local symmetry group (Gauge group) which defines a "gauge invariance". It applies well for phase transitions: usually the high temperature side is symmetric, and there is a symmetry breaking at the transition (upon decreasing T).

The term $i\hbar\vec{\nabla}$ takes into account the spatial fluctuations of ns. The odd terms are ≠ 0 only in presence of the field H (1rst order transition). At H=0, J=0 and $\psi$ is real and $\xi$ can be derived. The limit conditions at surface are J=0, B-H =0.

The minimization of Equation **(13**) requires long and complex calculation that will not be detailed here.

It results into the two Ginzburg-Landau equations:

$$\alpha\Psi + \beta\Psi|\Psi|^2 + \frac{1}{2m}\left[\frac{\bar{h}}{i}\vec{\nabla} - q\vec{A}\right]^2 \Psi = 0 \qquad (14)$$

$$\vec{J} = \frac{i\vec{h}q}{2m}\left[\Psi\vec{\nabla}\Psi^* - \Psi^*\vec{\nabla}\Psi\right] - \frac{q^2\vec{A}}{m}|\Psi|^2 \qquad (15)$$

These coupled nonlinear differential equations can be solved only numerically. Moreover they are valid only close to the transition (*e.g.* $H\sim H_{C2}$ *and* $T \sim 0$ *or* $H\sim 0$ *and* $T\sim T_C$). They accounts for Meissner state in type I et II SC (supercurrents), flux quantification, macroscopic quantic effects… and allow $\lambda$ and $\xi$ estimation.

The fact that F[$\psi$] is Gauge invariant has consequences on the shape $\psi$ can take: it must remain invariant under an arbitrary, constant phase multiplication: $\Psi = |\Psi|e^{i\Theta}$ with $|\Psi| \sim \sqrt{n_s}$.

When the applied field Ha = 0 and the current density =0, $\psi$ is real and $\xi$ can be derived. The boundary limits at the surface are: J=0, B-H =0.

**Non local effects:**

- Ginzburg- Landau proposes a local relation between current and potential vector $\vec{A}$, but in reality the exact relation is nonlocal, unless $\vec{A}$ or the current have slow variations on the scale of $\xi_0$, e.g. for superconductors type II with $\ell >> \xi_0$ (clean type II SC). In this case, one has the right to use this approximation. For dirty type II SC, a specific calculation has been developed initially by De Gennes et al [5].
- The magnetic penetration depth, named now $\lambda_0$, differs from the theoretical LONDON penetration depth $\lambda_L$ because of two very different mechanisms [6]:
  - One intrinsic, due to the nature of the Cooper pairs: the proportionality relationship between **j**(**r**) and **A**(**r**) is a local expression while the electrons of a Cooper pair can be several nm apart and "feel" a different local potential vector.
  - The other extrinsic, that adds to the previous one when the superconductor has a high concentration of impurities reducing the mean free path (dirty SC), described by De Gennes [5].

***Thermodynamic recalls:***

- *In absence of field, the system, is described by Helmotz free energy F= U-TS while in presence of applied field Ha, the system is described by Gibbs free energy G = H-TS-$\mu_0$HaM. T = Temperature, S = Entropy, U =internal energy, H = enthalpy.*
- *When Ha =0 => G ≡ F.*
- *When Ha ≠0 => Helmotz free energy becomes Gibbs free energy G = F-BHa where BHa term corresponds to magnetic forces work.*
- *At the transition Gs = Gn and free energy is minimal.*

***General approach for phase transitions:***

- *Close to transition temperature $T_0$, $\psi$ is small: one can develop/expand:*
  *$F(\psi,T) = F_0(T) + A_0(T)\ \psi + A(T)\ \psi^2 + B(T)\ \psi^3 + C(T)\ \psi^4 + \ldots$where F0 is the energy of the high temperature phase.*
- *For $T > T_0$, F must be minimal for $\psi = 0$ et $A(T) > 0$.*
- *For T < T0, F must be minimal for $\psi \neq 0$ et $A(T) < 0$ (concavity rule).*
  *The simplest choice is $A(T) = a(T - T_0)$ with $a > 0$.*
- *For a second order transition one can demonstrate that odd terms are equal to 0: $A_0(T) = B(T) = 0$, and $C(T) = c > 0$, so $F(\psi,T) = F_0(T) + a(T - T_0)\ \psi^2 + c\ \psi^4$, $a > 0$, $c > 0$*
- *For $T > T_0$, the only solution is $\psi = 0$. For $T < T_0$, there is 3 solutions: one maximum at $\psi = 0$ and two minima at $\psi = \pm\ a(T_0 - T)/2c$. Close to $T_0$, the order parameter behaves like $T_0 \sim T$.*

***Same approach is possible for all other phase transitions: for example***

- *• Ferroelectric transition: the parameter of order ⍰ = Ms (spontaneous magnetization).*
- *• Water vapor transition: the parameter of order ⍰= n the density of H2O.*
- *• In the case of superconductivity the electron correlation function is = 0 in normal conducting ste and ≠0 in the superconducting state. It can thus be chosen as the order parameter.*
- *• Also used for Higgs Mechanism in elementary Particles*

---

**<u>Linear ginzburg-Landau equations</u>**

This approach has been developed for high $\kappa$ superconductors for magnet applications at low temperature but high field (close to $H_{C2}$). In this case, $\xi$ (T) et $\lambda$ (T) are very large since they diverge for H~$H_{C2}$ and $\xi$, $\lambda$>> $\xi_0$(T=0). At $\lambda$ and $\xi$ scale one can assume that changes in $\psi$ and $\Delta$ at a Normal-Superconducting interface are ~linear (see Figure 20).

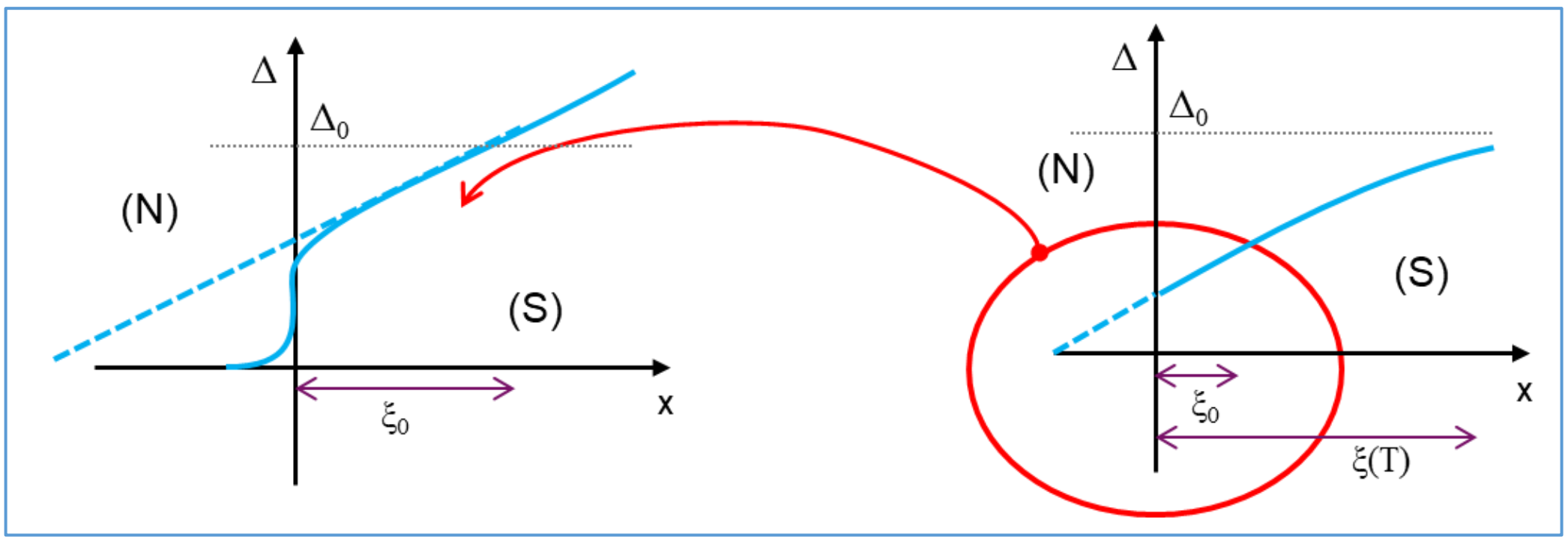

*Figure 20. Gap energy behaviour close to a N/S interface (for the recall $\Delta$ follows the superconducting electron density $n_S$).*

It applies only for very inhomogeneous cases (e.g. $H \sim H_{C2}$, or magnetic impurities…):

- When $H \sim H_{C2}$ (upon decreasing H), the SC phase just appeared, $\Psi$ is very small. Nonlinear terms in $|\psi|^4$ can be neglected, only $|\psi|^2$ terms remain
- Screening of magnetic field can be neglected (H>> $H_{C1}$) so $\vec{A} \sim \overrightarrow{A_{appl}}$ (potential vector term)

GL equations becomes decoupled and Free energy minimization becomes:

$$\xi^2\,(T)(\nabla + ik_A)^2\psi(\vec{r}) = -\psi(\vec{r}) \tag{16}$$

Equation (16) is analogous to Schrödinger equation for a particle with mass 2m, charge 2e. Quantum mechanics solutions can be applied, giving the shape of the wave function: $\psi(x) \sim Ae^{ikx}$.

In general, each specific case needs its own development, using continuities solutions, or variational methods. For instance, in "dirty" SC the translation invariance is lost but one add one more approximation (average value for all the impurities). The potential term due to impurities is then treated like a weak perturbation, and so on... There are many practical calculation/developments in superconductivity, and it is sometimes difficult to know is they are applicable to other particular cases.

Note :

Many "famous" textbooks on superconductivity are in fact focused on magnet applications. For instance "Superconductivity of metals and alloys" by Nobel Prize awarded P. G. De Gennes, starts with a typeI/Type II description, but nearly all the following mathematical developments apply then to very high κ SC (dirty SC) and are only applicable for magnet developments.

### 2.2.5 BCS theory elements.

The main difficulty to be overcome was that Coulombian repulsion, even if it was attractive, is so intense that the gap should not be so small; one had to find another mechanism of weak attraction. In 1950 Frölich showed that emission/absorption of a phonon can create an attractive interaction below a certain frequency, with the proper order of magnitude (see Figure 21).

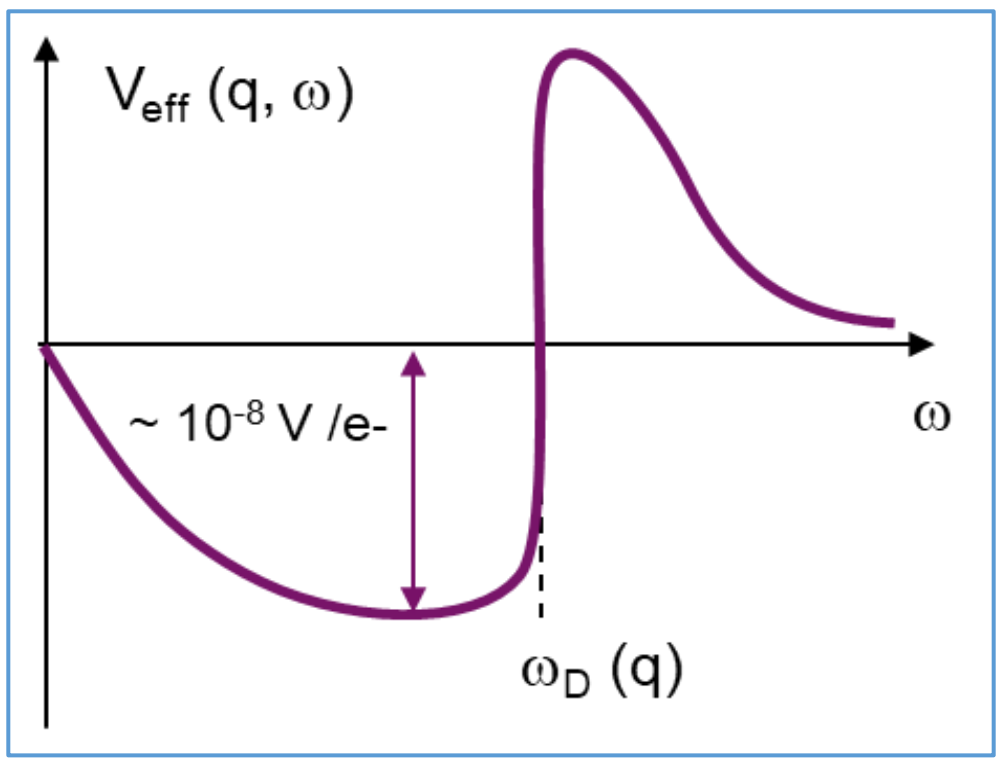

*Figure 21. Effective potential due to emission absorption of phonon vs the phonon frequency ω.*

Once this problem was solved, Bardeen, Cooper and Schrieffer proposed a variational approach based on the Jellium model:

First, one considers an effective potential due to all charged particles (ions + electrons), considered as a continuous media. It is equivalent to a Screened Coulomb Interaction.

Secondly, one considers delays due to phonons since ions move slower than electrons.

One builds an Effective Hamiltonian H = $H_0$ + V, where $H_0$ = Hamiltonian for system without interaction (jellium) and V is Perturbative potential that represents interaction with phonons.

As shown on Figure 21, only electrons close to the Fermi level do "feel" a positive interaction and can form a bound state (the Cooper pairs), so one can introduce a further simplification: only e- with energy Ev= $E_F \pm 2\hbar\omega D$ are submitted to a potential ~ –V/2 (Figure 22).

In the simplified model, one considers that the electrons from a small region of the phase space around the Fermi surface interact with each other by a constant attractive interaction (-V / 2).

The matrix diagonalization produces new wave functions with lower energy compared with the wave function without interaction (and the difference is $\Delta$).

At the end, one obtains a variational wave function that describes the creation of Cooper pairs with $c^{\dagger}_{k,\uparrow}c^{\dagger}_{-k,\downarrow}$operators:

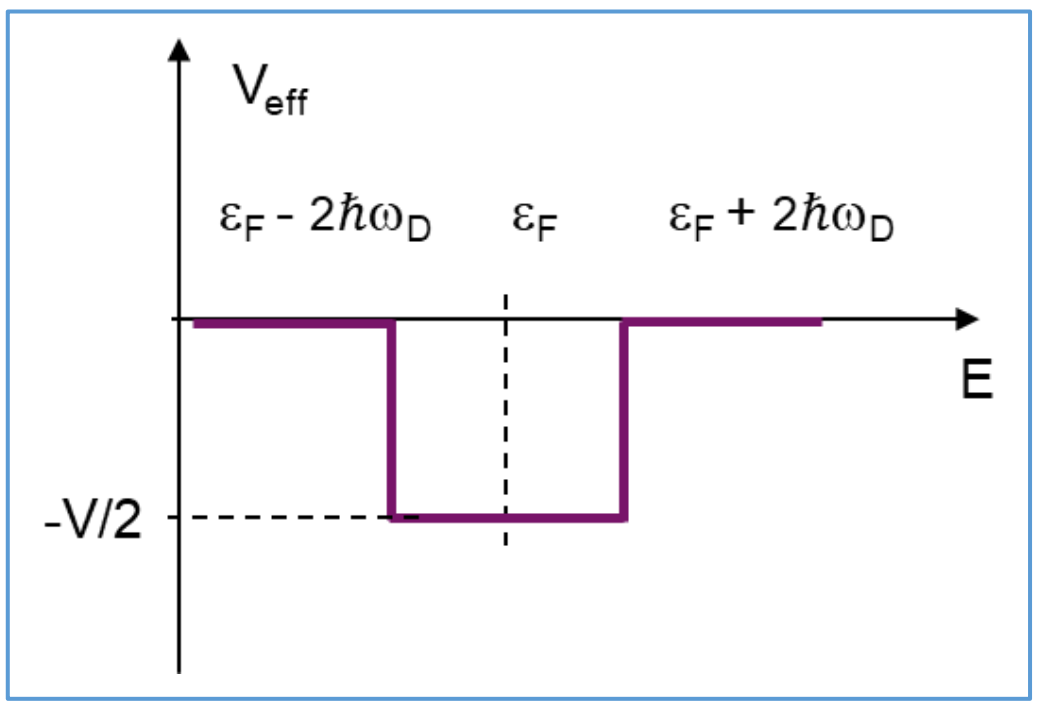

*Figure 22. Simplified attractive potential for electrons due to phonon interaction*

$$\psi\rangle=\prod_k(u_k + v_k c^{\dagger}_{k,\uparrow}.\,c^{\dagger}_{-k,\downarrow})|0\rangle \qquad (17)$$

Where $u_k$ are the $H_0$ matrix elements and $v_k$ are the V matrix elements

It explains the Cooper pair condensation, often described as a Bose-Einstein condensation of bosons, but it is not exactly one: the pairs are non-localized; they all have the same moment. It also explains the Phase lock-in and the gap existence $\Delta \sim\langle c^{\dagger}_{k,\uparrow}.\,c^{\dagger}_{-k,\downarrow}\rangle$[1]. The particle number is not fixed: one can create/annihilate pairs (uncertainty principle).

BCS theory was able to explain and confirm results obtained by the previous models, but in practice, it is too complex for practical applications where simpler models are still in use.

Notes:

- All Cooper pairs have the same phase if there is no current. A phase gradient appears only if a current appears
- The equations are invariable by reversal of time unless presence of magnetic field.
- In BCS approach, one has identical individual wave functions for each pair but a correlated system. The entire superconducting state has the same phase. When B $\neq$ 0, a non-uniform phase gives rise to a current flow. Now, because this current arises from minimizing the Ginzburg-Landau free energy, it must be an equilibrium property, and cannot dissipate energy.
- It also takes into account the interactions between pairs and cooperative effects, (analogy with superfluid He: Cooper pairs flow as a superfluid). The system is correlated because many Cooper pair overlap, which is the source of macroscopic phase coherence)
- Phase coherence is the same starting idea as G.L. approach, but in G.L., one has a single wave function; it does not account for exchanges between pairs.
- Gor'kov has shown that GL, although initially phenomenological, can be derived from the BCS theory. Near Tc GL and BCS give the same results, but GL simpler to handle.
- As with GL, it can be shown that changing the state of a pair is equivalent to destroying the state of all the system (and it would require enormous free energy: $k_BTc$)
- BCS concerns weak coupling superconductors like type I Pb or Hg. Numerical applications are not very good for superconductors with strong coupling (one moves away from the relation $2\Delta$ (0) = 3.54 $k_BTc$). It does not work either for HTC superconductors (YBCO and Pnictides)
- BCS approach also applies many other problems with n quantum bodies: e.g. ultra cold atomic gases, core of neutron stars, nucleons pairing , quarks, Anderson-Higgs Cooper pairs symmetry break (= 2 e- of spin and quasi-pulse opposite).

---

[1] $\Delta$ exhibit the same symmetry as the GL order parameter $\psi$ , but has not exactly the same value for the gap $\Delta_{GL}$

### BCS and quasiparticles

Quasiparticle notion is beyond the scope of this lecture. For those interested, we provide here some hints about its use in BCS formalism.

The formalism of BCS, although it concerns the coupling of electrons into Cooper pairs (Bosons), proposes an expression of the condensed state wave function as a linear combination of one electron states (Fermions) with a correlation among electrons of opposite moments and spin. The opening of the gap appears in the so-called "one electron energy spectrum".

At first order, one can consider that this formalism concerns only normal electrons from broken Cooper pairs, as was introduced in §2.2.1. In fact, when describing the normal electrons issued from "broken" pairs, one uses the term "quasiparticles", which is also used in band theory in electronics. It is a concept where a particle is described with an effective mass and effective momentum that takes into account its interaction with the lattice. In the case of Cooper pairs, it is even a little more complex: one Cooper pair breaks into two quasiparticles and each quasiparticle is a neutral combination of one electron and one hole. Since it is a quantum mechanics concept, the number of particles is not set, contrary to actual e- when considered as a classical particle [7].

The density of thermally excited quasiparticles is proportional to the density of states times the Fermi distribution function f(ε), where ε is the energy relative to the Fermi energy $E_F$ (see Figure 23). It might seems strange that the calculation concerning paired electrons is set as a function of the Fermi electronic density, but it is very effective.

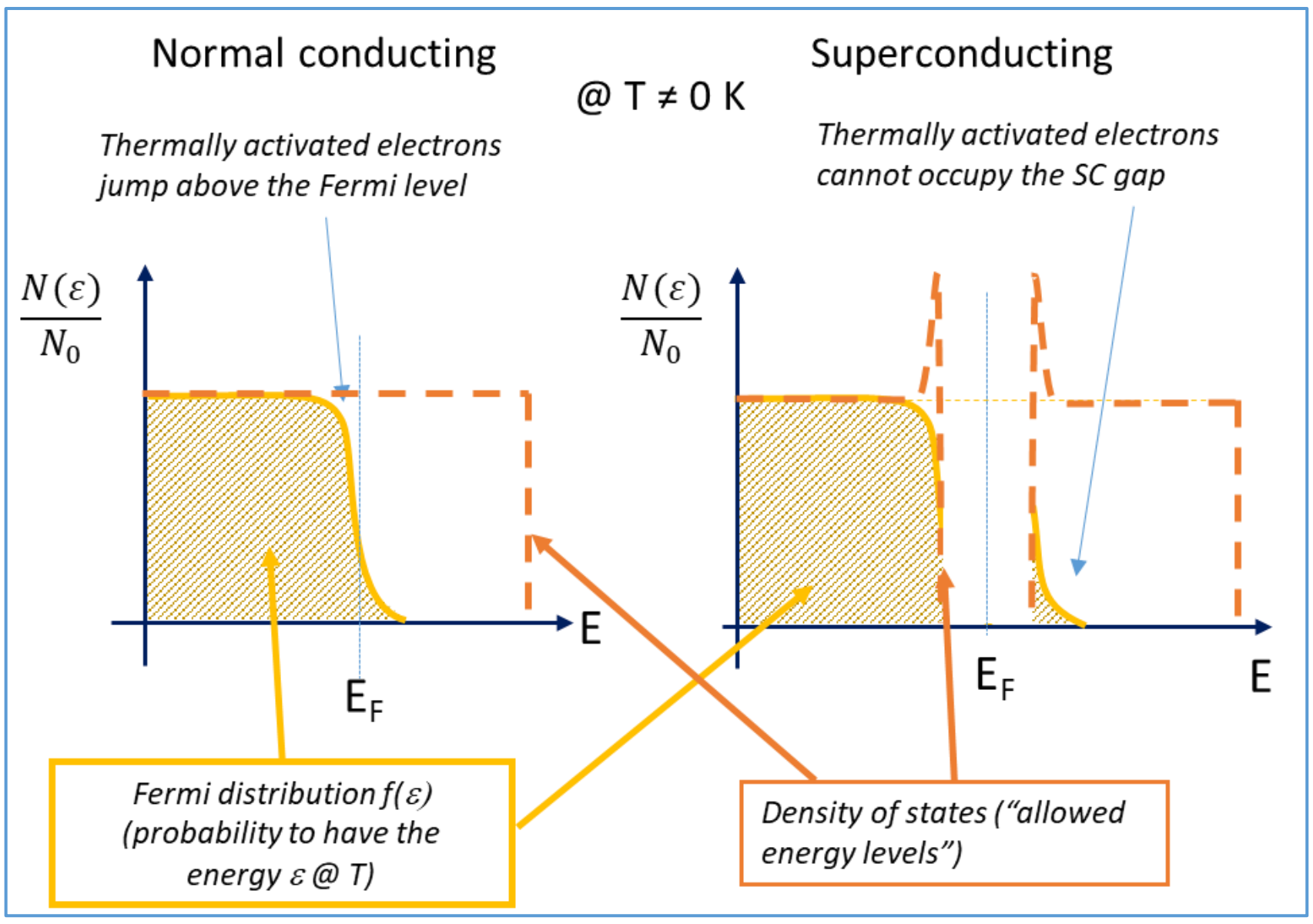

*Figure 23. Density of thermally excited quasiparticles (i.e. related to single electrons).*

In classical BCS, f(ε) only includes thermal activation and neglects other pair breaking mechanisms, like the influence strong current and/or strong field (which is the case in SRF), or the presence of magnetic impurities…. To treat other scattering mechanisms, one can play over modification (with due reasoning) of the shape of f(ε). Another approach to treat the influence of high current is to consider that the value of the gap is influenced by the displacement velocity of the Cooper pairs $V_s$ and their momentum $p_f$ (18):

$$\Delta\ (V_s) = \Delta - p_f\ /\ Vs\ / \qquad (18)$$

Thus at high current, the gap decreases and more and more pairs are broken by thermal activation [8].

NB. Breaking a Cooper pair by thermal activation can be considered as the absorption of a "thermal" photon. Table 2 gives the order of magnitude of the corresponding needed energy.

*Table 2. Thermal photon energy in function of temperature.*

| Thermal energy | 2K | 4 K | 80 K | Comparison w. 1 GHz photon | Comparison w. 100 GHz photon |
|---|---|---|---|---|---|
| $k_BT$ | 0.17 meV | 0.34 meV | 7 meV | 0.004 meV | 0.4 meV |

### 2.2.6 AC and RF

In dynamic mode (AC and / or pulsed), the current is carried by both types of charge carriers (Cooper pairs and normal conducting electrons) and the process is dissipative. The electrical properties of an AC superconductor can be seen as a perfect inductance in parallel with a resistance[1]. The normal electrons (issued from Cooper pairs broken by thermal activation) are accelerated and produce some dissipation[2].

This case has been treated by Mattis and Bardeen (MB) in 1958. They looked at the anomalous skin effect (i.e. the case where the field varies in amplitude over the mean free path) in presence of scattering centers, in a formalism that applies for both normal and type I superconductors [2]. The formalism was then extended to include effects of anisotropy and strong coupling [3]. Note that MB approach is valid only at low field. Several attempts have been proposed more recently to account for high field effects (see [9] and references inside)

As in the two-fluid model from London, one introduces the complex conductivity $\sigma_1$-i$\sigma_2$ in place of $\sigma_n$. The expression of surface impedance is derived based on BCS calculation (quadruple integral in energy, reciprocal space, averaging RF cycle and penetration depth!)

Their derivation produces two integrals ($\frac{\sigma_1}{\sigma_n}$ and $\frac{\sigma_2}{\sigma_n}$) that can be only numerically solved.

It is nevertheless possible to pursue London approximation to estimate the $R_{BCS}$ (for details, see e.g. [10]). As can be seen on Figure 24, the approximation is good enough for mean free paths below $10^3$ Å (100 nm). Note that London approximation for $R_{BCS}$ also deviates from experimental data for $\omega$ above 3 GHz [3].

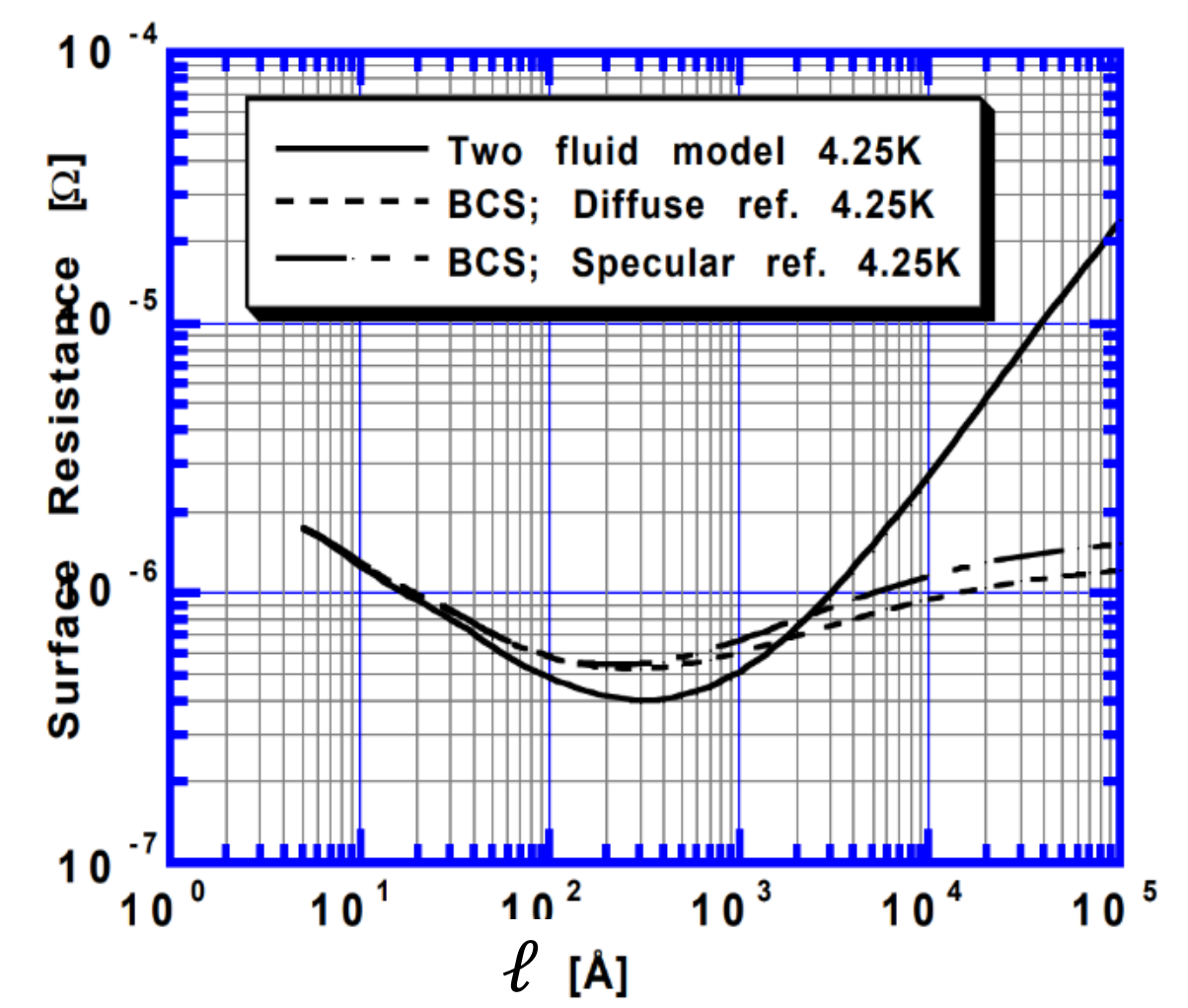

*Figure 24. Comparison between surface resistance evaluation with BCS and London approximation for Nb at 1.3 GHz and 4.25 K, in function of the mean free path ℓ. Picture reproduced with the kind authorization from K. Saito*

---

[1] An intuitive way to describe this is that the movement of Cooper pairs presents some inertia and they cannot fully screen the external field anymore.

[2] It's a simplified explanation. One can find a more subtle description based on quasiparticle excitations, as exposed in §2.2.5

The London model gives the following approximation of the Mattis and Bardeen integrals:

$$\frac{\sigma_1}{\sigma_n} \sim \left[\frac{\frac{2\Delta}{K_B T}}{\left(1+e^{-\frac{\Delta}{K_B T}}\right)^2}\right] e^{-\frac{\Delta}{K_B T}} \cdot \ln\left(\frac{\Delta}{\hbar\omega}\right) \tag{18}$$

$$\frac{\sigma_2}{\sigma_n} \sim \frac{\pi\Delta}{\omega} \tanh\left(\frac{\Delta}{2K_B T}\right) \tag{19}$$

and the expression of the surface resistance is:

$$\frac{R_s}{R_n} = \frac{1}{\sqrt{2}} \frac{\frac{\sigma_1}{\sigma_n}}{\left(\frac{\sigma_2}{\sigma_n}\right)^{\frac{3}{2}}} \tag{20}$$

At low and medium field, one can derive the expression of the RF surface resistance:

$$R_{BCS} = A\left(\lambda_L^4, \xi_F, l, \sqrt{\rho_n}\right) \frac{\omega}{T} e^{-\frac{\Delta}{kT}} \tag{21}$$

where A is a constant depending on $\lambda_L$, $\xi$, $\ell$ and $\rho_n$; $\omega$ is the RF frequency and $\Delta$ the superconducting gap.

- $\lambda_L$ is the London field penetration depth, it relates to the depth where supercurrents develop to screen the external field,
- $\xi$ is the Cooper pair coherence length; its value at O K is similar to the "size of a cooper pair" (interparticle spacing). A high $\xi$ makes the material less sensitive to defects. A small $\xi$, will on the contrary favor pinning (magnet application). For SRF: Nb has a very large $\xi$ about 40 nm.
- $\ell$ is the mean free path, and is a very important parameter. It can totally modify the properties of a superconductor compared to its theoretical one, and it is used to monitor the properties required for specific applications. We will see later that for SRF applications one has to trade off between two opposite requirement for $\ell$.
- $\rho_n$ is the normal conductivity. This parameter tends to announce that only good normal conductors will have low surface resistance when SC, but the situation is more complex and cannot be shortly described here.
- The last important parameter is the superconducting gap: $\Delta$, which is related to the transition temperature of the superconductor. This parameter is dominant for the surface resistance: the highest $T_C$ the lower the surface resistance…
- There is a limit frequency in frequency (~ 100 GHz) where the photons produced by the electron accelerations (Bremsstrahlung) can excite the super electrons and break the Cooper Pairs (this is related to the IR absorption band, see also Table 2 above).

Typically the surface resistance of Nb it is 100 000 times less than that of normal conducting Cu.

**<u>Experimental surface resistance</u>**

In fact, experimentally, one observe that another term of surface resistance must be added to $R_{BCS}$:

$$R_S = R_{BCS} + R_{Res} \tag{22}$$

$R_{Res}$ (sometimes quoted $R_0$) is not predicted by BCS. It contains all what initially was "unknown". It can include:

- Flux trapped during cooldown, especially in presence of pinning center (e.g. remnant of damage layer, thermal strain, poor recrystallization…). The mitigation consists into using a well recrystallized material, but also a high *magnetic hygiene* in the cryostats (non magnetic steel for nuts and bolt, active and passive magnetic shielding, cooldown procedures…). With these procedures, RF surface resistance of accelerator cavities have been divided by ~20 over the past decades.
- Other pair breaking mechanism:
  - Existence of magnetic impurity (e.g. vacancies in the oxide layer)
  - Proximity effect (there a NbO monolayer at the metal oxide interface, and it is metallic. It becomes SC by proximity effect but it affects the gap value, thus the resistance)
  - Other suspects *(*not yet identified). Indeed the metal oxide is difficult to model accurately.

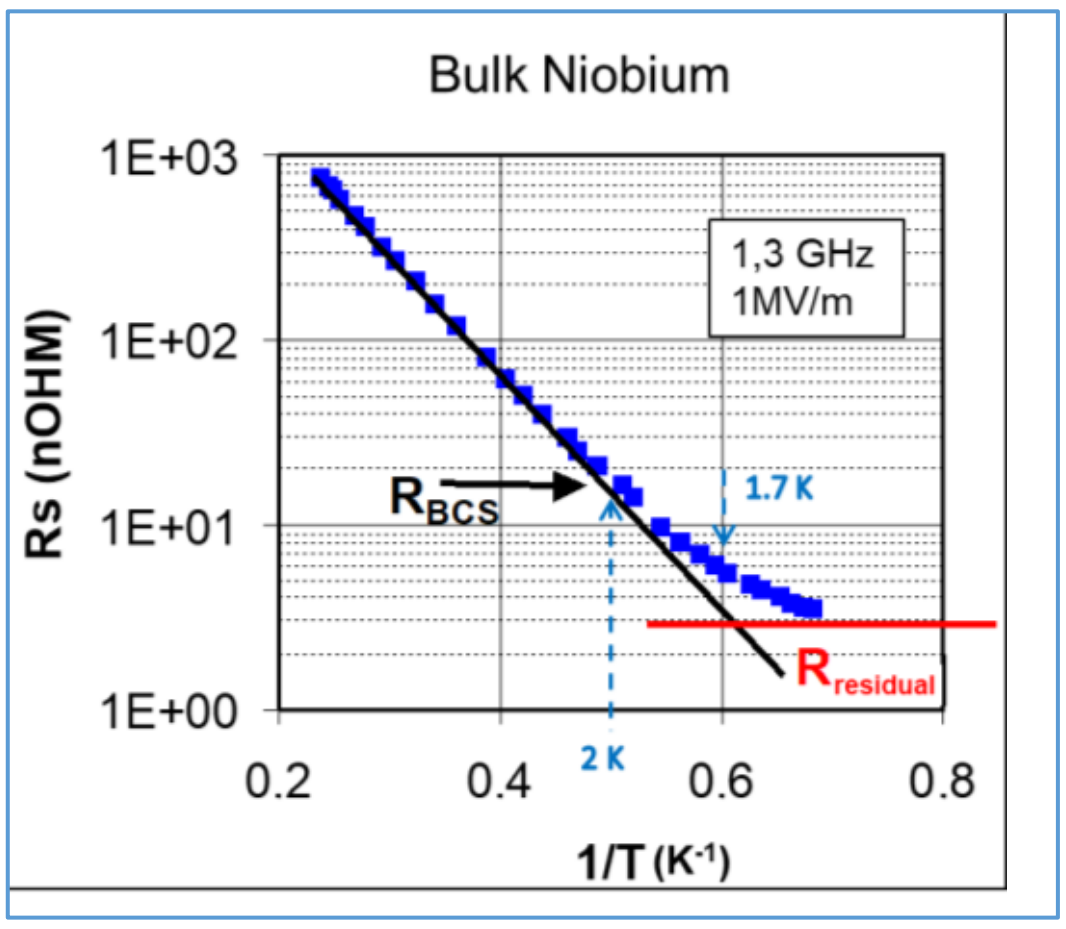

*Figure 25. Experimental surface resistance for a 1.3 GHz cavity at 1MV/m vs 1/T*

At higher field, with $H^{RF}$ close to $H_C$, the screening current of the Cooper pairs causes a reduction of the effective gap in the spectrum of the quasi-particles, the density of normal electrons increases and therefore the surface resistance increases as well. This non-linear component of $R_{BCS}$ has been determined only for type II superconductors, in the clean-limit and low frequency ($\hbar\omega << \Delta$) [11]. The non-linear corrections increase exponentially with increasing field (and temperature) [4, 12]. More recently, there has been attempts to integrate other possible origin of Cooper pairs breaking (magnetic and non-magnetic impurities, presence of a normal conducting layer on the surface combined with proximity effect...) in one general formula [13, 14].

## 2.3 Most common applied superconductors

| Material | $T_C$ (K) | $\rho_n$ ($\mu\Omega$cm) | $\mu_0 H_{C1}$ (mT)* | $\mu_0 H_{C2}$ (mT)* | $\mu_0 H_C$ (mT)* | $\mu_0 H_{SH}$ (mT)* | $\lambda_L$ (nm)* | $\xi_0$ (nm)* | Type |
|---|---|---|---|---|---|---|---|---|---|
| Pb | 7,1 | | n.a. | n.a. | 80 | | 48 | | I |
| **Nb** | **9,22** | **2** | **190** | **400** | **200** | **219** | **40** | **38** | **II** |
| NbN | 17,1 | 70 | 20 | 15 000 | 230 | 214 | 200-350 | < 5 | II |
| NbTi | | | 4-13 | >11 000 | 100-200 | 80-160 | 210-420 | 5,4 | |
| NbTiN | 17,3 | 35 | 30 | | | | 150-200 | < 5 | II |
| 2H-$NbSe_2$ | 7,1 | 68 | 13 | 2680-15000 | 120 | 95 | 100-160 | 8-10 | II- 2gaps |
| **$Nb_3Sn$** | **18,3** | **20** | **50** | **30 000** | **540** | **425** | **80-100** | **< 5** | **II** |
| $Mo_3Re$ | 15 | | 30 | 3 500 | 430 | 170 | 140 | | II |
| **$MgB_2$** | **39** | | **30** | **3 500** | **430** | **170** | **140** | **10** | **II- 2gaps** |
| **YBCO/Cuprates** | **93** | | **10** | **100 000** | **1400** | **1050** | **150** | **< 3** | **II d-wave** |
| **Pnictures $Ba_{0.6}K_{0.4}Fe_2As_2$** | **38** | | | | **900** | **756** | **200** | | **II s-wave** |

# 3 Mixed state and applications

As stated before the frontiers of mixed state is what condition the applications of a superconductor: magnet vs RF cavities. The needs are specifically different for both applications:

For electromagnets one aims at very high current densities with 0 resistance (in DC). It implies that they are in the mixed state (also called vortex state), with non-moving trapped vortices, medium fields ($H_{C1}<H_{operation}<H_{irr}$, $H_{irr}$ will be defined below). Defects are voluntarily introduced to enhance pinning. The effect is to decrease mean free path $\ell$ and $H_{C1}$ and to increase $H_{C2}$. Magnets operate below a critical current density $J_C$ about 10-15% of the depairing current density $J_D$ (see below).

For cavities, one aims at very high field with minimal dissipation but as showed in §2.2.6, the surface resistance cannot be 0. Vortices submitted to variable field cannot keep pinned at high frequency, they would cause very high dissipation. So one has to prevent vortex entering in the material and work in the Meissner state. So one needs to reduce the number of defect that might promote early vortex penetration and use a material with very high $H_{C1}$ and/or $H_{SH}$ if accessible.

In summary, superconductors fitted for magnet applications are bad for cavities applications… And vice-versa !

## 3.1 Vortices: penetration inside the superconductor

In this paragraph, we will describe the mechanism of vortex penetration and how some defects can promote early penetration.

### 3.1.1 Free energy at the surface

If one considers an interface between a superconductor and a normal conducting region (or vacuum), to get the free energy, one has to take into account the magnetic energy, where the induction B decreases over $\lambda$ and the condensation energy where the density of superconducting electron grows over $\xi$.

**Magnetic energy:**

$$\delta g_m \sim \mu_0 \frac{H^2}{2}\left(1 - e^{-\frac{x}{\lambda_L}}\right) \tag{19}$$

**Condensation energy**

$$\delta g_C \sim m\mu_0 \frac{H_C^2}{2}\left(\frac{n(x)}{n_s}\right) \tag{20}$$

Figure 26 shows the typical interface free energy for type I and type II superconductors at relatively high field (H~0.7 $H_C$). One can observe that for type II superconductors, the free energy is negative close to the surface. It means that it is energetically favorable to nucleate a normal zone at high field. That is the origin of the apparition of vortices.

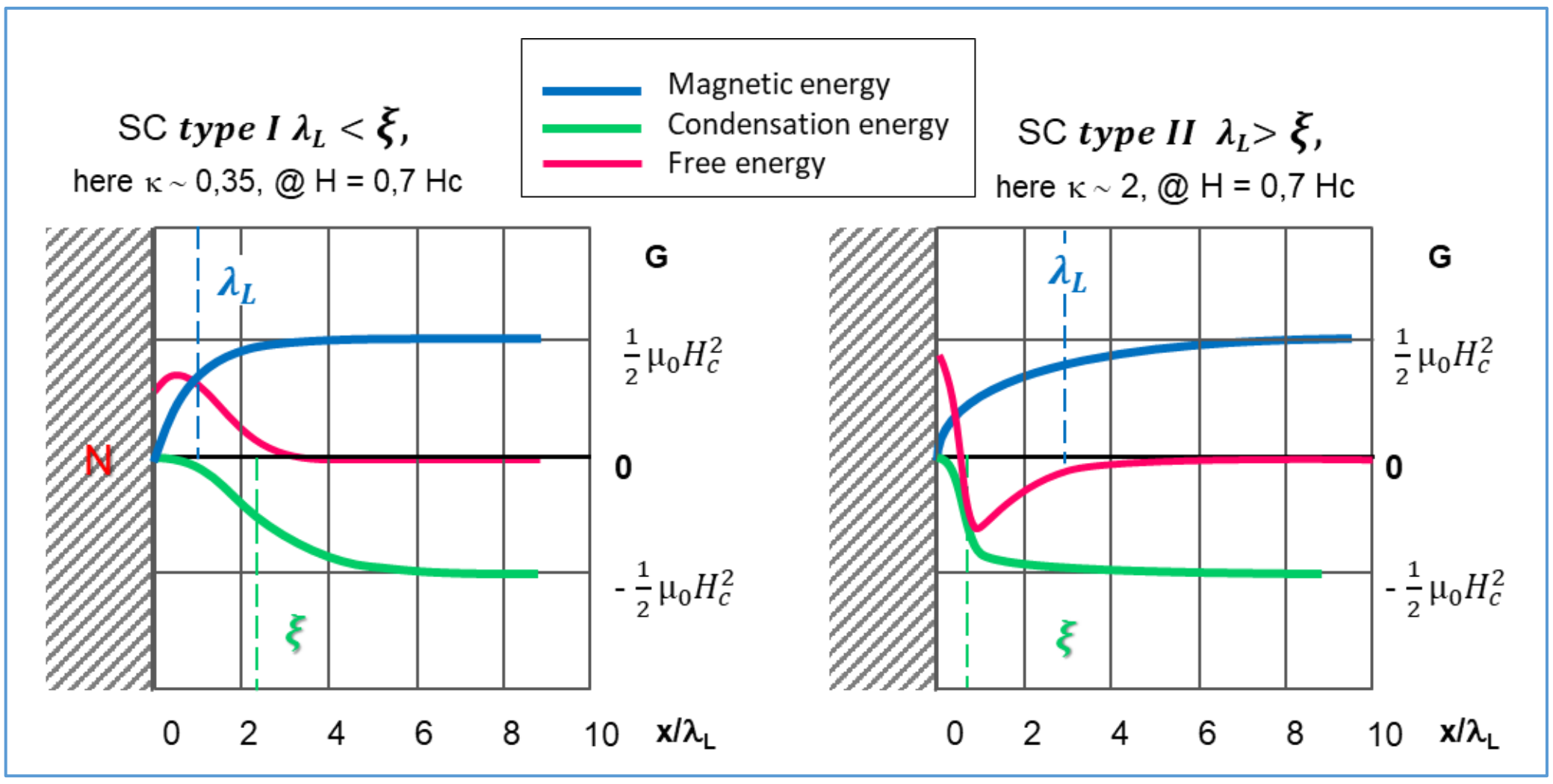

*Figure 26. Variation of the magnetic energy and the condensation energy that compose the free energy. For type II SC with $\lambda>\xi$ the free energy is negative, which means it is energetically favorable to create a NC/SC interface.*

The minimization of free energy leads to several predictions:

- The normal zone must be as small as possible (ø~ 2ξ); this holds only for $\xi<\lambda$, i.e. Type II SC.
- The magnetic size of the system is ~2 λ (a vortex is a normal region with one single flux line inside and screening currents around).
- The number of Vortices is the one that minimize ΔG/L, it depends on applied H and sample dimension
- Surface stabilizes the apparition of the SC mixed state, and nucleation always occurs at surface, Vx always emerge perpendicular to the surface (see Figure 27)
- Vortices repel each other (but attract antivortices). In absence of defect they form a regular hexagonal centered[1] lattice so-called Abrikosov lattice (see Figure 29)

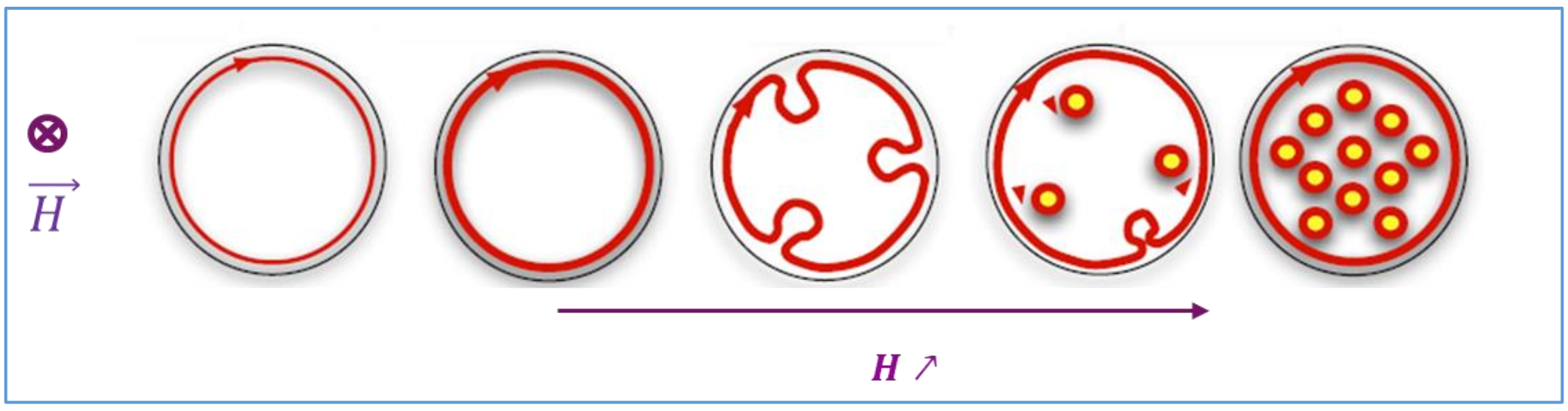

*Figure 27. Schematic description of the vortices nucleation in a thin sample perpendicular to the magnetic field. In the Meissner state, current line flow (in red) around the sample according to Lorentz laws. As field increases, the current lines incurve to enclose a field line in the normal zone. The number of vortex depends on the applied field.*

## 3.2 Parallel versus perpendicular field

### 3.2.1 Demagnetization effects

In perpendicular field, Field lines enter easily on thin samples due to demagnetization factor. If one exposes an ellipsoidal sample to uniform field, the field on surface is the same everywhere, whereas if it has an arbitrary shape one has to consider the local deformation of flux line (antenna effect). Equation (1) becomes

[1] The Abrikosof lattice is often described as a "trigonal" lattice, but the exact term from the crystallographic symmetry group point of view is "centered hexagonal".

$$\overrightarrow{B} = \mu_0 \ (\overrightarrow{H} + (1\text{-}D)\ \overrightarrow{M}) = 0 \quad (21)$$

Where D is the demagnetization factor. For a spherical sample, D= 1/3, for an infinite cylinder D= 0, for an infinite strip, D= 1 (field cannot go around) but for a square strip size, D~3/5 (For more details see [15]). At lower scale, roughness can also play a similar role. As a result, some region of the sample (or device) see a local field different from the external field and may transit earlier. This effect makes it very difficult to measure accurately $H_{C1}$ by classical magnetometry where the sample is immersed in a uniform field. It is a large source of discrepancies in the literature. Only the first penetration field can be measured, which is a convolution of $H_{C1}$ and shape factors that can vary a lot the preparation of the sample. For more detail see annex on magnetic measurements (§6).

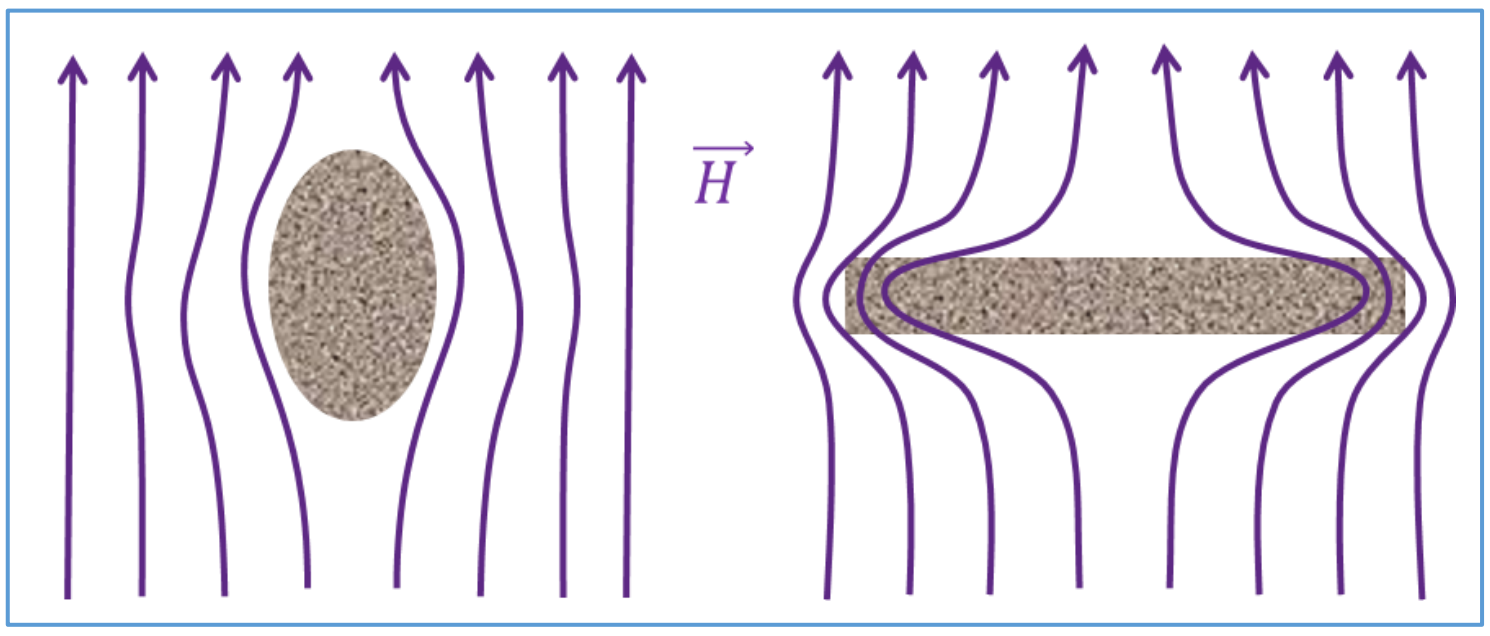

*Figure 28. Effect of the edge of a thin slab sample on the deformation of the field line*

**Magneto-optics to visualize field**

One very interesting technique to visualize field penetration and geometric effects is magneto-optics (MO). The sample is installed inside a cryostat equipped with a transparent window, surrounded by magnetic coils liable to provide parallel or perpendicular magnetic field. On the top of the sample, one puts an indicator (typically a material which color depends on the presence of magnetic field lines). The whole system is placed inside an optical microscope equipped with a camera.

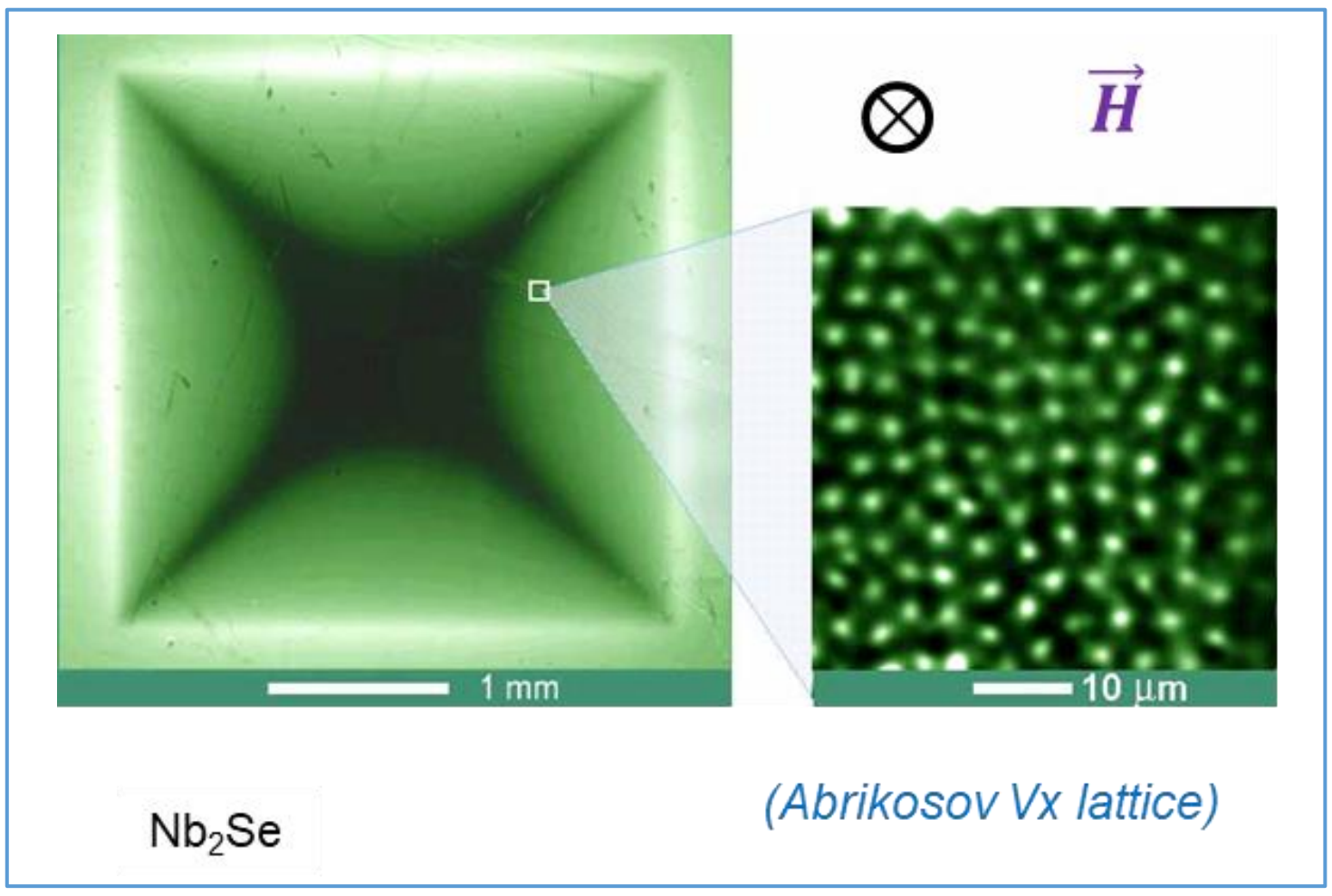

*Figure 29. MO image of the field penetration features in a square sample. At larger magnification, individual vortices can be resolved. Image extracted from the former website of Oslo university, not in use anymore: (http://www.mn.uio.no/fysikk/english/research/groups/amks/superconductivity/mo/)*

### 3.2.2 Surface barrier

In parallel field large samples), the situation is different. One has to take into account the deformation of the local potential introduced by the rupture of periodicity caused by the surface. This was formalized by Bean and Livingston in 1964 [16], by introducing an "image " vortex[1] that allows to respect the $J_{\perp} = 0$ boundary condition. While supercurrent tends to push vortices inside the SC, Image vortices tend to pull them out. Therefore, before entering the material, vortices have to cross a surface barrier. The Vortex thermodynamic potential can be expressed as:

$$G(x) = \phi_0 \left[ H_0 e^{-x/\lambda} - H_v(2x) + H_{C1} - H_0 \right] \tag{22}$$

The first term in the bracket corresponds to Meissner induction decreasing at the surface, the second term corresponds to image vortex and $H_{C1} - H_0$ helps to take into account the fact that one must exceed $H_{C1}$ to obtain stable vortices. Figure 30 shows the modification of the expression (26) when $H_0$ increases.

When H~ $H_{C1}$, G = 0 in the depth of the material (vortices are stable), but there is still a barrier that prevents them to enter. On an "Ideal surface", the barrier disappears only at $H_{SH}$~f($H_C$ ) >$H_{C1}$. So having parallel field stabilizes the superheating field, and that is the rationale used to predict SRF limits. Unfortunately, if there exist local defects, with for instance $H_c^{Local} \ll H_c^{bulk}$ (or $T_c^{Local} \ll T_c^{bulk}$), one can observe early penetration of several vortices there and the superheating state cannot be maintained. In this case, only material with a high $H_{C1}$ can prevent vortex to enter. That is why Niobium, with the highest $H_{C1}$ of known superconductors has been a material of choice for SRF applications.

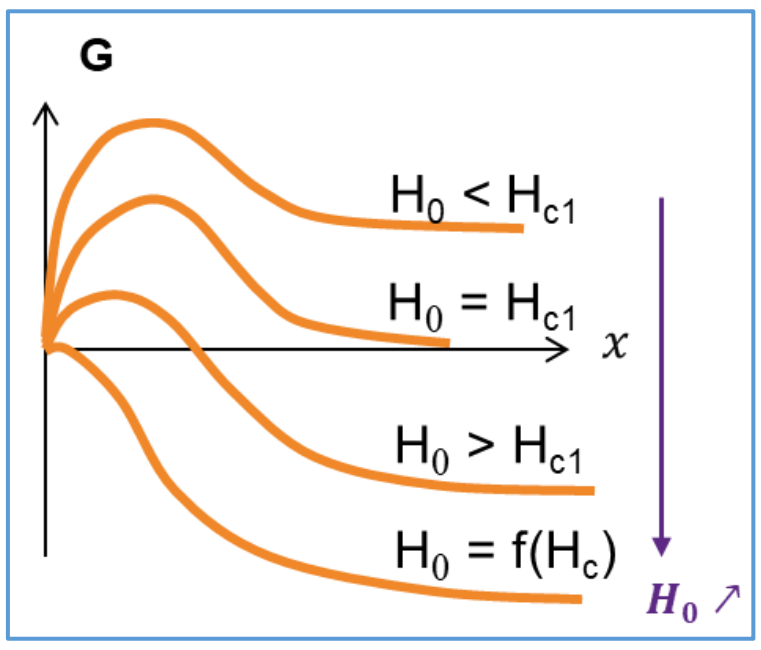

*Figure 30. Vortex thermodynamic potential vs depth for various applied field values.*

### 3.2.3 Morphologic defects

Realistic materials often exhibit morphologic defects: general shape, scratches, natural roughness, but also at smaller scale: crystalline facets, inclusions, atomic steps… When parallel field encounters a morphologic defect, the magnetic field lines are compressed and the field is locally enhanced by a factor β, which is the inverse of the demagnetization factor exposed in§ 3.2.1. Even if the field enhancement on a defect is small, the first point reaching the transition field is obviously there. It is not the height of the steps that matters but the angle of curvature on the step edges [17]. Typically an etching pit which has roughly a half-ellipsoid shape (50 µmx100µm) has a β around 1.5-2. A crack in a thin film can go as high as a factor 5 (see Figure 31).

In magnet, surface imperfection can be at the origin of flux trapping and flux jumps (see §3.4).

In SRF, roughness at the µm and the nm scale matters in the high magnetic field parts of the cavity: it can trigger an early quench.

[1] A similar approach exists in electrostatics with image electrons to get the actual potential close to the surface.

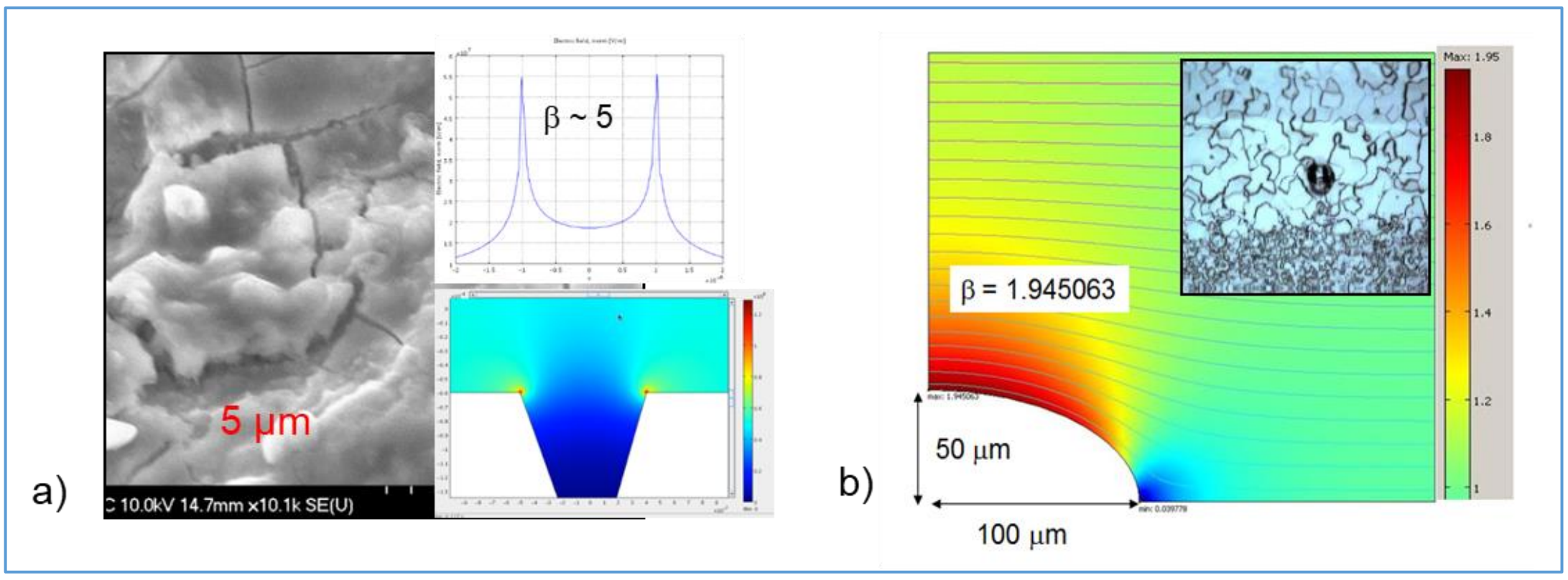

*Figure 31. 3D-Modelling of field enhancement for a) Crack in a thin film b) an etching pit. Reproduced with the kind authorization from J. Norem Argonne Nationnal Laboratory.*

Note that similar features will also result in electric field enhancement on parts exposed to high electric field, but the consequences are not so drastic. The field enhancement required to draw electrons from the surface is one or two orders of magnitude higher. These two points will be further detailed below.

### 3.2.4 Grain boundaries

Magneto-optics show that field can preferentially enter at grain boundaries (GB), but only if it is parallel to the GB plane. Conductivity experiment show that GB exhibit degraded superconducting properties in presence of external magnetic field. For example, reference [18] shows measurement on Nb at 4.5 K where GB becomes normal conducting at 32 mT, while monocrystalline Nb becomes normal conducting at ~150mT. The role of GB and other crystalline defects has been extensively studied in the magnet wire development as they act pinning centers, as will be detailed below in § 4.1. Any defect with degraded superconductivity when present at the surface is also liable to lead to premature entrance of vortices.

Nevertheless, it does not necessarily affect RF applications. For instance GB in Nb present weaker penetration field, but monocrystalline cavities exhibit the same RF behavior as polycrystalline ones [19, 20]. In Nb, GB are not at the origin of the observed high field losses.

### 3.2.5 Flux jumps and avalanche penetration

Flux jumps are observed when a group of vortices is suddenly detrapped, usually from the surface, and start to enter deeper in the material[1]. It is a very common phenomenon observed in magnets when coils are put under charge. If not well controlled it can become destructive. Indeed, if for some reason, the field, or current, or temperature is locally increased, vortices start to move, causing some energy dissipation and a slight temperature rise. This temperature rise helps to detrap more vortices, which cause further dissipation and temperature rise. The phenomenon will rapidly lead to an avalanche penetration of vortex followed by such high dissipation that superconductivity is lost. The avalanche stops propagating when heat removal to the system eventually balances the heat production, hence the importance of the cryogenic systems for stabilization. Figure 32 shows magneto-optic picture of such an avalanche penetration inside $NbSe_2$. After 1 ns (typically the duration of an RF period), several thousands of vortex have penetrated on a depth ~200 µm, i.e. 3 orders of magnitude deeper than the field penetration depth $\lambda_L$. For more details and movies, see [21]

[1] Several interesting videos are available by searching "superconducting vortices movement" in an usual browser.

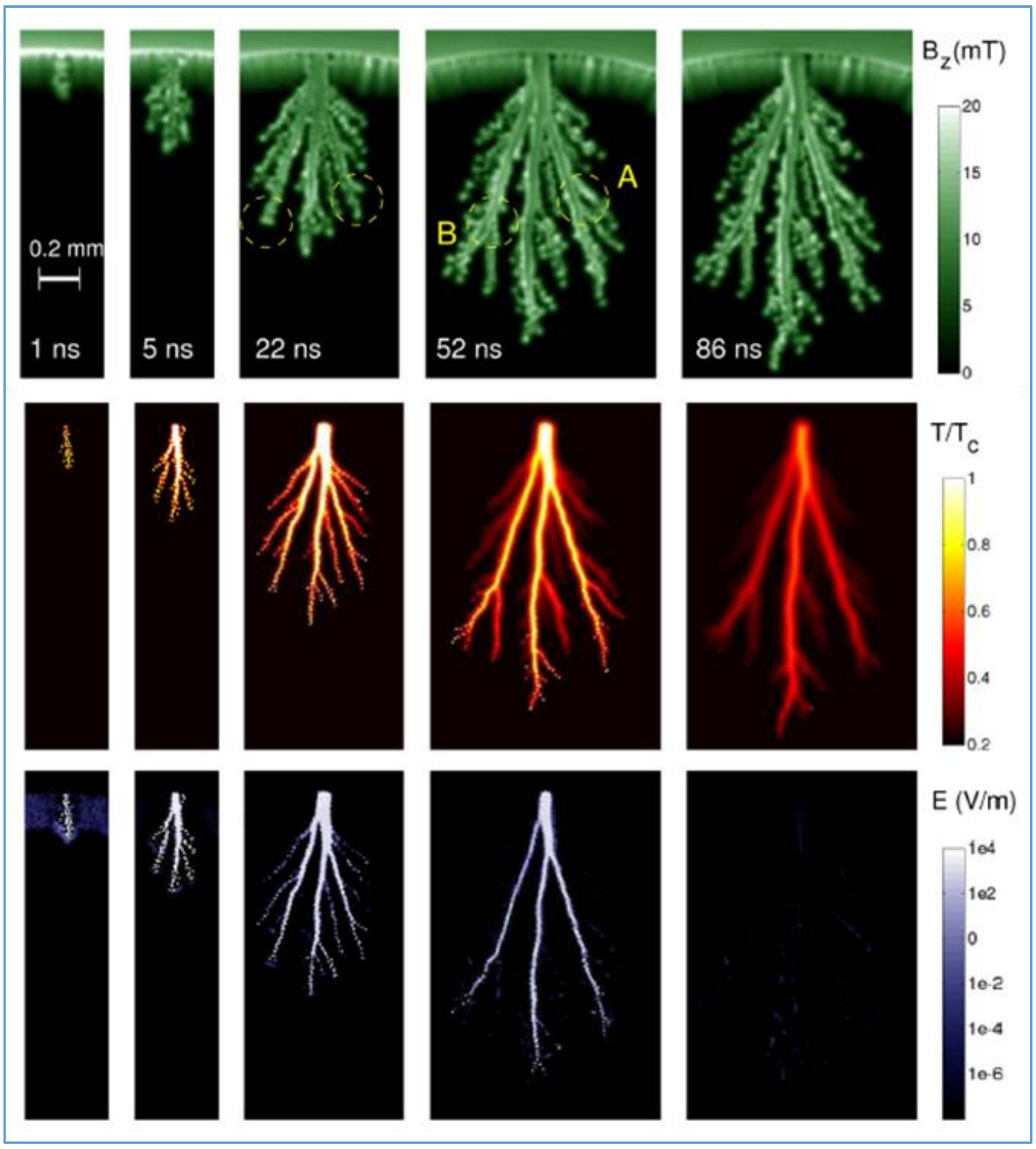

*Figure 32. Magneto-optics view of avalanche penetration of vortices in $MgB_2$ at T= 5K. One can observe that the phenomenon is fast: a few 100s nm deep within one ns, and that temperature and electric field reach high values. The complete burst corresponds to the entrance of several millions of vortices. Reproduced with the kind authorization from J. I. Vestgården (University of Oslo).*

For cavities, this point is of paramount importance. Typically, one ns is the period of a RF field pulsation. It means that within a period, several hundred thousand vortices can enter the material in a ~200 µm depth and dissipate accordingly, to be compared with λ ~50 to 250 nm for typical superconductors. This type of incident severely limits the use of a SRF cavity, and extra care must be taken to prevent the existence of surface defects that could promote early penetration.

## 3.3 Vortices in presence of current

### 3.3.1 Flux flow resistance

In a perfect material, in presence of current, vortices are submitted to Lorentz force $\vec{F} = \vec{J} \times \vec{B}$, and start to move collectively. $\vec{F} = \vec{J} \times \vec{B}$ at a constant speed v. It generates an electrical field $\vec{E} = -\,\mathrm{v} \times \vec{B}$, parallel to $\vec{J}$. Because of Ohm law, if there is a potential difference, then the resistance cannot be 0.

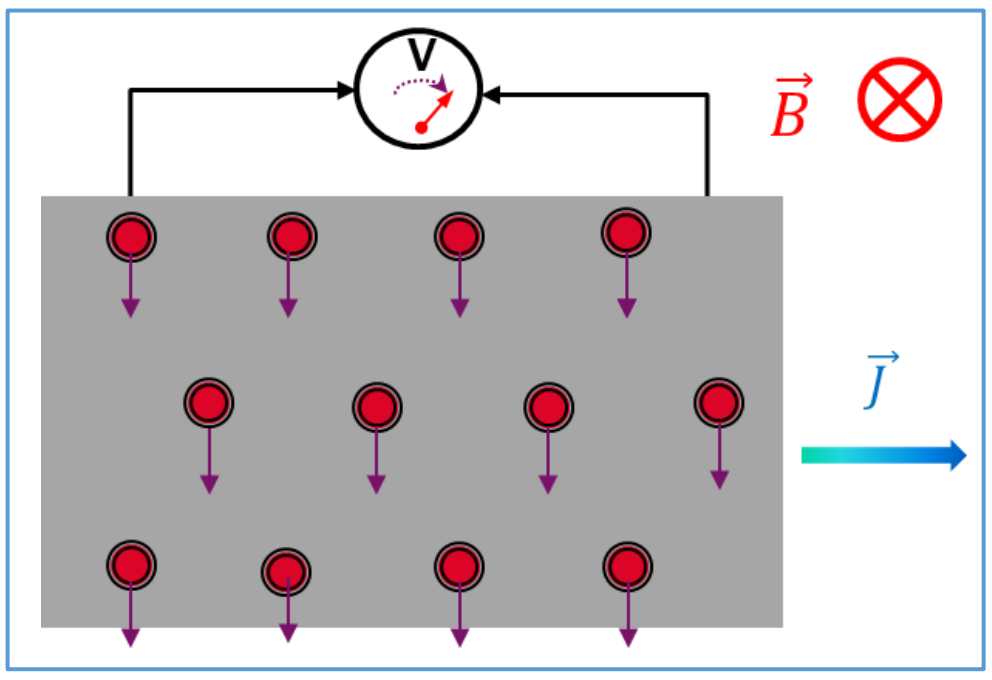

*Figure 33. Movement of vortices in presence of magnetic field and current density*

Indeed, one observe experimentally a non-negligible resistivity called "flux flow resistivity" $\rho_{ff}$ . The movement of the vortices is limited by a viscous force (Magnus force), which originates from the normal zone dissipation. It is easy to determine that $\rho_{ff}$ is proportional to $\rho_n B/B_{C2}$, since only a fraction of the surface is normal conducting: the resistance is proportional to the number of vortices, which depends on the intensity of field (Figure 34). When B reaches $B_{C2}$, the vortices overlap and the surface becomes completely normal conducting. The viscosity $\eta = \phi_0 B/B_{C2}$ and the constant speed v = E/B. The elastic energy of the system tends to keep the vortices equidistant.

Therefore, a perfect material in the mixed state the DC should never have resistance equal to 0. Fortunately, actual materials always exhibit defects, especially if their composition is complex, and the first experimentalists obtained 0-resistance below a certain current density $J_C$, without knowing anything about pinning mechanism.

### 3.3.2 Critical current density $J_C$

In practice the resistance R = 0 only for $J<J_C$ for type II SC in the mixed state, when vortices are pinned on pinning centers. At a certain point, the Lorentz forces overcome the pinning forces $F_P$ and the vortices start to move again, reinstating the flux flow resistance (Figure 34). $J_C$ is the current density at which flux flow starts. $F_P$ and therefore Jc depend on the temperature T and on the field B

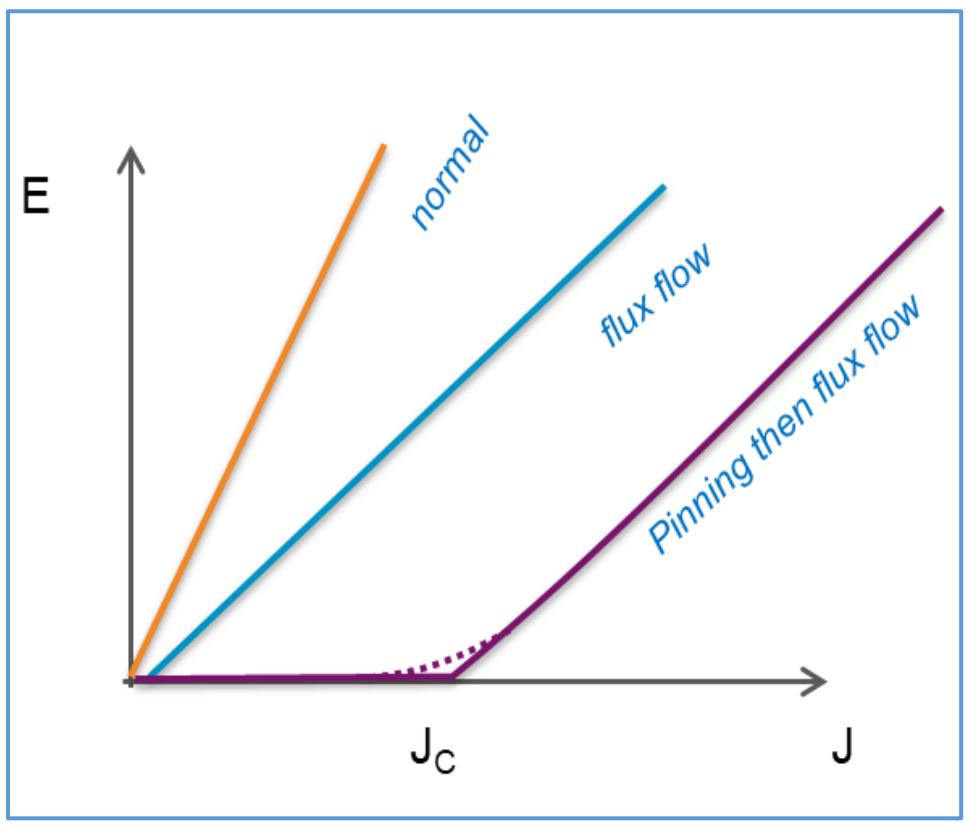

*Figure 34. Electric field E vs current density J. he slope of the curve is the resistance. Flux flow resistance is only a fraction of the normal conducting resistance, and in case of pinning, the apparition of non 0 resistance is delayed. Close to $J_C$ one can observe some flux creep or flux jumps, before the collective movement starts.*

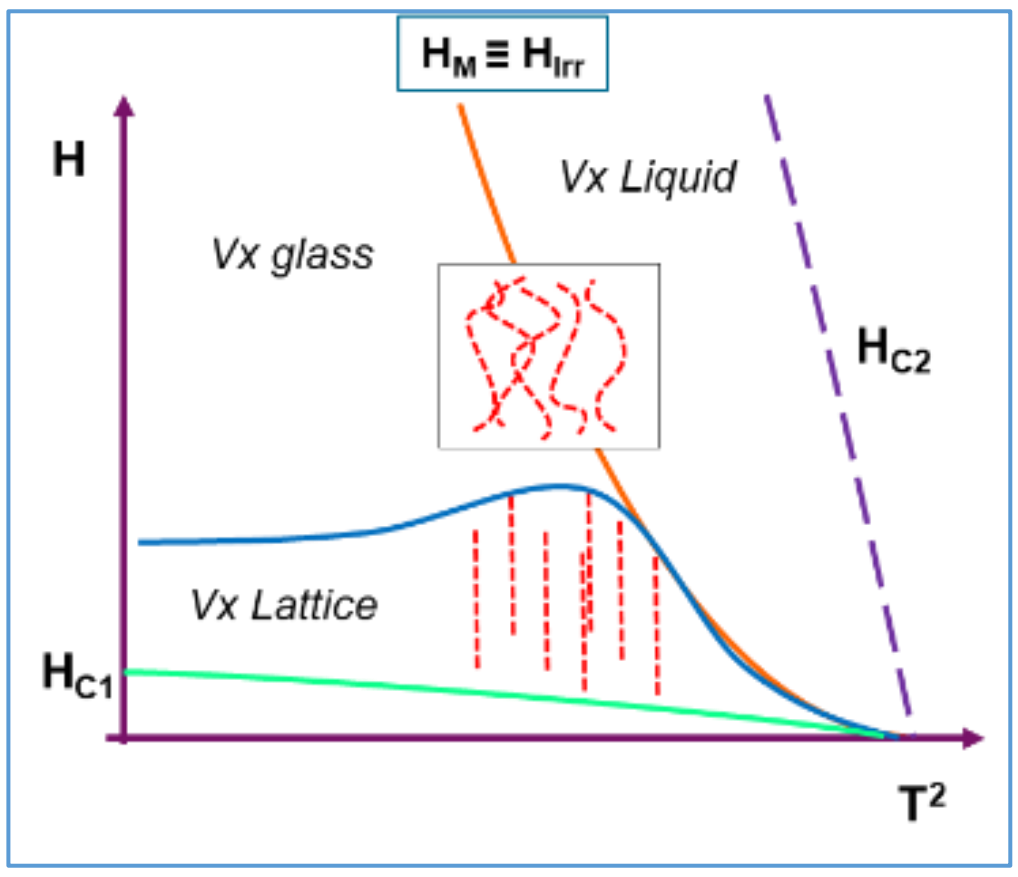

*Figure 35. $H_M$ correspond to the melting of the vortex network. When vortices start to move freely marks the beginning of thermal dissipation with a non 0 resistance* $H_M$ defined as the H=f(T) line @ $J_C$ (H,T) = 0 in the (H,J,T diagram)

The projection of Jc in the H vs T plane (or H vs $T^2$ plane in Figure 35) corresponds to $H_M$ (like “Melting”), as the individual vortices move “liquid-like” beyond this field. It is also called $H_{irr}$ (like irreversible) as it is the field at which the signal becomes fully reversible upon increasing field (or fully irreversible upon decreasing field) in DC magnetometry which is a common technique used to evaluate $J_C$ (see § annex 6.1).

To increase $J_C$, one has to keep vortex trapped, so one artificially incorporates defects (inclusions, grooves, alloying, damaging …) as will be detailed in §4.1. The electrons mean free path ℓ is reduced, which in turn decreases $\xi$ and $H_{C1}$, and increases $\lambda$, $\kappa$ and $H_{C2}$. Meissner state is marginal in these kinds of type II SC, they transit readily into the mixed state.

$J_C$ is a technical limit defined for magnet purposes, $J_C$ (extrinsic) << $J_D$ (intrinsic). It has no signification at high frequency where pinning is inefficient (see §3.4.4).

## 3.4 Pinning on crystalline defects

### 3.4.1 Crystalline defects at the atomic level

Figure 36 gives some recalls on crystalline defects and their dimensionality. We will see below that dimensionality influences the distortion of the crystalline lattice and plays a large role in the local modifications of superconductivity. Not that any well-recrystallized material still contain an equilibrium concentration of each type of defects, which depends only on the temperature and purity.

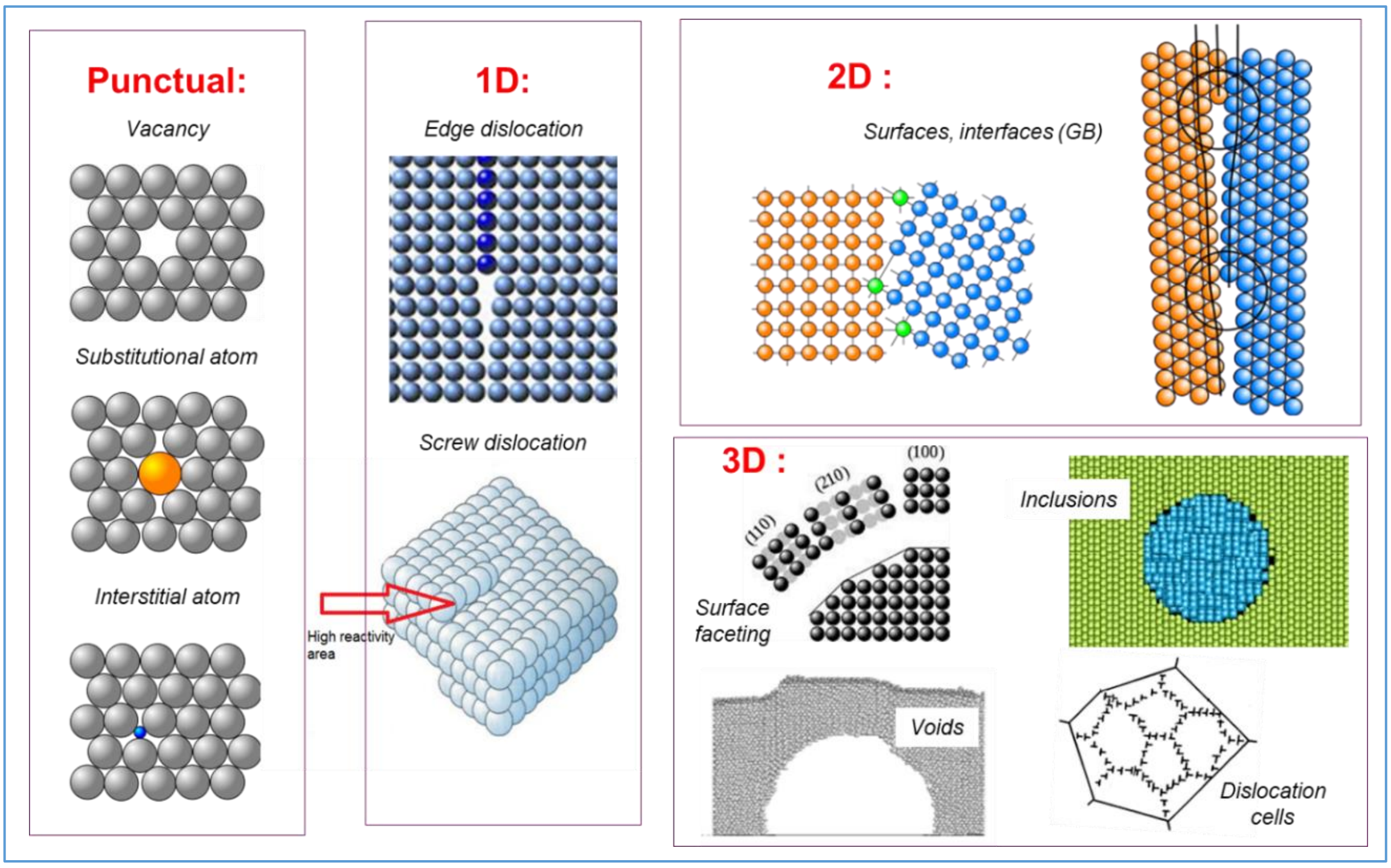

*Figure 36. Examples of crystalline defects. 2D defects like Interfaces (e.g. grain boundaries) can be considered as an array of 1D defects for some orientations. Seemingly, a crystal can be subdivided into subcrystals with slightly different orientation separated by dislocation walls, forming 3D dislocation cells. Inclusions or voids are often encountered in industrially produced materials. Surfaces tends to reconstruct to exhibit crystalline planes with lesser energy. Even 1D defects tend to gather to form larger dimension defects.*

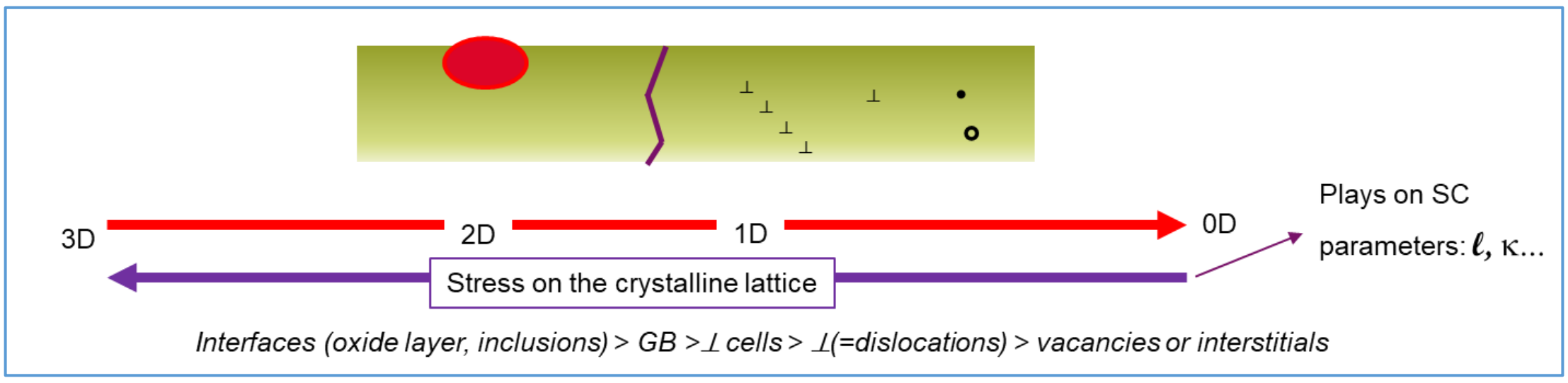

*Figure 37. Relation between dimensionality of crystalline defect and the stress they impose locally on the crystalline lattice.*

Note: the elastic strain due to defects also plays a strong role in impurity segregation. For instance, H tends to gather around dislocation cores to form so-called "Cottrell clouds". In Nb, one also observe H, C and O segregation at the oxide-metal interface.

### 3.4.2 Simplified view, or "saving condensation energy"

If normal conducting regions already exist, it is more favorable for a vortex to bend to go through this normal area because it saves condensation energy (remember that it requires a lot of energy to break one single Cooper pair). At the equilibrium, the cost in elastic energy (one has to bend the vortex) compensates the gain in condensation energy.

It is intuitive to understand that the resulting pinning force depends on the diameter d of the defect: for very small defect, the gain in condensation energy is minimum, while if the defect is much larger than the vortex core, the vortex becomes a simple flux line inside a normal conducting material (Figure 38).

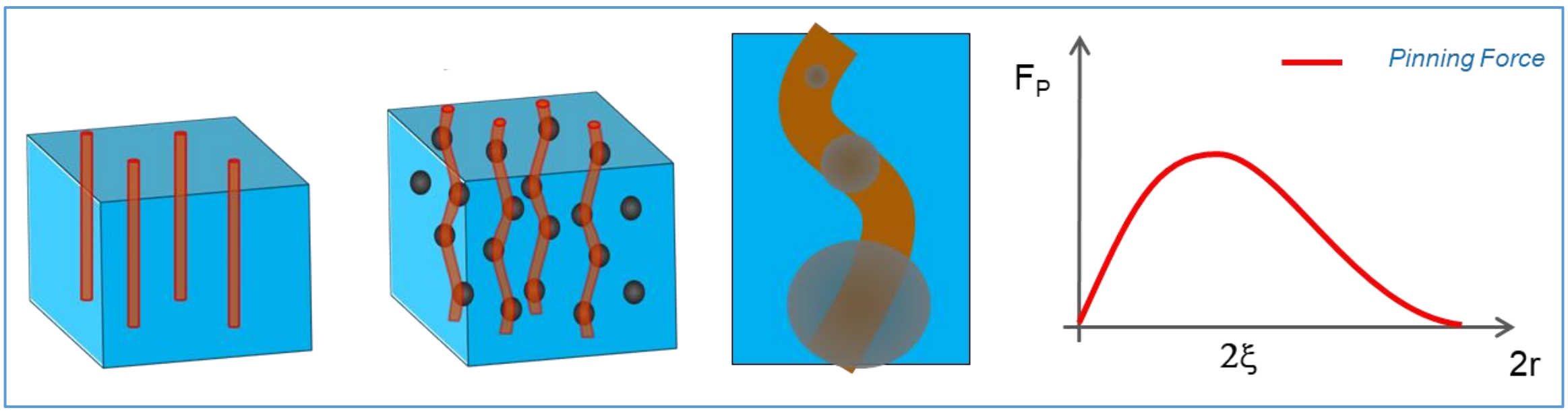

*Figure 38. Left: without defects vortices are regularly spaced (Abrikosov Lattice). Center: effect of normal conducting inclusions (diameter 2r) on the shape of vortices, effect of the size of the NC particles. Right: shape of the pinning force curve. See text for details.*

In presence of pinning center, the periodicity of the Vortex Abrikosov lattice is lost: one gets Vortex "polycrystals", even Vortex "glass" or "liquid"… In some high Tc materials, a new material state has even been discovered: Bragg glass [22].The phase diagrams of Vortex states is complex, with weird dynamic properties and their behavior is the object of the majority of publication in the domain.

**Pinning force**

As stated before, the pinning force $F_P$ (r) and $J_C$ are maximum for defects with a lateral size ~2ξ parallel to the vortex axis (columnar defects, plane interfaces).

For inclusions with radius r ~ ξ, interspaced by d, one can define $J_C$' ~$J_C$xL/d, where L is length fraction occupied by pinning centers with d interspace. The more general case is when a volume fraction f is occupied randomly by defects with radius $a_0$. The gain in free energy is : $\Delta g \sim (H_C^2 \mu_0/2) f\pi x^2$ per length units $\mathcal{L}$, the pinning force $F_P \sim \Delta g/\mathcal{L} \sim (H_C^2 \mu_0/2) f\pi\xi^2/\mathcal{L}$ and $J_C = f/\Phi_0 \sim (H_C^2 \mu_0/2) f\pi\xi$ (for $\mathcal{L} \sim \xi$) or $J_C = \phi/\Phi_0 \sim (H_X^2 \propto_0/2)\phi\pi\xi^2/\alpha_0$ (for $\mathcal{L} \sim a_0$). These formulas are ad minima; in practice, $J_C$ is always very difficult to predict and need to be asserted experimentally.

**Elastic force**

The elastic potential goes like: $E_{elastic} \sim \frac{C_{ii}}{2} \int (\nabla u)^2 d^3 r$ , where u is the displacement of the vortex from its ideal position, and Cii the elastic constant per unit length. The main elastic constants vary with field : $C_{11}$ (compression) ~ $B^2/\mu_0$, $C_{44}$ (torsion) ~ $B^2/\mu_0$, $C_{66}$ (shear) ~ $B^2/\mu_0$, $B\Phi_0/16\pi\mu_0\lambda_L$ and usually $C_{44} >> C_{66}$.

Meanwhile, elastic forces are also affected by the elastic stress imposed on the lattice by crystalline defects. We will see in the next § that in fact several mechanisms of pinning co-exist.

In summary, a pinning center can be any defect with degraded superconductivity (including normal conducting inclusions or voids).

### 3.4.3 The four actual components of pinning

In his comprehensive book on flux pinning in superconductors [23], Matsushita describes four mechanisms:

A- **Condensation energy variation:** one saves condensation energy if a normal zone already exists close by.

B- **Elastic interactions**: there are several aspects:

- Bending a vortex requires elastic energy
- The lattice elastic moduli in SC state smaller than the elastic moduli in the normal state [24]
- The elastic deformation due to the presence of crystalline defects-interplays with the local elastic energy. The higher the dimensionality of the defect, the higher its influence on SC properties (See Figure 37).

C- **Magnetic interaction**: if a defects has dimension >>$\lambda$, one can treat it like an interface, with image vortex approach, the apparition of a surface barrier etc… (as seen in § 3.2.2) It is a very strong effect

**D- Kinetic energy interaction**, it concerns areas with different $\xi$ and vortex velocities, and concerns very particular materials. It will not be detailed here.

What is important is that all these mechanism concern **local effects**, where the material exhibit non uniform properties. If one recall §3.2, models like the 2 fluids model apply for a uniform density of Cooper pairs. Similarly, Ginzburg Landau approach concern slowly varying potential (see §3.2.4). In presence of Pinning, classical models need to be revised to produce accurate prediction.

**Surface magnetic pinning vs volume core pinning**

There are two ways to produce efficient pinning:

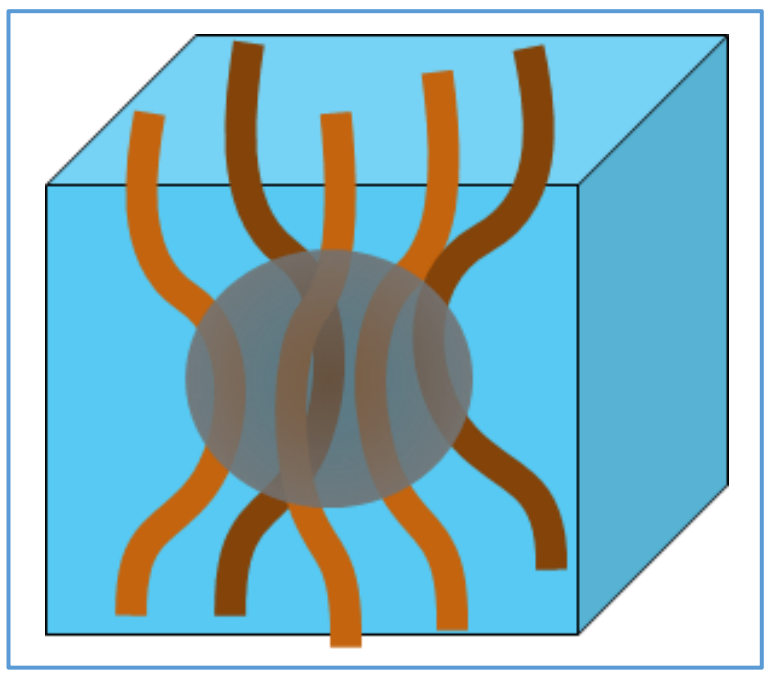

*Figure 39. Several vortices pinned on the same large defect*

- "Surface magnetic pinning": one need a few strong Pinning Centers: twinning planes, voids, non-SC aggregates, irradiation defects (columnar), nano-indentations. In that case, the dominant mechanism is **A** then **C** then **D**. If the defect is large enough, several vortices can get pinned on the same defect (Figure 39).
- "Volume core pinning" can be achieved with numerous weak pinning centers. In that case, the dominant mechanism is **B**. **B** less efficient than **C** but if they are many pinning centers, it also results into strong pinning. As said previously, the efficiency of the **B** mechanism increases from 0-D defects (e.g. interstitial atom) to 3D defects (e.g. small inclusions or voids). Large 3D defects with dimension >> $\lambda$ fall under "surface magnetic pinning".

More details can be found in [20] for Niobium.

**0D-Vacancies, interstitials…**

When ponctual defects are uniformly distributed one get a typical weak pinning situation. If V is the pinning potential and $\rho$ is the vortex density, the pinning energy $E_{pinning}$follows:

$$E_{\text{pinning}} \sim \int V(r)\rho(r)d^3r \tag{23}$$

It is mostly studied with irradiation defects.

Usually the pinning force Fp tends to increase with density of defect, but not monotonically, an indication of non uniform distribution. Indeed at high defect density, defects tends to regroup in specific areas.

**Continuum model:** generally, if $a_0$ is the diameter of the pinning center and $\lambda > a_0$, one can work with mean values over $\lambda^3$ (Larkin volume) [25]

### 1D- Individual dislocation to 3D dislocation cells

The elastic stress field around a single dislocation is the order of ≤ 1 GPa. It is responsible for many physical phenomenon such as electron scattering, alteration of resitivity, or the apparition of an hysteresis in DC magnetometry[1].The dislocation density can increase up to 10 orders of magnitude upon deformation. Here again one observe non linear behavior, an indication that the repartition of dislocation is not uniform at high deformation rate (see Figure 40 and [26]). Indeed dislocation tend to accumulate in walls (2D defects annalogous to Grain boundaries), then in cells surounding areas of untouched crystal (3D defects, see Figure 41). The stress measured on a dislocation pill-ups can reach ~20 times the value of an isolated dislocation.

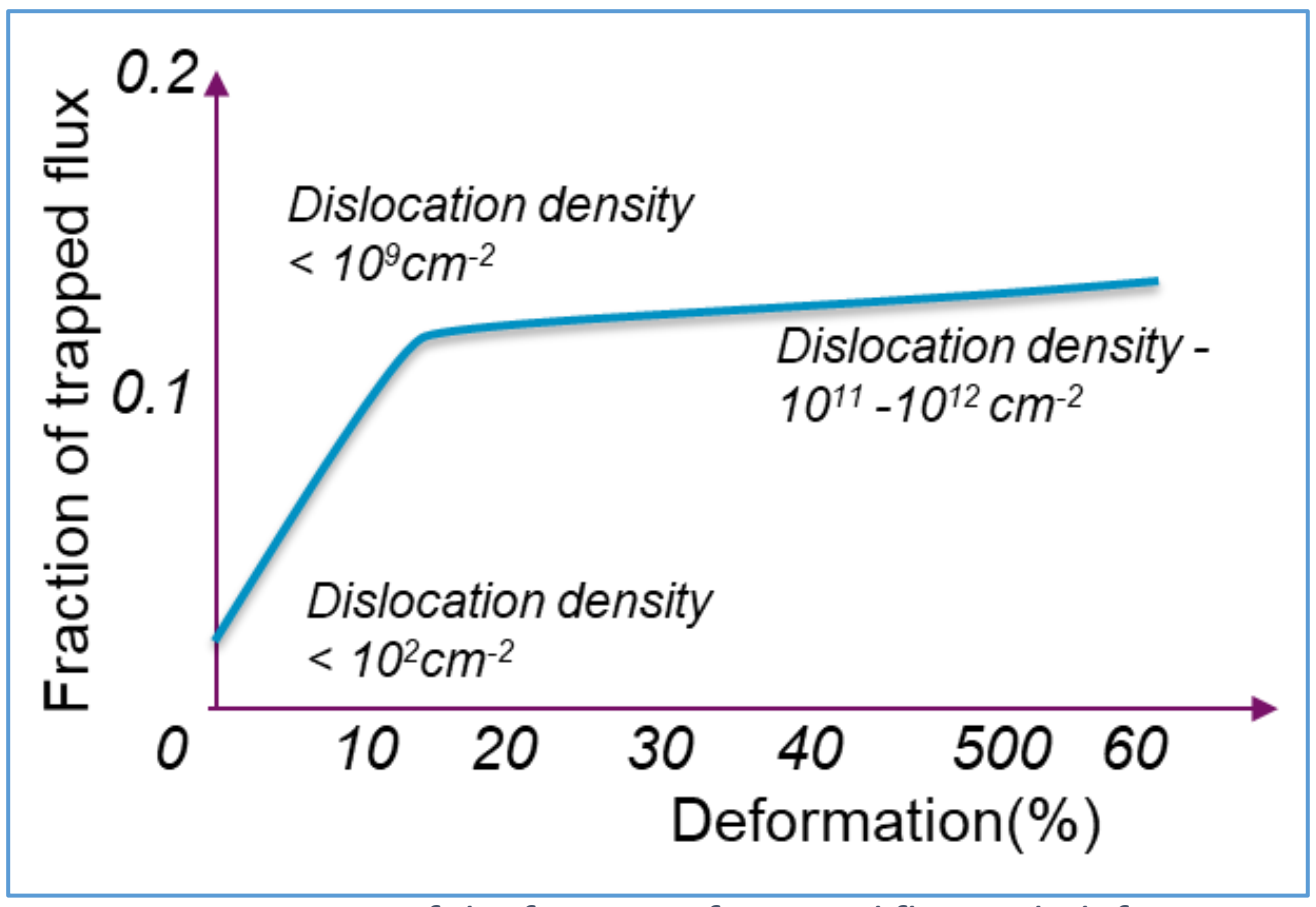

*Figure 40. Increase of the fraction of trapped flux with deformation rate. The change of slope corresponds to the moment where dislocation start to interact to each other and to accumulate locally (After [22]).*

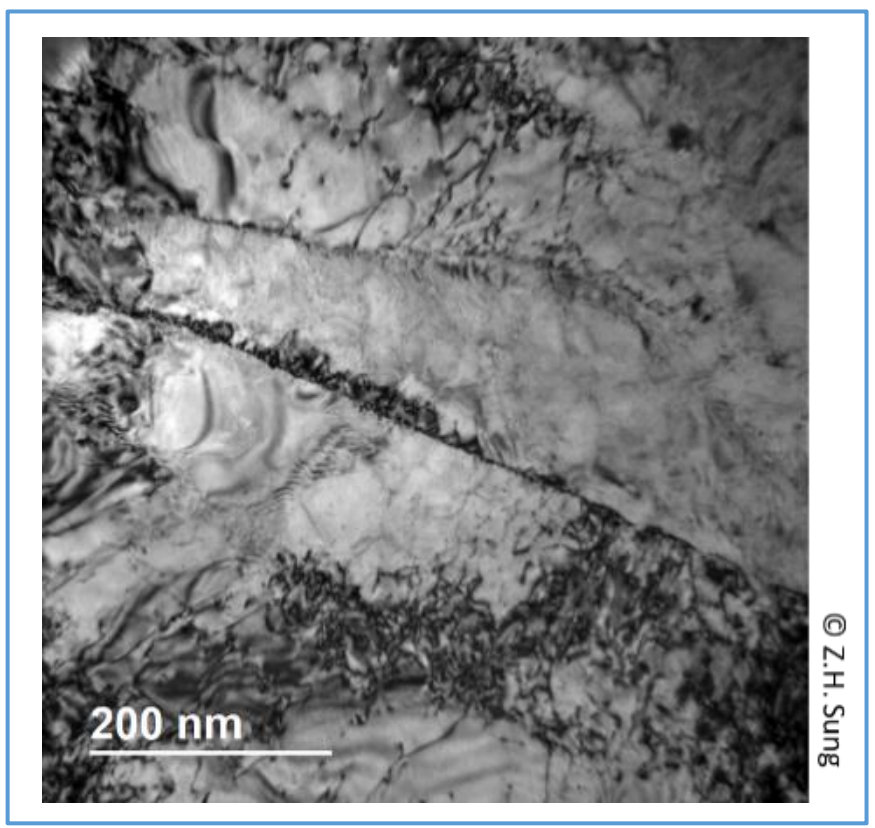

*Figure 41. Dislocation cells in 5% deformed Nb. Dislocation tend to regroup, while other areas of the crystal remain nearly intact. (Image reproduced with kind authorization from Z.H. Sung).*

### 2D- Grain boundaries

Grain boundaries (GBs) are often presented as a typical efficient pinning center by most lectures on superconductivity, but in fact, it once again depends on applications. GBs constitute a loss of the crystalline order over n atomic distances; n depending on the crystalline structure and particular grain orientations. In a superconductor like bulk Nb, n~2-3, so the disordered area is no larger than ~1nm. Compared to the coherence length of Cooper pairs $\xi^{Nb}$ ~ 40 nm, one can see that GB in Nb cannot play a strong role in pinning. Indeed, in Nb, dislocation cells which size ~ 100nm (so the same order of magnitude as 2 $\xi^{Nb}$) constitute a far more efficient pinning source. In Nb thin films[2], on the contrary, grains are very small (~100 nm ∅) and the mean free path is much lower. The coherence length decreases accordingly (Equation (2) in §2.1.5), and GBs become effective pinning centers.

For the HTC, YBCO family, n can reach up 10 in some direction, (~3 nm) which is the same order of magnitude of $\xi^{AB}$. In that case, the risk that a Cooper pair is scattered by the GB is very high, and GB constitute de facto very efficient pinning centers. Most of the SC used for magnets applications have usually a small $\xi$ and fall under this situation.

### 3D- and Surfaces/Interfaces: strong pinning

As stated before, large inclusions or even voids are very effective pinning centers because they introduce an interface between a superconducting area and a NC (or vacuum) area. We have seen in § 3.2.2 that this configuration is energetically favorable. We will see in § 4.1 that this type of defects are introduced on purpose inside SC wires destined to generate magnetic field. Another aspect of pinning is related to the external surface. As we have seen before, boundary condition states that no current exit the surface: $\vec{J}_S \times \vec{n} = \vec{0}$. Which means that vortices must always exit the surface perpendicularly and have to curve to follow this condition (see Figure 42). If a vortex is curved, its length increases and thus the elastic return force also Increases. Since at equilibrium the line tension results from the balance between pinning force and the elastic force, if the elastic force is increased, then the pinning force is also increased. It is true at macroscopic scale (Figure 42a) but also at microscopic scale (Figure 42b): the surface roughness also plays a role. Since the spacing of vortices depends on the applied field, the scale of roughness of concern depends on the magnetic field: µm level at low field, nm at field higher than 0;1 T [27].This effect is mostly appreciable on thin films or tapes. Surface corrugation has been successfully used to increase Jc on tapes for magnet applications [27]. Thins films are also more and more considered in RF applications and this aspect will probably have to be explored in this context.

---

[1] The hysteresis is in fact proportional to the trapped flux

[2] RF applications, but also SC electronics, detectors… see Introduction, Figure 2.

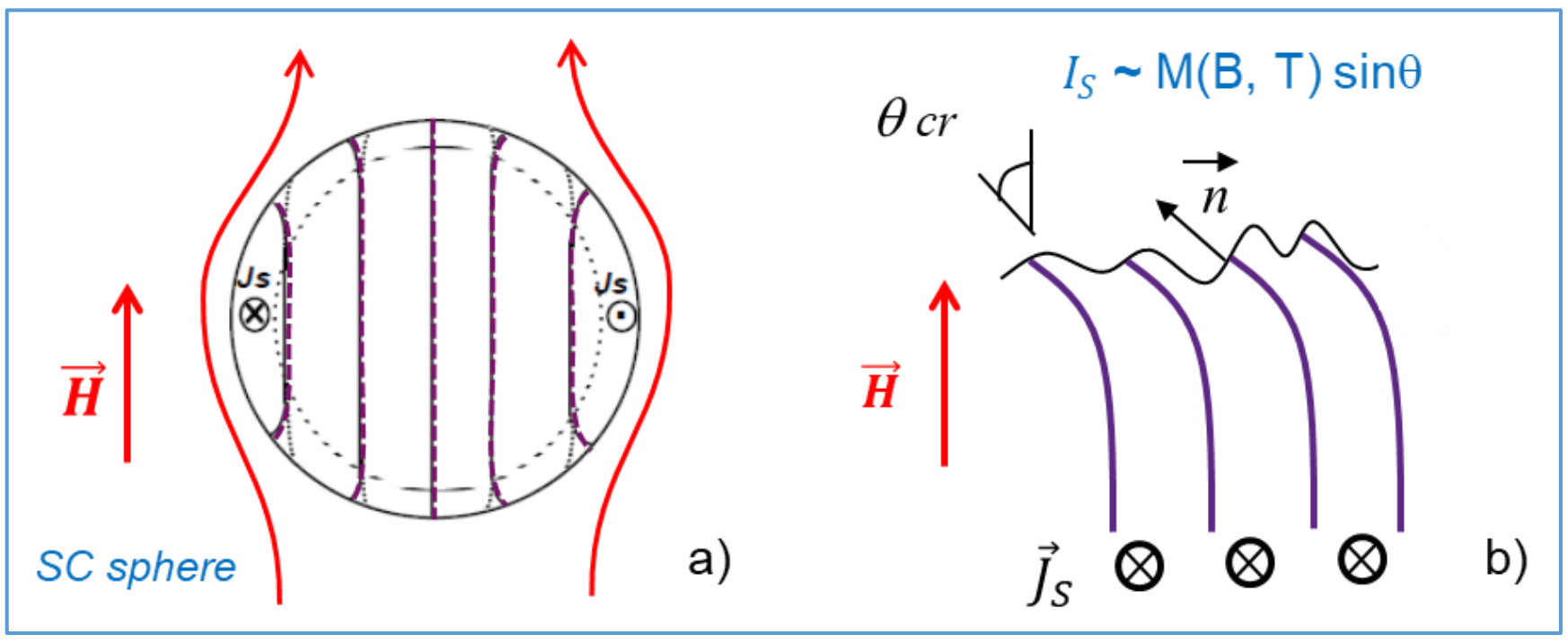

*Figure 42. Because of boundary conditions, vortices always emerge perpendicular to the surface at macroscopic scale (a) as well as at microscopic scale (b). The critical current $I_C$ increases with roughness, but the scale in concerns depends on the field (µm at low H, nm at high H).*

### 3.4.4 Pinning at high frequency

Complex impedance measurement on SC samples allows determining the behavior of complex penetration depth $\lambda_{AC} = \lambda' + i\lambda''$ vs frequency, as shown on Figure 43. The measured flux is $\phi_{ac} = \int b_{ac}\, dS \sim 2\lambda_{ac}\ell_b b_0$. At low frequency, when vortices are trapped: $\lambda'' \sim \delta_{AC}$ (normal conducting penetration depth), whereas at high frequency, when the vortices cannot keep pinned, $\lambda' \sim \lambda'' \sim \lambda_L << \delta_{AC}$. Depinning frequencies range from $10^6$Hz for very high quality Nb to several 10s of GHz for thin film materials.

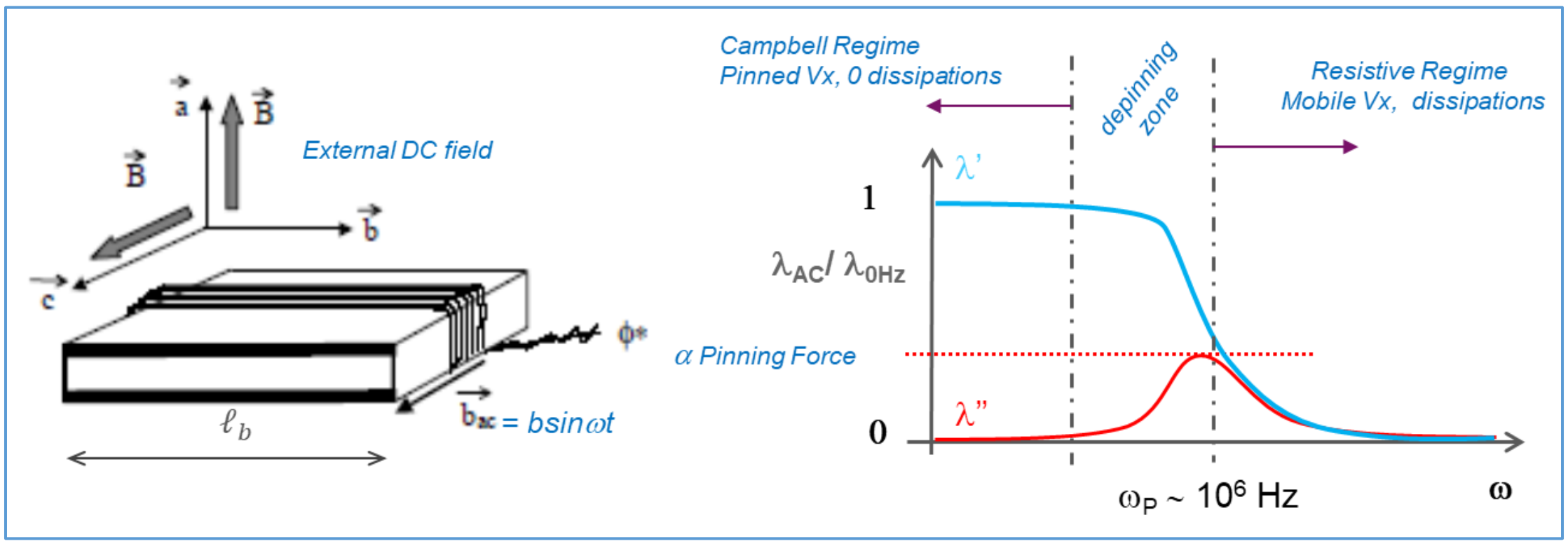

*Figure 43. Complex penetration depth vs frequency. At low frequency vortices (Vx) are pinned and the resistance is 0 (so-called Campbell regime), while at high frequency they move freely and the regime is resistive. The frequency where l'' is maximum is called the "depinning frequency". It can also be used to evaluate the pinning force.*

At very low frequency, in the Campbell regime, losses are negligible. Nowadays, in particular in SC electronics, materials with high depinning frequency are used in microwave or RF. At such a high frequency, even the oscillations of vortices inside their pinning well produce noticeable dissipation [28].

## 3.5 Conclusion on pinning

**Magnets:**

Pinning efficiency drives the critical current density Jc (and R=0 range of application). Whenever possible, pinning features are amplified (impurities, deformation, alloying, voids, inclusions, topological defects …), which leads to a modification of the mean free path and a large increase of $J_C$. Theoretical prediction of $J_C$ is far from straightforward, as several phenomenon occurs conjointly at different scales. Experimental measurements are still necessary.

**SRF cavities:**

There is no pinning at high frequency even if they are pinning centers in the material. Nevertheless they can prevent flux lines from being properly expelled during cooldown. Those trapped flux lines become oscillating vortices within ~λ and can be at the origin of hotspots [29].

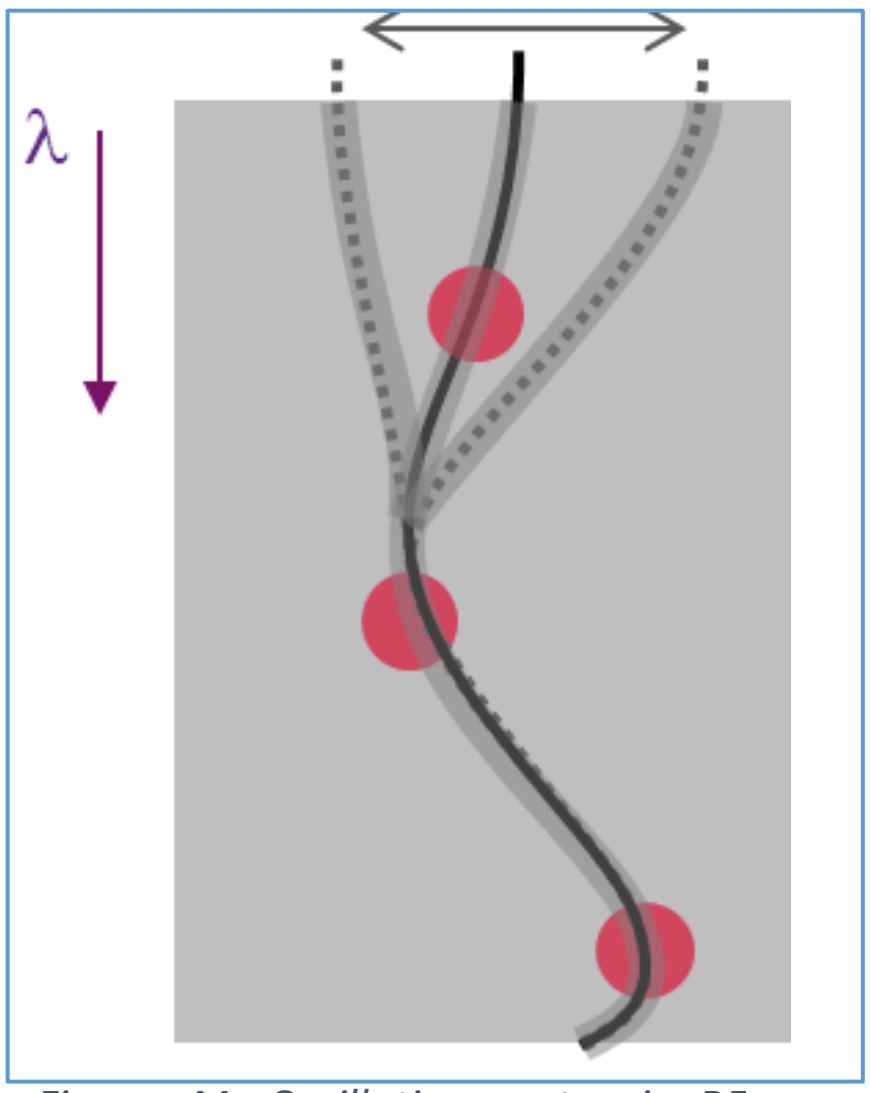

*Figure 44. Oscillating vortex in RF field*

Thus, superconductors good for magnets application are terrible in RF conditions… And vice-versa !

# 4 Optimization of Superconducting materials

As we have seen previously, the optimization of superconducting material depends a lot on its application: Magnet vs RF cavities. Magnet is a much more mature technology; most of the wires are available in commercial companies. Some optimization of the cables can be conducted corresponding to a particular project, but the laboratory work consist mostly in optimizing the design from the mechanical and safety point of view.

Cavities technology is more recent. Ultrapure Nb based technology is now available in 1-2 supplier worldwide, but improvements of the technology as well as R&D on successor materials are still actively under development in SRF labs.

In the following paragraphs, we will provide a short description of some common materials. For more details, one need to go to advanced literature (see § 6).

## 4.1 Magnets

Figure 45 shows some examples of cable developed for various applications. The structure of the cables is complex, since it has to ensure several functions in addition to resistance-less conductivity: mechanical load and/or thermal transfer, uniformity of current flow, reduction of demagnetization effects… The full detail of cable design is beyond the scope of this lecture, we will just provide a few examples of the development approaches the following paragraphs.

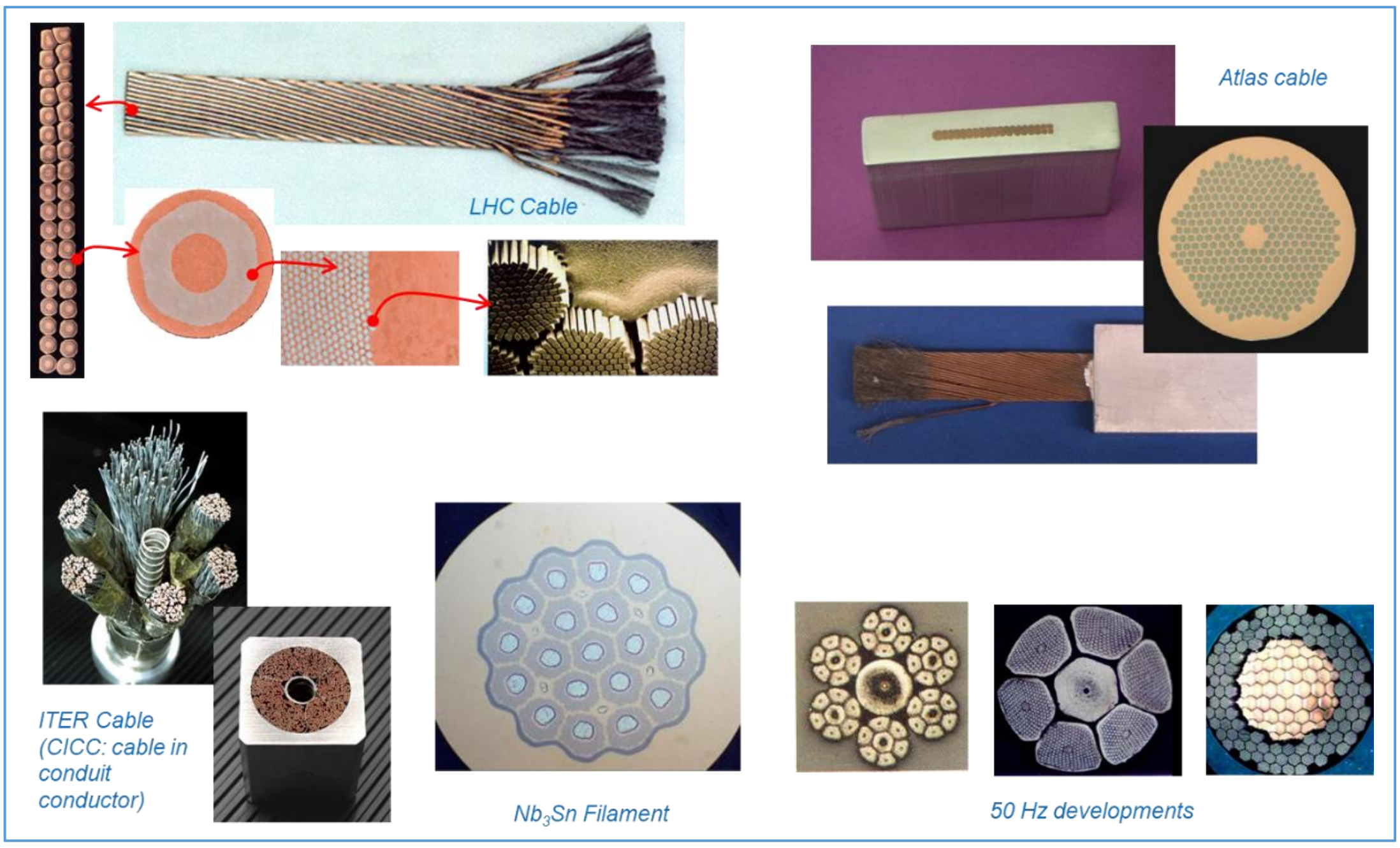

*Figure 45. Examples of cable developments*

### 4.1.1 Wire size influence. Bean Model

In the mixed state, but below the critical current, the superconductor is in the so-called "critical state". At equilibrium vortex are perfectly pinned. If one increases the field at the surface, locally the field induces a force that is higher than the pinning force: new vortices enter the material and the vortex front is progressively redistributed. The field inside the superconductor decreases progressively from the surface until it equals 0, but it is different from the screening current observed in the Meissner state: here the vortex front plays the role of surface current screening. Bean model proposes that the field decreases linearly with the slope $-\mu_0 J_C^{Bean}$, with a constant current density $J_C^{Bean}$ (see Figure 46).

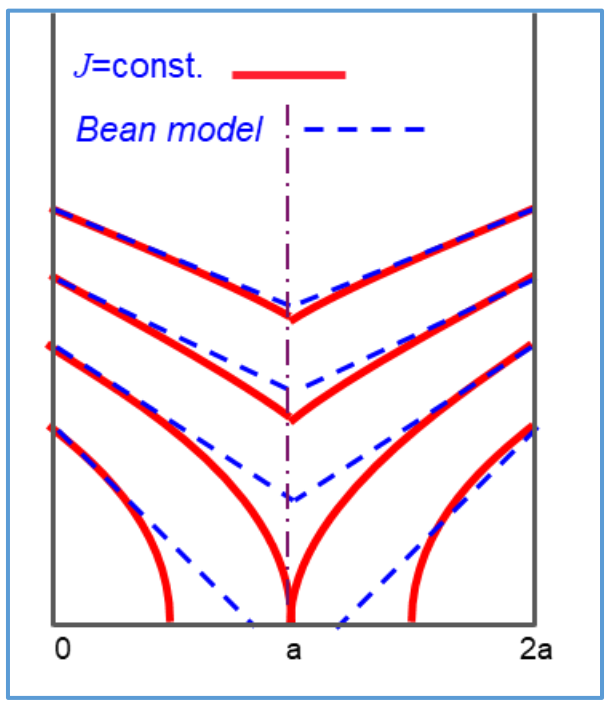

*Figure 46. Bean model for the decreasing of the field inside a thin SC slab.*

If the slab or wire is large enough, the current density will be much less at its center. This has consequences on the cable design: for a same total section and same applied field, it is more efficient to have many small filaments than one large wire. Moreover, large diameters favor instabilities (flux jumps, avalanches instead of progressive penetration of the field).

In addition, the filaments must be twisted and transposed because, in variable current or during charging/discharging, inductances control the current distribution. Twisting decouples the wires with respect to the external field while transposing decouples them with respect to their own self-field and helps homogenizing the distribution of currents (by minimizing the size of the current loops). It prevents local high current density to cause premature quench [30].

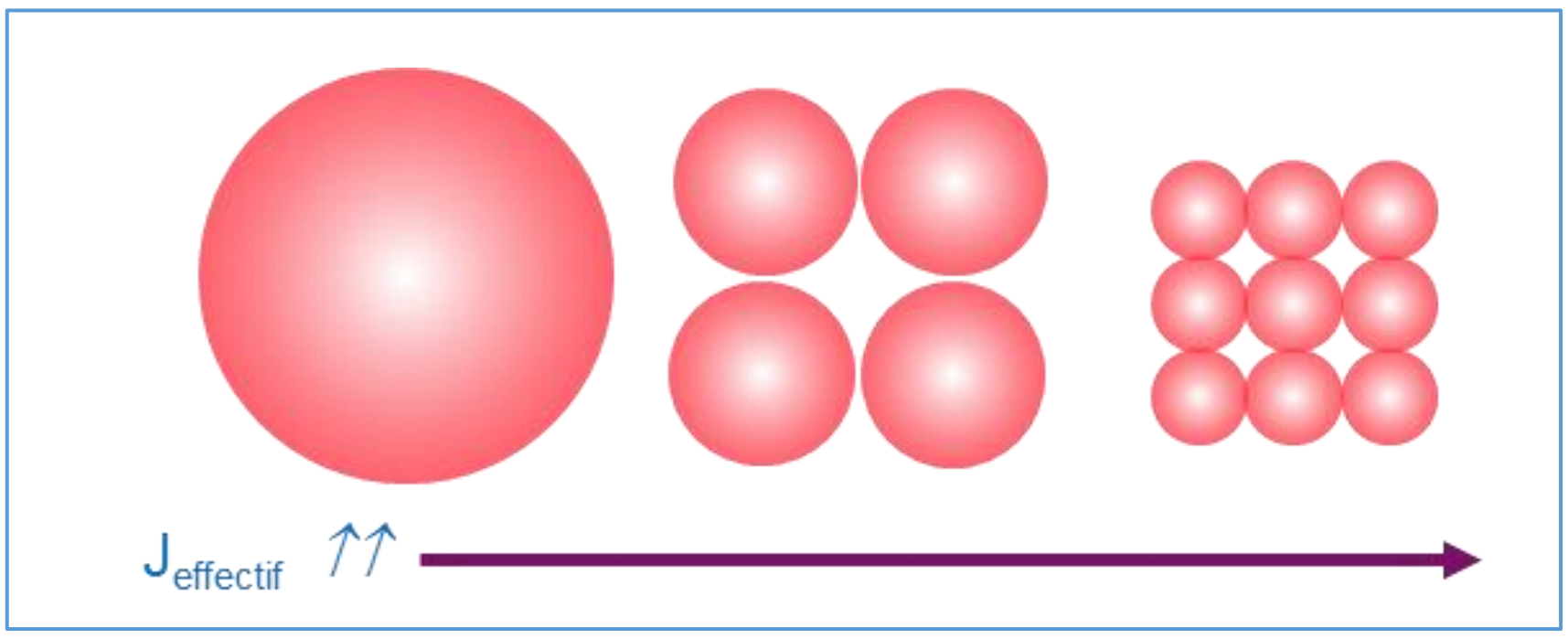

*Figure 47. The effective current density increases if the section of the wire is divided into several filaments*

### 4.1.2 NbTi: compromise between Jc and Tc

For magnet application, the alloy with composition 46-48% Ti present an optimum for $H_{C2}$ with an acceptable degradation of $T_C$. Ti has the same atomic radius as Nb, so for this composition at room temperature, the matrix consists into Nb with a few substitutional Ti (with a Tc that keeps close to the Nb one) along with metallic Ti. Upon wire forming (extrusion, wire drawing, intermediate heat treatment) those precipitates are elongated along the drawing (wire) direction and end up in the form of thin flakes with thickness the order of a few nm [31].

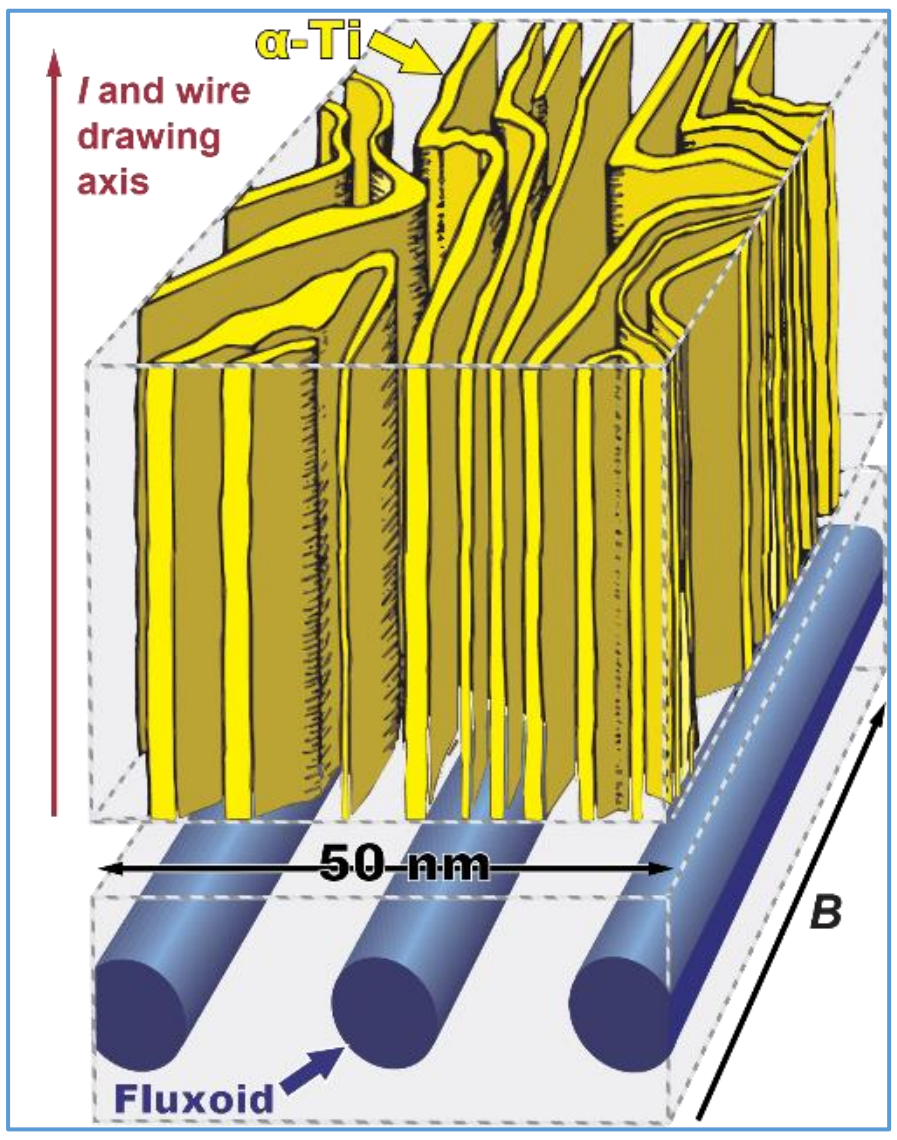

*Figure 48. Schematic structure of the Ti "flakes" obtained in a NbTi wire and their comparative size with vortex core (Image reproduced with the kind authorization from P. J. Lee, Florida State University).*

As seen in the previous chapter, best pinning sites are a normal, insulating or void region with dimension ~ $2\xi$ and parallel to the flux lines. Therefore, those "flakes" have the perfect dimension and orientation to constitute very effective pinning sites. The saving in condensation energy is maximum: Energy/unit length $= 1/2\mu_0 H_{C2}\pi\xi^2 \sim 10^{-11}$ J/m. In DC supercurrent short circuit the normal areas, but too many pins result in Tc suppression and current blocking. The exact quantity of pinning center needs to be adjusted to optimize Jc.

On the other hand, those normal conducting precipitates proved very dissipative if submitted to RF, even at very low field [32].

*NbTi and RF cavities. Like every alloys, NbTi is harder than pure Nb. It can be used as rigidification elements or flange material in SRF cavities provided that it is not submitted to RF field. The normal conducting precipitate would provoke huge dissipation otherwise.*

### 4.1.3 $Nb_3Sn$, $MgB_2$: brittle material handling

Brittle material like $Nb_3Sn$ cannot be drawn directly as a wire. Several fabrication processes have been established, where precursor material is first drawn and coiled in place and a thermal treatment allows obtaining the final product (Figure 49). One issue is devising an insulating system that can stand the heat treatment. In $Nb_3Sn$ wires, the critical current is inversely proportional to grain diameter, indication that grain boundaries are the main pinning center.

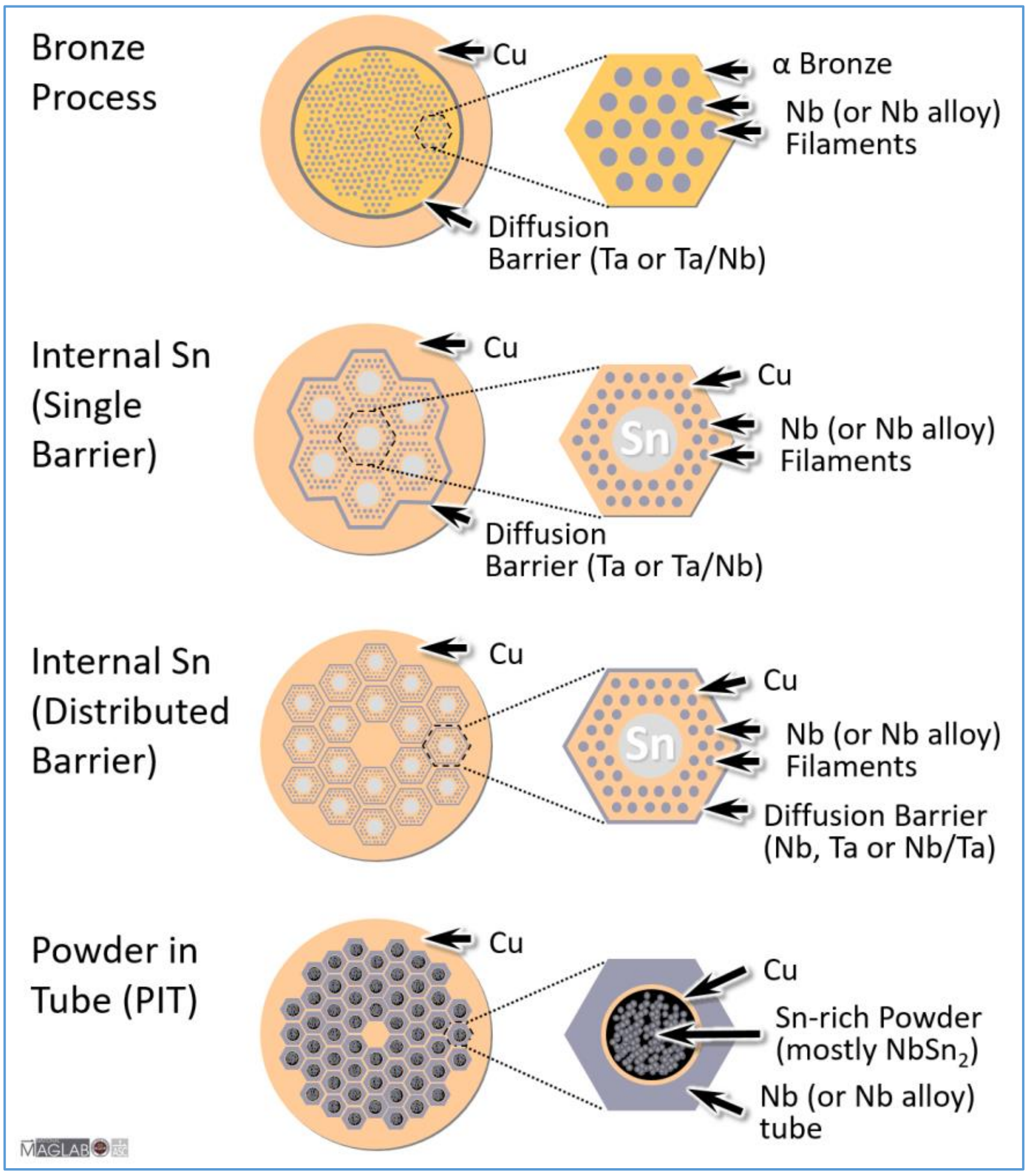

*Figure 49. Four major designs of commercial $Nb_3Sn$ starnds. Reeproduced with authorization from the Applied Superconductivity Center Image Gallery, Florida State University [33].*

$MgB_2$ is prepared in a similar way: either $MgB_2$ powder, or Mg Rod + B powder in a Fe tube, then drawing into a monofilament wire, and eventually reactive heat treatment to achieve the proper structure and composition.

### 4.1.4 HTS: anisotropy, weak links

YBCO family was discovered in the end of the 90s, and very quickly, compounds with Tc above liquid nitrogen were found, auguring a revolution in the use of superconductors, in particular for electric engineering. Unfortunately, this family of material presents two main drawbacks: they are strongly oriented and the superconducting properties (**Erreur ! Source du renvoi introuvable.**) differ a lot with orientation: $J_C$ is maximum for (a,b) planes and minimum when in the direction parallel to c axis, and $\xi_c << \xi_a$, $\xi_b$. Thus, a particular orientation needs to be imposed during synthesis. The second drawback is that they are ceramics by nature making them impossible to draw like a conventional metal. It took several decades for consortium including states, research laboratories and companies to be able to develop industrial processes able to produce mostly tapes (only Bi-2212 can be drawn into wire). Cost (typically 100 €/kA/m) is still a huge obstacle to the development of industrial realization. The tape shape has its own disadvantages since demagnetization factor needs to be taken into account, as well as collective effect due to the wiring scheme.

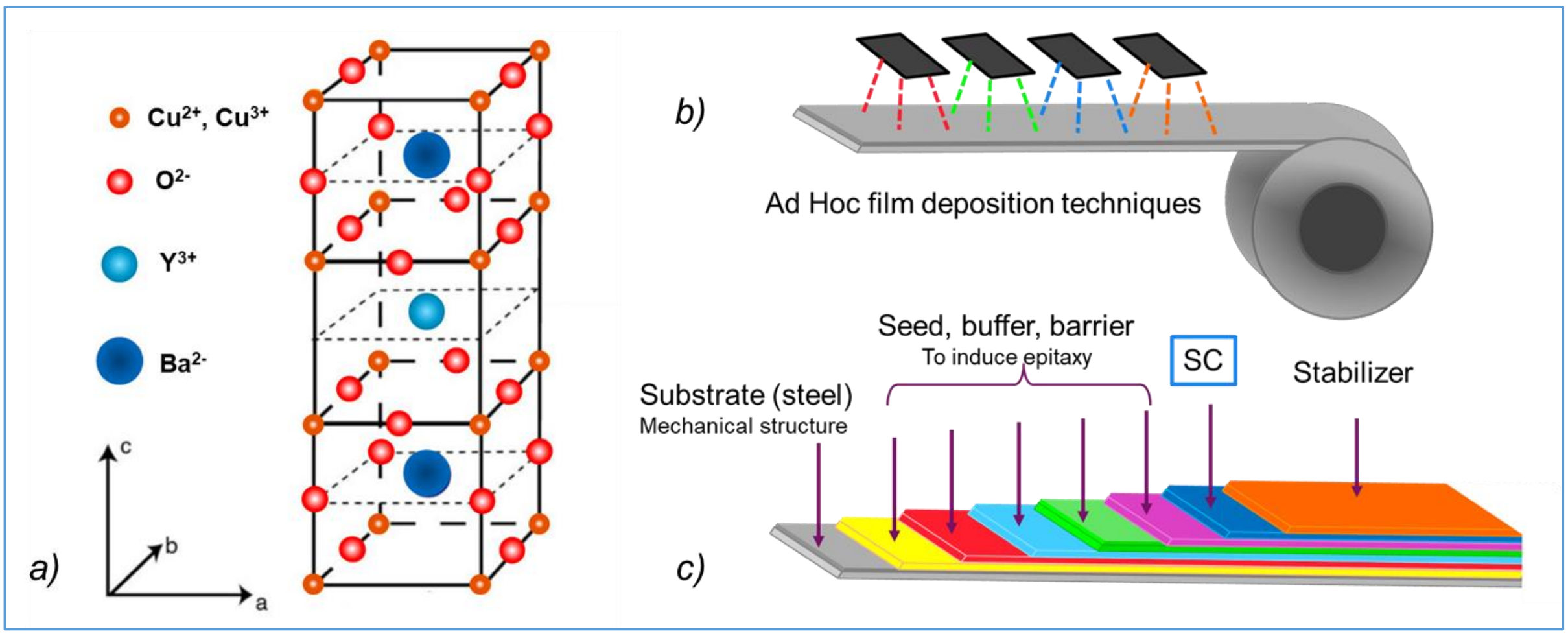

*Figure 50. a) Layered structure of YBCO family, b) Fabrication mode of high Tc tapes: thin films are deposited with several different techniques on a support tape. c) Structure of the final product: the support tape (50-100µm) plays mainly a mechanical role. Seed, buffer and barrier layers (a few 10s of nm each) aim at obtaining a specific orientation of the superconducting material so that the plane with highest superconducting properties are aligned with the current. Stabiliser is usually copper and/or silver and plays mainly a thermal role.*

YBCO family are not conventional superconductors. Cooper pairs are formed but the coupling mechanism is different from the one proposed by BCS. Actually, the coupling mechanism is still not fully elucidated yet. Another difference with conventional superconductors is the symmetry of the gap (d-wave instead of s wave). This symmetry of the gap implies that they are "directions" where the gap is 0. In RF, the surface resistance dependence is a power-law dependence: $R_S(T) \sim R_i + CT^{\alpha}$
[11, 34], contrary to BCS behavior where the decreasing of $R_S$ is proportional to 1/T. At high field, grain boundaries tend to transit in the normal state and are considered as "weak links". Early penetration of vortex is favored [35].

In RF, these high $T_C$ materials exhibit surface resistances 1 or 2 orders of magnitude better than copper and current density about one order of magnitude higher, so they are used in the form of thin films for superconducting electronic devices in the 10 GHz-1 THz range, although in the mixed state and the presence of vortex. At high field (accelerator cavities), the dissipation would still be prohibitive [36].

## 4.2 Examples of large magnet realizations.

As stretched in the conclusion, superconducting is not the main optimization parameter

### 4.2.1 Example of ATLAS Detector Toroid

The magnetic system of the ATLAS detector in the Large Hadron Collider at CERN consist of several superconducting magnets, operating at in liquid helium. It comprises a central toroid, two end toroids and a central solenoid.

***Main features of central toroid (mechanical structure + cryostats):***

- *8 5 m x 25 m coils , mean field = 2T*
- *30 km NbTi cable @ 4,2 K*
- *Field: 0,4 T toroïd center and 3,9 T inside SC*
- *20 500 nominal Amperes (~Jc:/2)*
- *Stored magnetic energy : 1,1 GJoules*
- *~1400 tons including muons detector*
- *Global structure deformation < 30 mm*

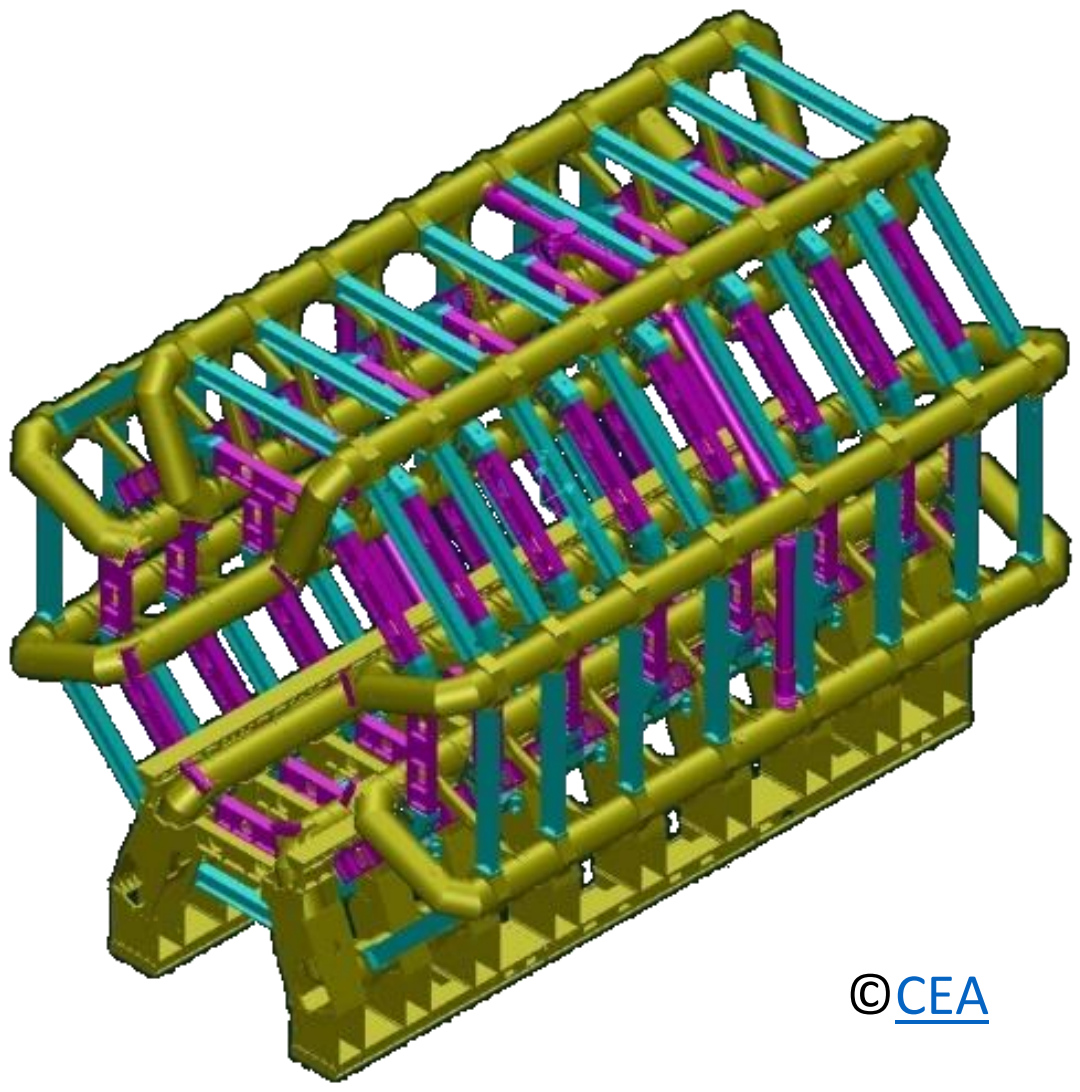

©CEA

Several parameters need to be optimized together:

**Mechanical strength**

The structure needs to resist mechanically to colossal magnetic forces, while being as "transparent" as possible. The radial magnetic forces are in the order of ~1100 tons per coil.

Thermal shrinkage during cooldown needs to be taken into account, which oblige to put the whole structure under pre-tension.

**Protection)**

Up to 1.1 GJ is stored in the whole structure when under tension. In particular, in case of transition in one single magnet among 8, the huge magnetic forces become unbalanced and other coils must be stopped in emergency. It requires electronic protection and the design of a fast discharge system.

**Precision and/or field uniformity**

The volume included inside is ~10 000 m$^3$; the geometry of the magnetic field is warranted within 10 µm, with the coils cold and under tension, taking into account cooling shrinkage.

**Thermal constraints**

The cooling of a cryogenic system is a paramount safety issue. Indeed, when liquid He transits to gaseous He, its volume multiplied by a factor ~ 700, which is comparable to an explosive situation. The cooling of the Atlas toroid occurs by thermal conduction with the complete structure immersed in liquid helium. The cooling efficiency depends strongly on the wire design. Several cooling modes (forced circulation, thermosiphon…) coexist.

For more information: [37].

### 4.2.2 Example of ITER

Iter is a Tokamak dedicated to complete the research stage still necessary to enable the construction of a prototype producing electricity by 2050 from nuclear fusion.

The plasma temperature is as high as 150 000 000°C. At this temperature, the only way to “hold” the plasma is magnetic confinement. The only way to get such a high induction in a large volume is to use superconducting magnets.

Operating regime consists into repeated cycles with high current and field. The mechanical strain reaches ~ 60 MN (twice the energy developed by the spatial shuttle take off!). The mechanical resistance of the system is one of the main concern.

The other main risk is plasma disruption, which would induce very fast variation of the field inside the conductor. Transitory regime produce inductive losses and the cooling system must be able to face them.

**Cable In Conduit Conductors**

A cable-in-conduit conductor (CICC) is a type of conductor used in high-field magnet applications (see Figure 45 in § 4.1). It consists of twisted-stacked coated-conductor tapes arranged around a helically slotted core where pressurized helium circulates. It allows providing fast reaction to rapid thermal excursion, for instance in case of plasma disruption.

***Main features of toroid mechanical structure + cryostats:***

- *Cryostat Volume =16 000 $m^3$ - Plasma Volume= 837 $m^3$*
- *10 000 tons of superconducting systems*
- *~ 80 000 km $Nb_3Sn$ cable (twice the earth circumference)*
- *Total stored magnetic energy: 51 GJoules*

***Central Solenoid:***

- *6 coils 13 m x 4 m : $B_{max}$ 13 T- 6,4 GJ stored*
- *Current inside the plasma: $I_{plasma}$ =15 MA for ~400 s*

***Toroidal coils***

- *18 coils 17 m x 7 m, toroidal field = 5,3 T*
- *Maximum field on the conductor 11,8 T*
- *164 MA.t (Amperes/ turn) in total*
- *Stored energy 41 GJ – Discharge time constant 11 s*
- *Centering force pre coil = 403 MN*
- *Vertical force per coil = 408 MN*
- *Nominal current: 68 000 A*
- *Weight= 5362 tons*

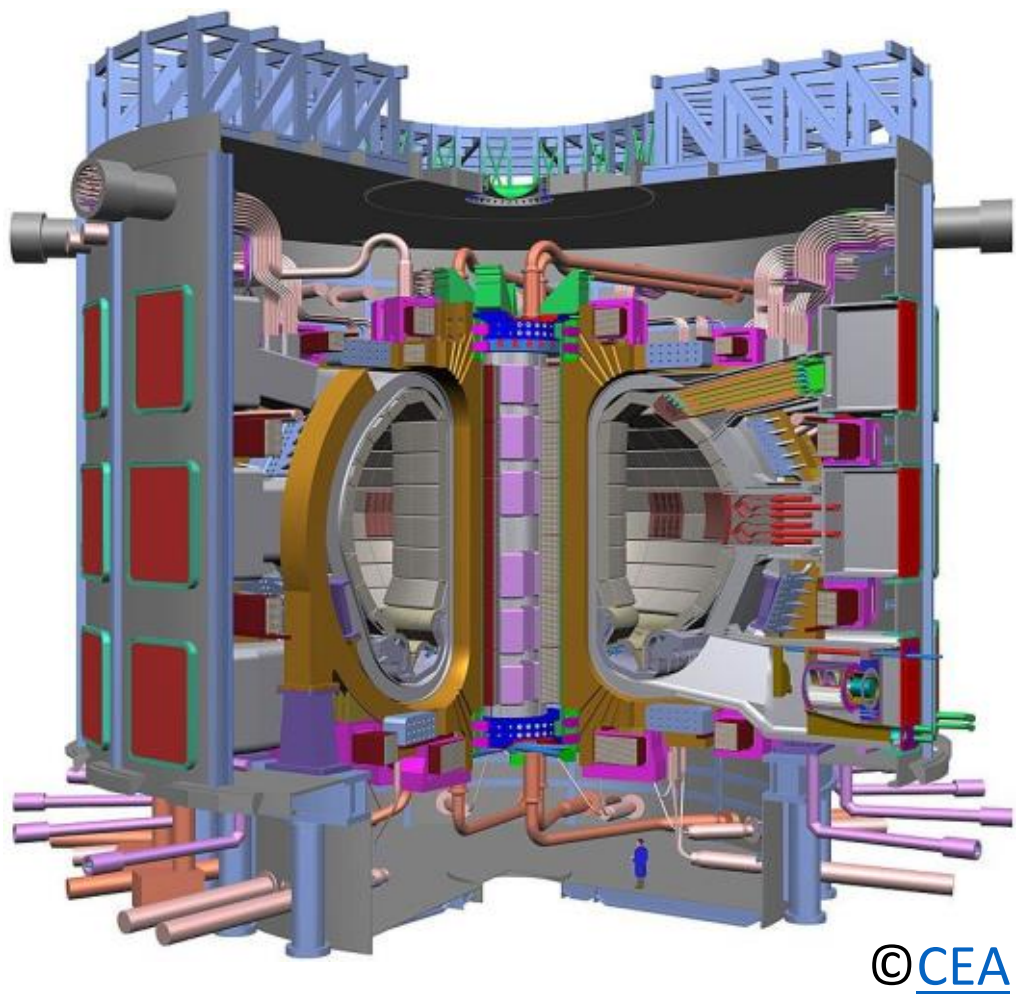

©CEA

### 4.2.3 Example of ISEULT

Iseult is a full-body 11.7 T MRI facility that was commissioned at Neurospin in 2017. It is dedicated to research in cognitive sciences and neurodegenerative illnesses diagnostic, with a resolution 10 times higher than what preexisted in the domain of brain imaging.

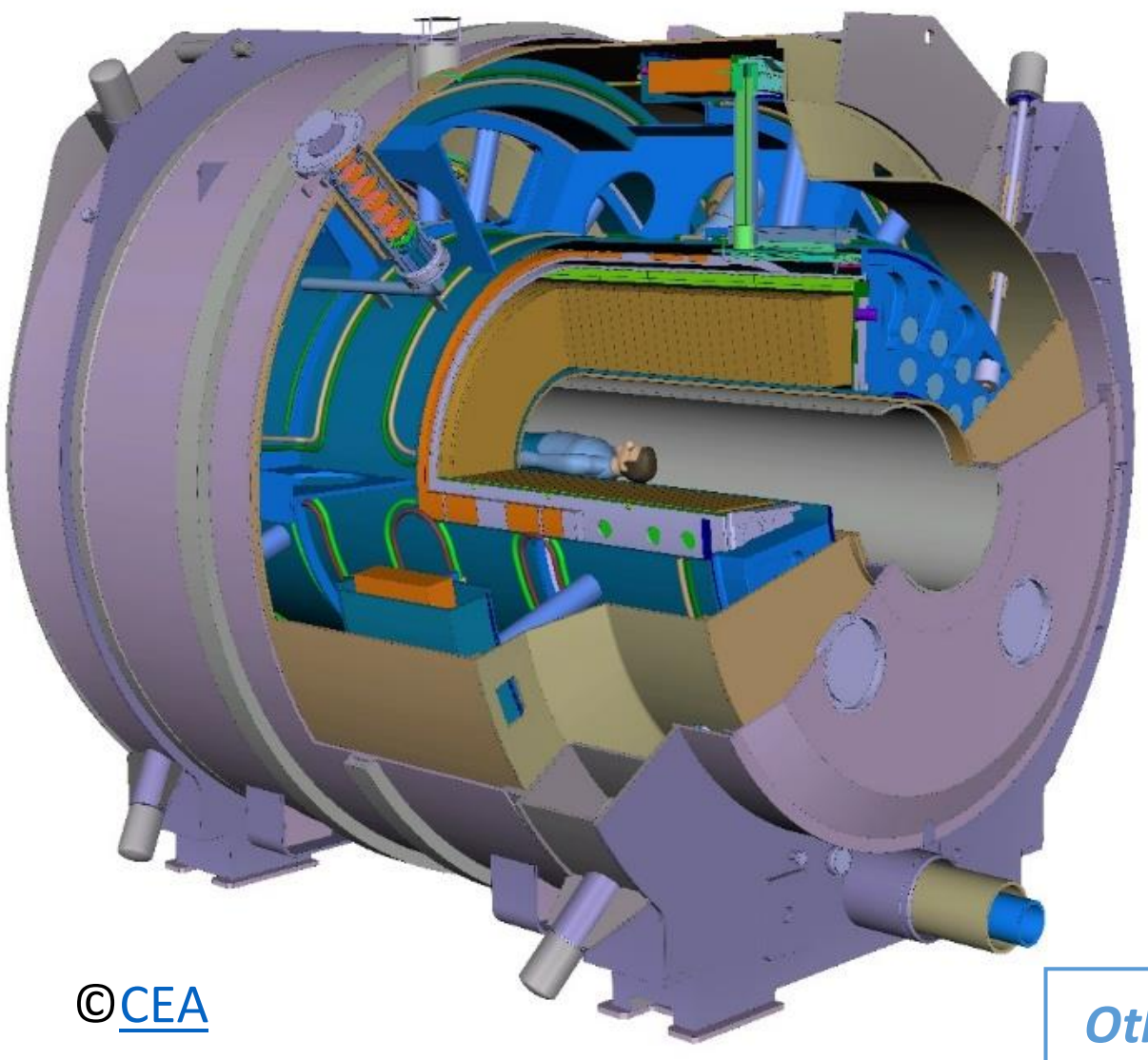

©CEA

Its main challenge is to provide a high intensity field (> 200 000 times the earth magnetic field) in a large duty volume (whole human body)[1].In order to provide high resolution brain imaging, one need a very homogeneous field. The variation of the field must be less than $5.10^{-7}$ T over some $dm^3$ (about the brain volume). Its design is based on double pancake coils, a structure that was initially developped for fusion magnets and was applied for the first time on an MRI facility.

***Other figures of merit:***

- *180 km NbTi cable coiled in170 double pancakes*
- *4 m diameter and 4 m in length*
- *Alignment better than 1 mm/full length*
- *132 Tons including 2 active coil shielding, 10 Tons each (Field must be confined in the experimental hall)*
- *Protection resistance: 1 Ton*
- *Nominal current 1483 A, stable at some 0,05 ppm/hr*
- *High temporal stability: $\Delta B < 10^{-9}$ T/10 minutes ~scan duration*
- *7000 L of He @ 1.8 K (Claudet bath), 24 h/day and 365 days/ year*

[1] At Neurospin, there is also an commercial 17 T MRI set-up for small rodent… with a duty volume 100 times smaller.

## 4.3 RF cavities

As noted in § 3.2.6, the RF surface resistance is not 0.

Nowadays RF cavities are mostly fabricated out of high purity, well-recrystallized, bulk Niobium with dedicated surface preparation. Nb is the material with the highest known $H_{C1}$, so such material is easily maintained in the Meissner state –as required for RF cavities, even in presence of small surface defects.

### 4.3.1 Thermal conductivity vs superconductivity

This material is optimized for thermal conductivity rather than superconducting properties, so that, in effect, small dissipative defects at the inner surface can be thermally stabilized by transporting the heat across the thickness of the cavity wall to the helium bath.

Superconductors are intrinsically bad thermal conductors, because many of the conduction electrons that usually transport the heat are in the form of Cooper pairs. To increase thermal conductivity, one has to decrease the density of electrons scattering centers, which is done by using high purity material with reduced light interstitial elements (H, O, C, N…) concentration.

$$R_S = R_{BCS} + R_{\mathrm{Res}}$$

$$R_{BCS} = A(\lambda_L^4, \xi_F, \ell, \sqrt{\rho_n})\frac{\omega^2}{T}e^{-\Delta/kT}$$

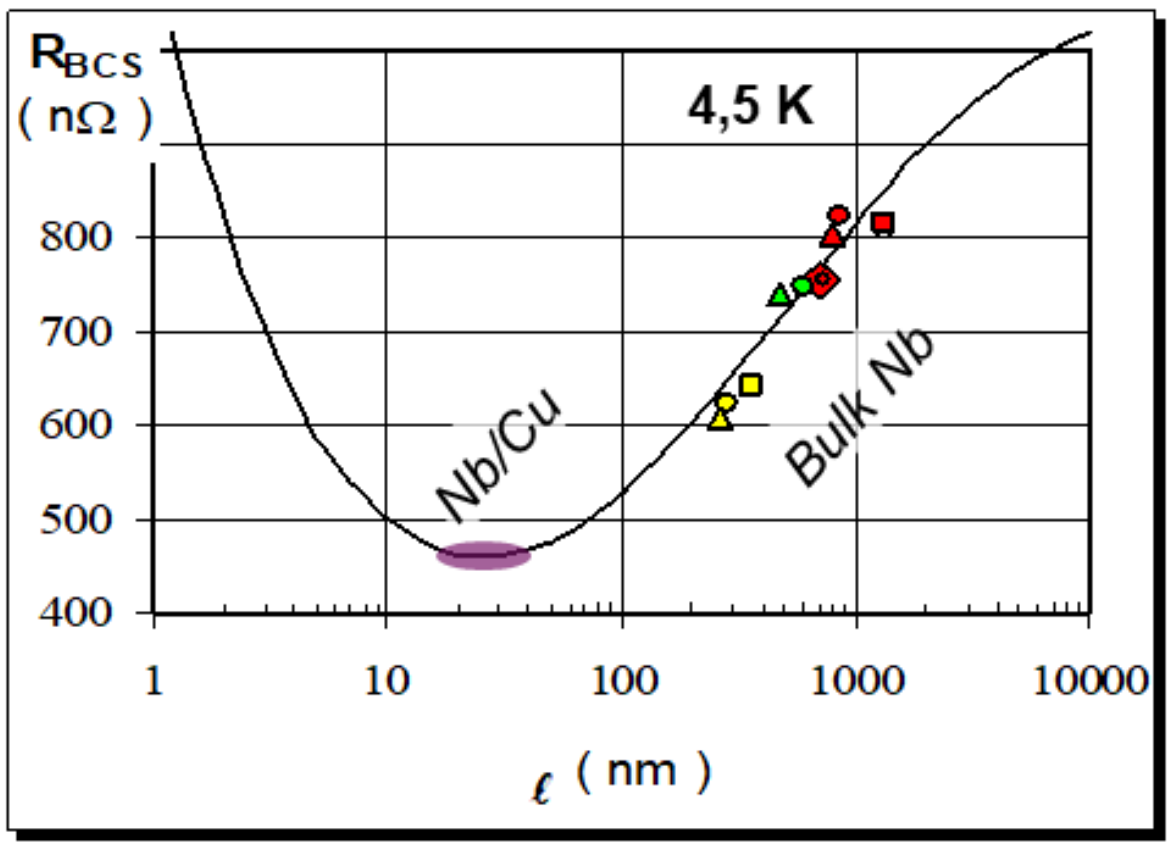

*Figure 51. Surface resistance $R_{BCS}$ vs mean free path $\ell$.*

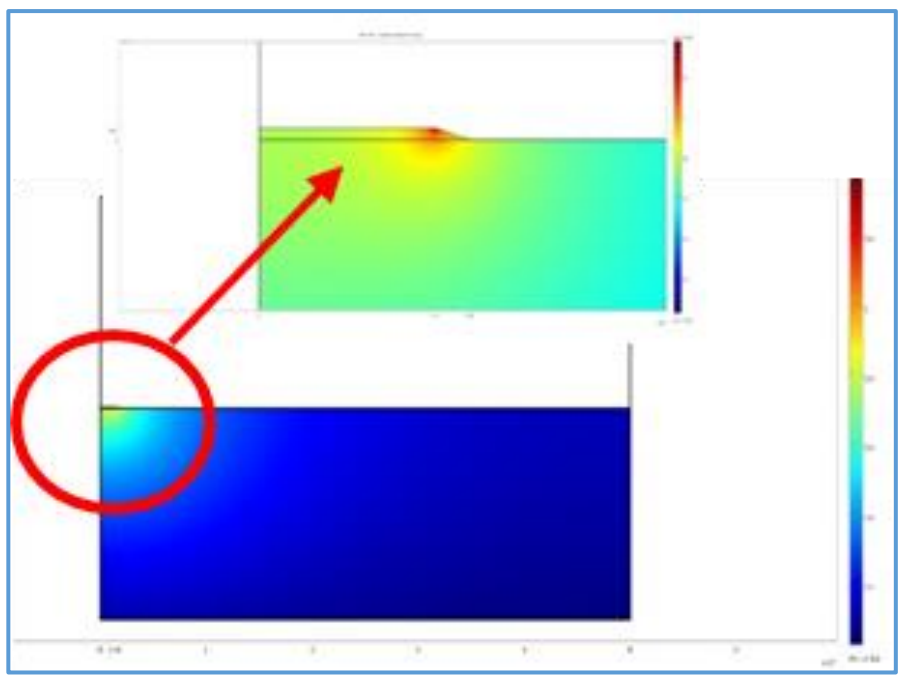

*Figure 52. Thermal transport calculation example*

### 4.3.1 Quasi perfect material on the surface

Because of electric peak field of close to 100 MV/m on the surface, cavities are very sensitive to dust contamination. It has been shown that particles tend to accumulate along electric field line and form "antennas" that locally enhance the field with a factor $\beta$. If $\beta$ reaches a factor 50 to 500, the local field can surpass ~10 GV/m, a field high enough to extract electrons form the surface by tunnel effect. Those electrons are then accelerated by the RF field stored in the cavity, disturb the beam and can provoke heating and X-Ray emission when they hit the opposite cavity walls.

On way to prevent this field emission phenomenon is to work in very clean condition, including a filtered water, high pressure rinsing that is able to dislodge the smallest dust particles. Cavities are assembled inside clean rooms with the same cleanliness level as used in microelectronics. Note that due to their fractal structures scratches can also be at the origin of field emission (tip on the tip model [38])

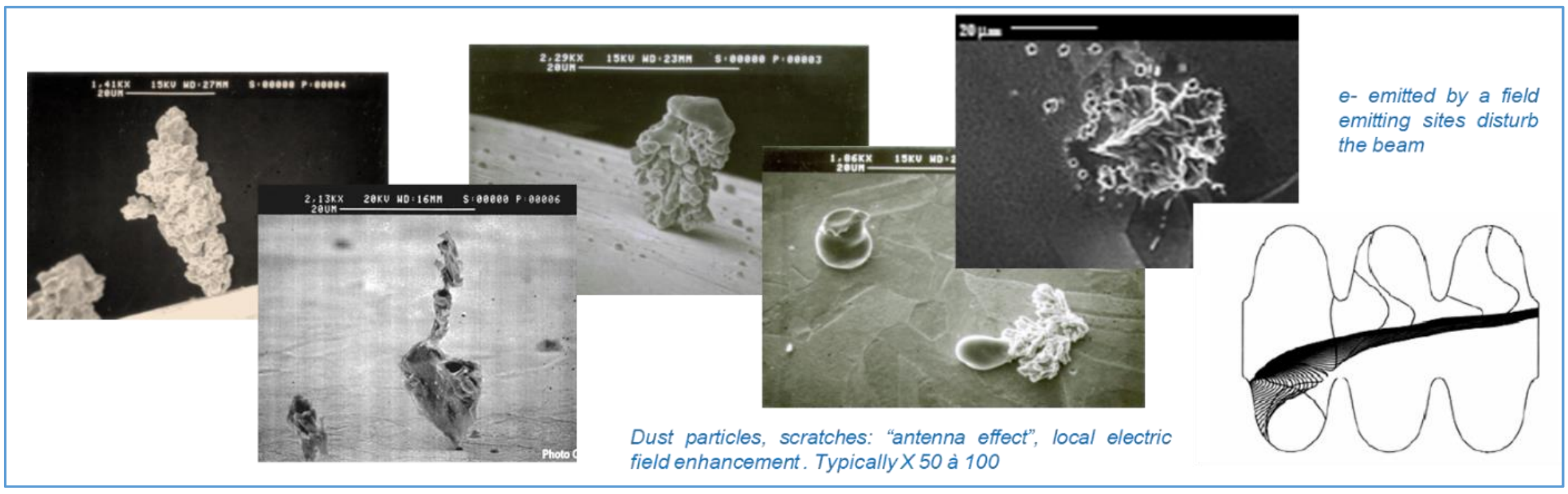

*Figure 53. Example of dust particle accumulation and conductive "antenna" formation. Due to the intense current some of the tips can be processed (melted away), but in some other condition a electrical breakdown occurs and leaves an immovable emitting site.*

### 4.3.2 Quasi perfect material under the surface

When one switch from copper where skin depth is the order of a few micron to Niobium where the penetration depth is about 40 nm, the quality of the surface becomes preponderant. The damage layer due to contact with tooling during the fabrication phases (rolling equipment to prepare Nb sheets, deep drawing dyes, thermal strain during welding…) are all liable to leave crystalline defects and affect superconductivity. Therefore, the damaged surface (typically 150-200 µm) needs to be polished away. It is usually done either by electropolishing (EP, etching rate ~ 0.3 µm/min) or, when the structure of the cavities are too complex by buffered chemical polishing (BCP, etching rate ~ 1 µm/min) which is in fact more an etching process than real polishing and produces some roughness.

Niobium at room temperature is a very stable material, so dissolving it requires to use concentrated acids including hydrofluoric acid which makes this step quite hazardous and requires specific chemical safety installation.

Recrystallization treatment (800C, 2h or 650 C, 8-10 h) is also mandatory. Not only it improves the crystalline quality of the surface, but it also reduces the density on nucleating site for niobium hydrides on the surface. Those hydrides NbHx forms pyramidal crystals on the surface and exhibit degraded superconducting properties (or even normal conducting behavior depending on the hydrogen fraction x). So the provide early entry point for vortex and trigger early quenches on untreated cavities [17].

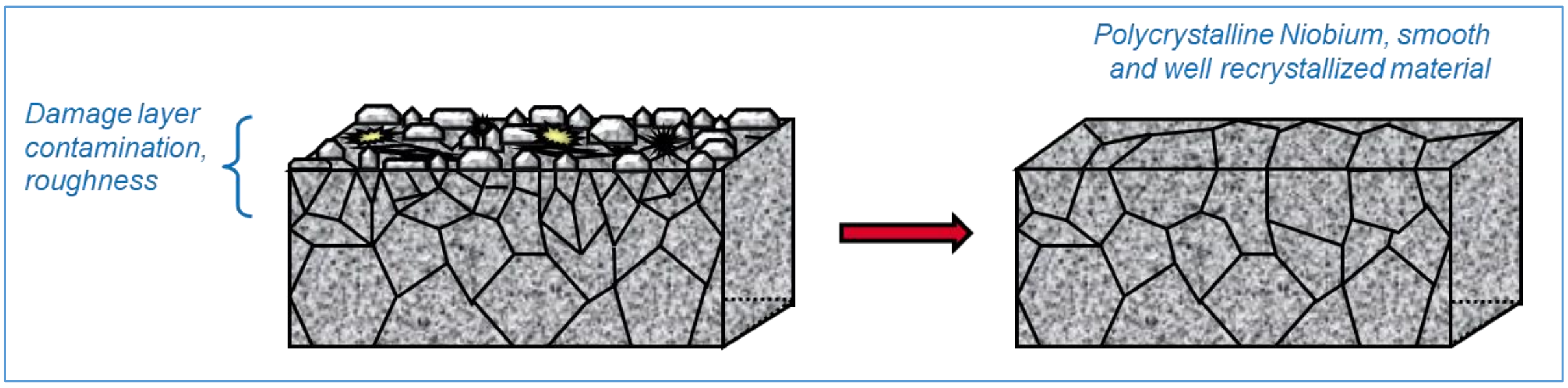

*Figure 54. Representation of the damage layer: crystallites in contact with tools are deformed, accumulating high density of crystalline defects (see §3.7). Not only superconducting properties are affected, but the risk of poorly expulsing flux line during cooling is increased.*

### 4.3.3 Improving surface resistance without affecting thermal conductivity

With various heat treatments, it is possible to affect the surface composition and the distribution of light elements on a thickness ranging from ~10 nm to a few penetration depth of the material. Since only the surface is affected, it is a way to monitor the surface superconductivity without affecting the thermal conductivity of the bulk of the cavity wall.

Very high $Q_0$ with a somewhat limited $E_{acc}$ can be reached by nitrogen or oxygen doping (see Figure 55). Nitrogen doping is obtained after the cavity has been heat treated at 800°c, 3 h, and then introduction of nitrogen gas for several minutes, and subsequent 800°C (several recipe exits [39]). A light electropolishing is necessary to remove surface pollution. The various recipes result in a flat distribution of some 100 ppm N inside Nb over several 100 nm depth. The mechanism of the decreasing of the surface resistance with field are still debated, but there are some indications that surface Tc is affected by the presence of a large quantity of interstitial, which could explain why the accelerating field is limited.

More recently it was discovered that medium T-baking around 300-400 °C [40], with no nitrogen was producing the same type of behavior. In this case, Oxygen from the oxide layer diffuses from the surface and occupies the same interstitial spaces as nitrogen. Both N and O are known to affect the diffusion rate of H inside Nb[1], and as shown on Figure 51 the surface resistance is minimum for a mean free path about 10 nm. More complex mechanism need to be implemented to fully explain the observed behavior [39].

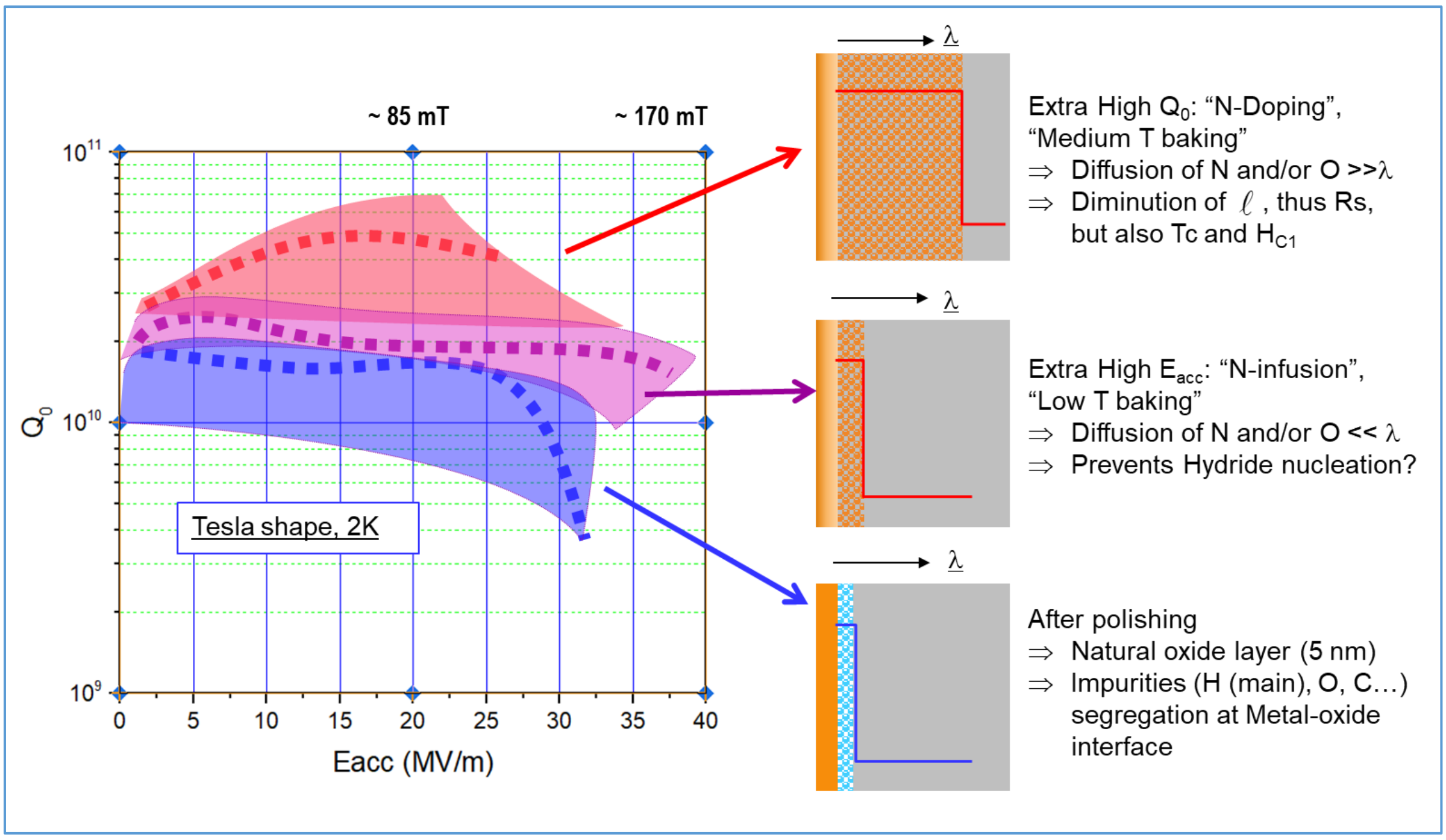

*Figure 55. Schematic behavior of cavities after various surface preparation recipes.*

Lower temperature treatment (120°, 48h, with or without nitrogen) also affect the behavior of cavities, enhancing the accelerating field, without necessarily increasing the $Q_0$. The profile of impurities keeps concentrated in the first 10 nm of the surface. Knowing that hydrides are very good candidates to early vortex penetration, the main effect seems to be the prevention of large hydrides precipitates.

The important results is that one can monitor the behavior of the cavities to their applications, without changing the overall bulk Nb technology. High $Q_0$ are required for high duty cycle and CW machines, while high $E_{acc}$ rather apply to longer machines such as linear colliders.

[1] H move relatively feely inside Nb at room temperature, but due to elastic interaction with defects, in particular the oxide –metal interface, surface segregation is observed (see §3.4.3).

### 4.3.4 Future: thin film cavities

Bulk Nb is a rather expensive technology, so very early the possibility to depose thin film of Nb inside a copper cavity has been explored. Unfortunately, the crystalline quality of thin film is far lower than that of bulk cavity. During more than 50 years an active R&D has been conducted without real improvement although many directions have been explored. This technology has been confined to circular machines where accelerating field is not paramount. Only recently, by combining better surface preparation of copper substrate cavities with higher density deposition techniques the quality of Nb thin films could be improved and come close to bulk Nb.

High Tc material like NbN, $Nb_3Sn$ or $MgB_2$ have also been studied. Here again, these materials, which exhibit very small $\xi$ compared to Nb, have shown to be very sensitive to surface defects. The challenge for the near future is to achieve deposition onto copper cavities and reaching operating temperature of 4.5 K instead of 2 K with no increase in dissipation. The simplification of cryogenic installation and consummation decrease is expected to reduce investments as well as operating costs. Today active R&D is conducted in this domain, but no cavity production has been completed. $Nb_3Sn$ layers prepared by Tin thermal diffusion inside bulk Nb cavities show the same performances at 4.5K as bulk Nb at ~2 K but are limited in accelerating field around 25 MV/m (~100 mT) whereas record cavities reach 50 MV/m (210 mT) [41].

**<u>Multilayers to access more realistic materials</u>**

As explained before, the main difficulty is to maintain the superheating state in presence of defects. In 2006, a theoretician proposed multilayer structures to help to overcome this difficulty [42]. By inserting a dielectric layer (transparent to RF) a few 100nm under the surface, one can block the avalanche penetration of vortex. The initial vortex loop converts into a small portion of vortex plus a small portion of antivortex that readily coalesce after a few RF periods (Figure 56 c)).

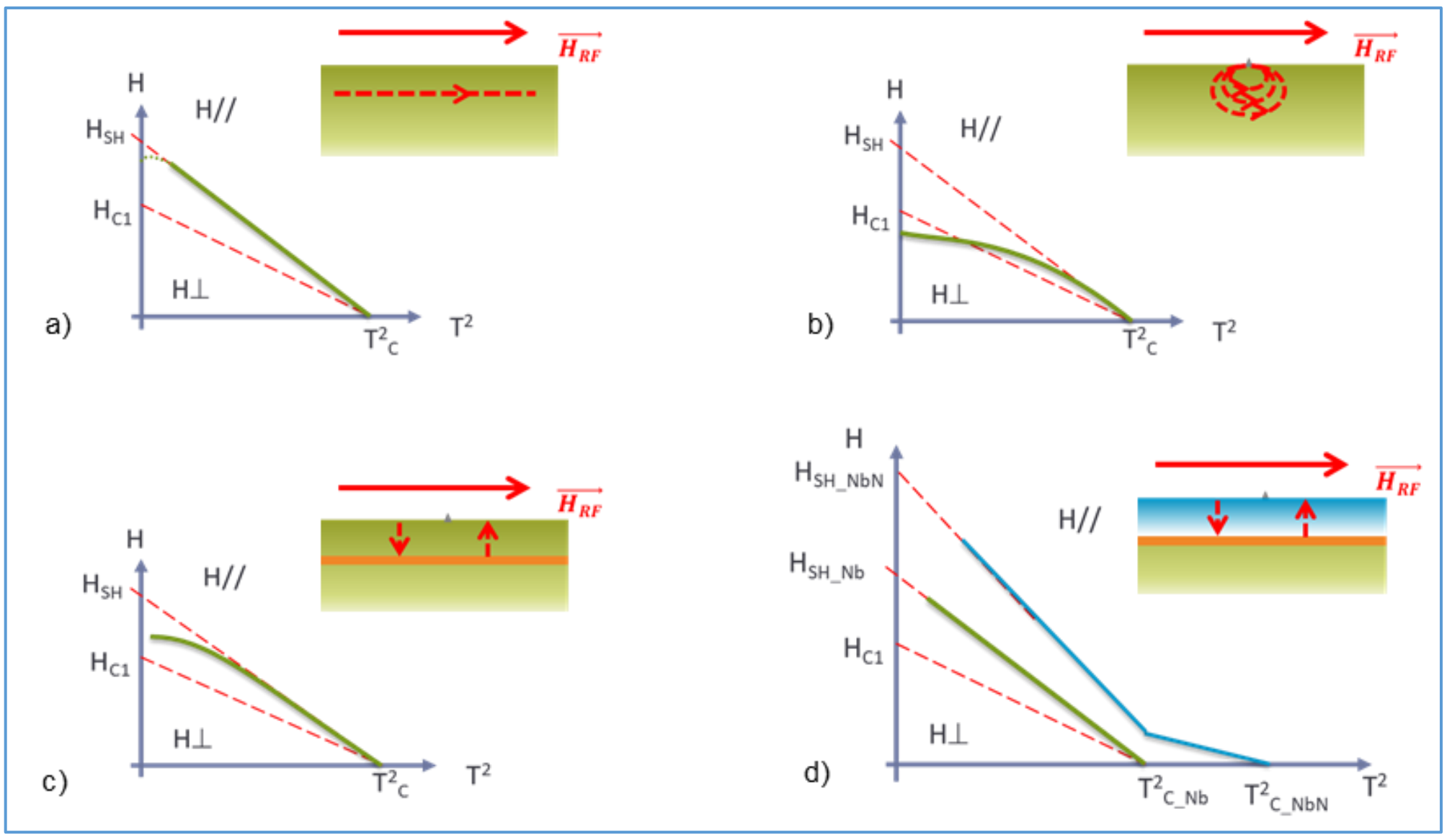

*Figure 56. Principle of multilayers. a) In the ideal case, Bean-Livingston barrier stabilize the superheated state and Meissner state can be maintained above $H_{C1}$. b) In presence of a surface defect, vortex loops can enter the SC and at high field the dissipation are too high to maintain the superheated state. c) In presence of a dielectric layer, the initial vortex loop converts into a small portion of vortex plus a small portion of antivortex that readily coalesce after a few RF periods, preventing local heating and avalanche penetration. The use of a higher Tc superconductors with thickness ≤ $\lambda_L$ allows to access high accelerating field with decreased surface resistance.*

Moreover, if one use a higher $T_C$ SC with thickness ~$\lambda_L$ as a top layer, then its $H_{C1}$ is artificially increased, allowing enhancing the RF field inside the cavity. The BCS surface resistance of a higher $T_C$ material is expected to be smaller, so overall performances of a coated cavity is expected to be improved in accelerating field as well as in quality factor. These structures have been explored with magnetron sputtering material, showing that the multilayer structure is much less sensitive to defects [43].

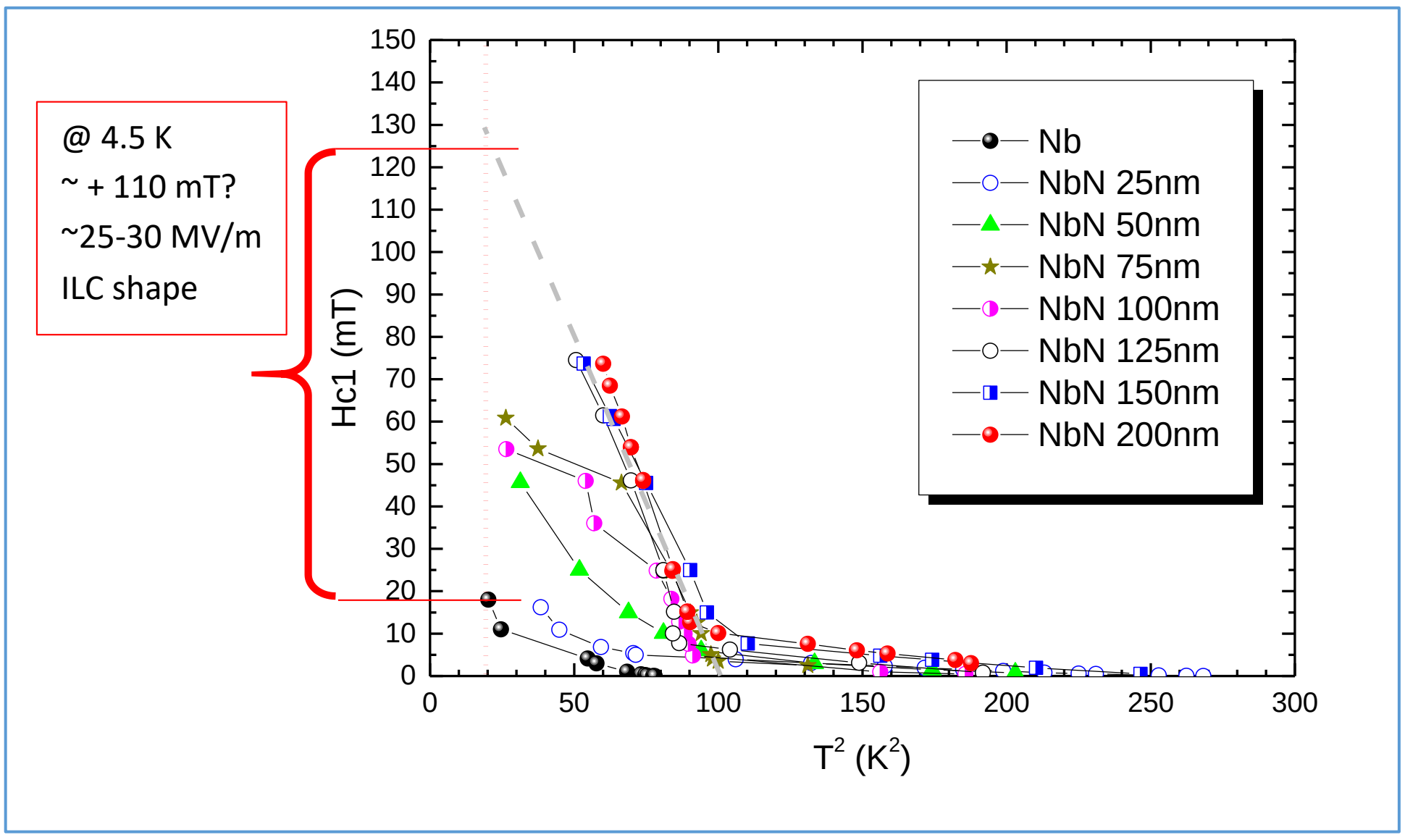

*Figure 57.Field penetration of samples deposited by magnetron sputtering, with ~500 nm of Nb (to mimic bulk Nb), 12 nm of dielectric (MgO), and various thicknesses of NbN, showing the HC1 increase compared to magnetron sputtered Nb layer alone (black curve) [43].*

## 4.4 Examples of large accelerators realizations or projects

### 4.4.1 High gradient applications and spin-of

Initially, high gradient applications have been developed for collider in high-energy physics or nuclear physics. We will focus here on Higgs factories projects. The technology developed for such application is now widely used for other large instruments in service to a larger scientific and industrial community.

**Circular colliders**

The use of superconducting cavities in high energy storage rings offers the advantage of considerable power savings over normal conducting structures. The main advantage of SC cavities are its large openings which allows low beam emittance, low wake field, high gradients and high duty cycles or even continuous beam. For instance superconducting cavities have been used at CERN in LEP (1989-2000) then LHC (2008- ). Few cavities are needed since one has only to replenish the energy lost by synchrotron radiation. As these losses go as $E^4$, at very high energy, interaction of the beam with the accelerator walls may become disruptive. Although the problem is not insurmountable, as evidenced by the development of FCC [44]or CEPC projects [45], linear machines are becoming competitive at high energy. Accelerating gradient is not an issue, but Q0 is, as main concern is cryogenic losses.

> ***Cavities for circular accelerators.*** *In circular machines, one can use lower frequency (lower beam impedance R/Q), but it induces cost issues: cavities are larger, they cannot be man held, one need clean room compatible handling robots. Everything needs to be bigger: cleanrooms, fixtures, tooling …*
>
> *Because of high beam current, they are issues with radiation induced desorption (vacuum); issues with losses in beam tubes (capacitive/inductive effects); issues with beam life time that can be mitigated with HOM effective removal and shaping of the bunches*
>
> *Same requirement are needed for other type of accelerators like proton or heavy ions sources*

## International Linear Collider (ILC)

ILC is an example of an electron-positron linear collider for which R&D started while the LHC was still being built, under the assumption that the LHC would spot the Higgs and physicists would want to study its properties in more details. Its technology is mature, and based on bulk Nb 9-cells 1.3 GHz cavities as shown on Figure 58. That kind of cavities can reach routinely accelerating gradient between 40 and 50 MV/m (170-210 mT) at 2 K with quality factors $Q_0$ above $10^{10}$.

In the initial 0.25 TeV collider design, one could need about 15 000 cavities, in an effort supported by the three main world regions: Americas, Europe and Asia. Over the past 20 years, an international collaboration has been very active into bringing this technology from lab prototype to the level of standardized industrial production [46]. Even if this project has not be realized yet, mastering this technology found several "spin-off" applications, in particular in the domain of light sources.

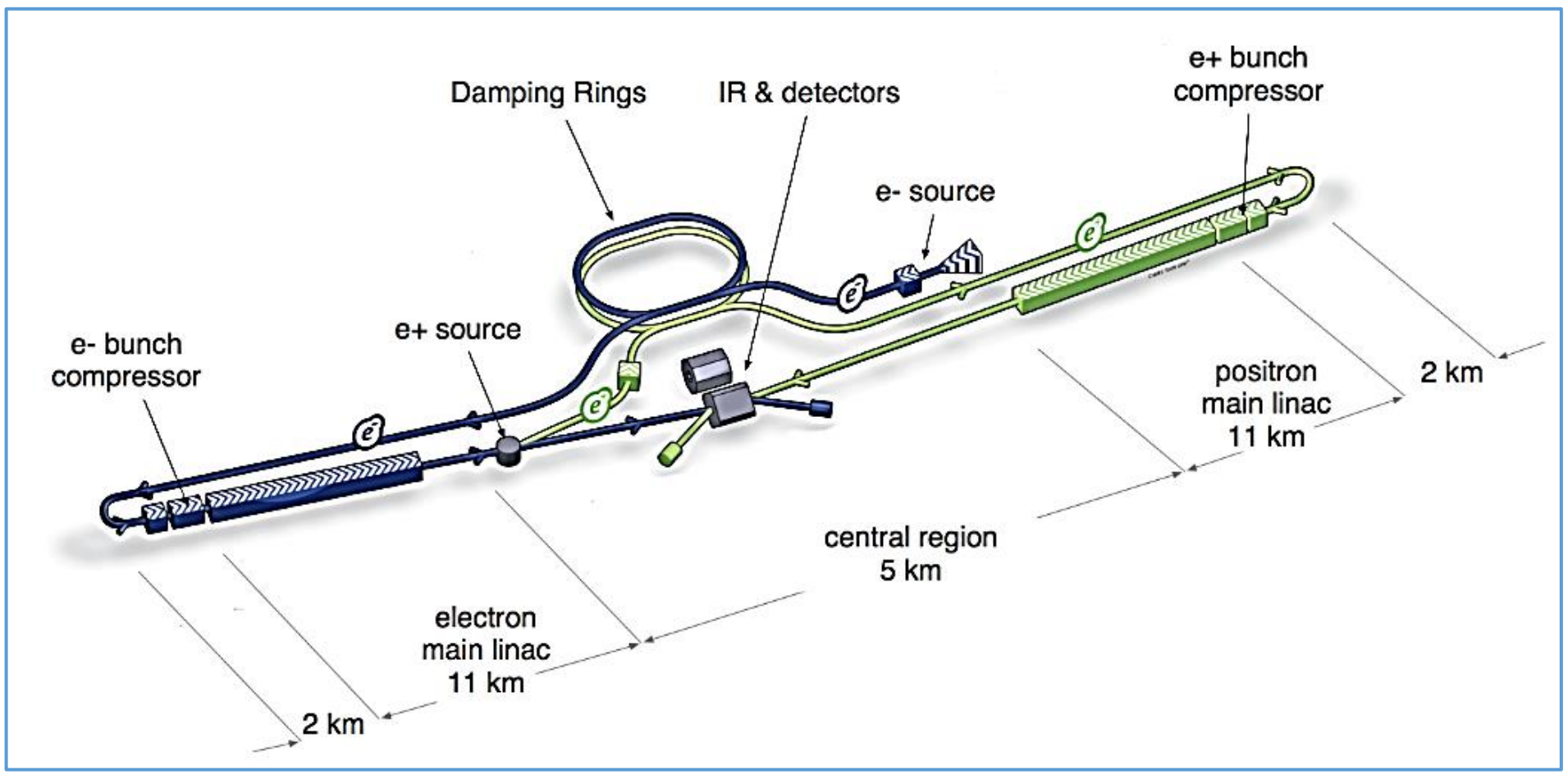

*Figure 59. Layout of the ILC project (© https://linearcollider.org/)*

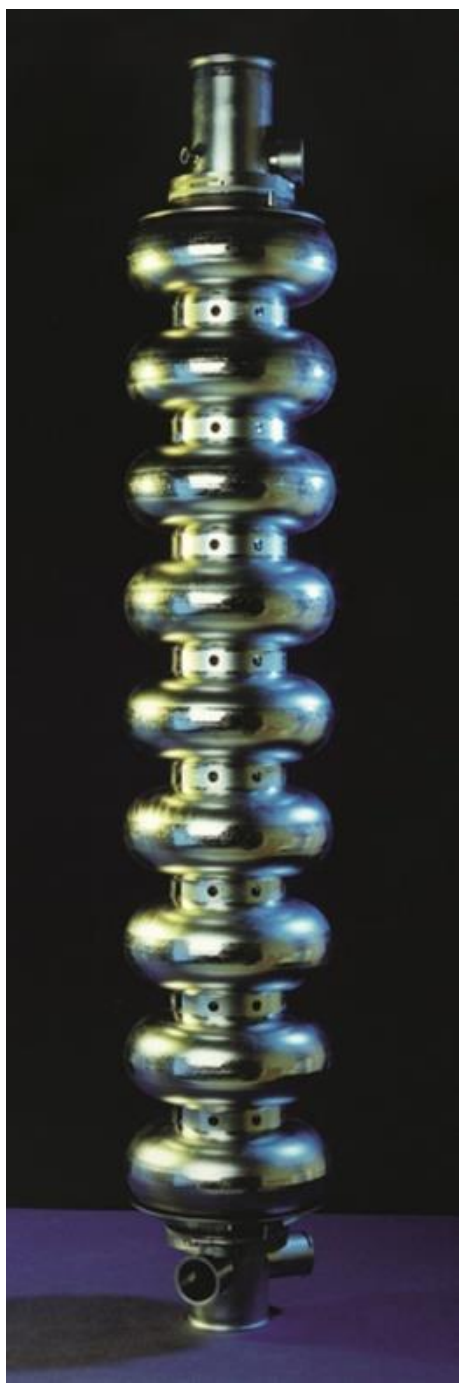

*Figure 58. Bulk niobium 9-cells RF cavities developed for high gradient application*

Note that other accelerator technologies are still being explored in parallel to built such a "Higg factory". For instance, like FCC (a circular collider with superconducting cavities and magnets) or CLIC (a linear accelerator with hot copper cavities at 12 GHz). Each technology has its advantages and disadvantages in term of performance vs cost.

## Electron Light sources

Light Sources take advantage of the initially parasitic synchrotron radiation when the beam is curved. They can be either based on storage rings or linacs for the highest energy. Recently, new design like ERLs (Energy recovery Linacs) have been proposed, which take the advantages of the both circular and linear designs [47]. Compared to collider applications, light sources are even more demanding, since rather than high gradients one aim to very high beam current and brightness, with very high duty cycle or even CW [48]. In particular Free Electron Lasers allow obtaining high energy radiation (e.g. X-Rays) when optical mirror do not exist for a conventional laser.

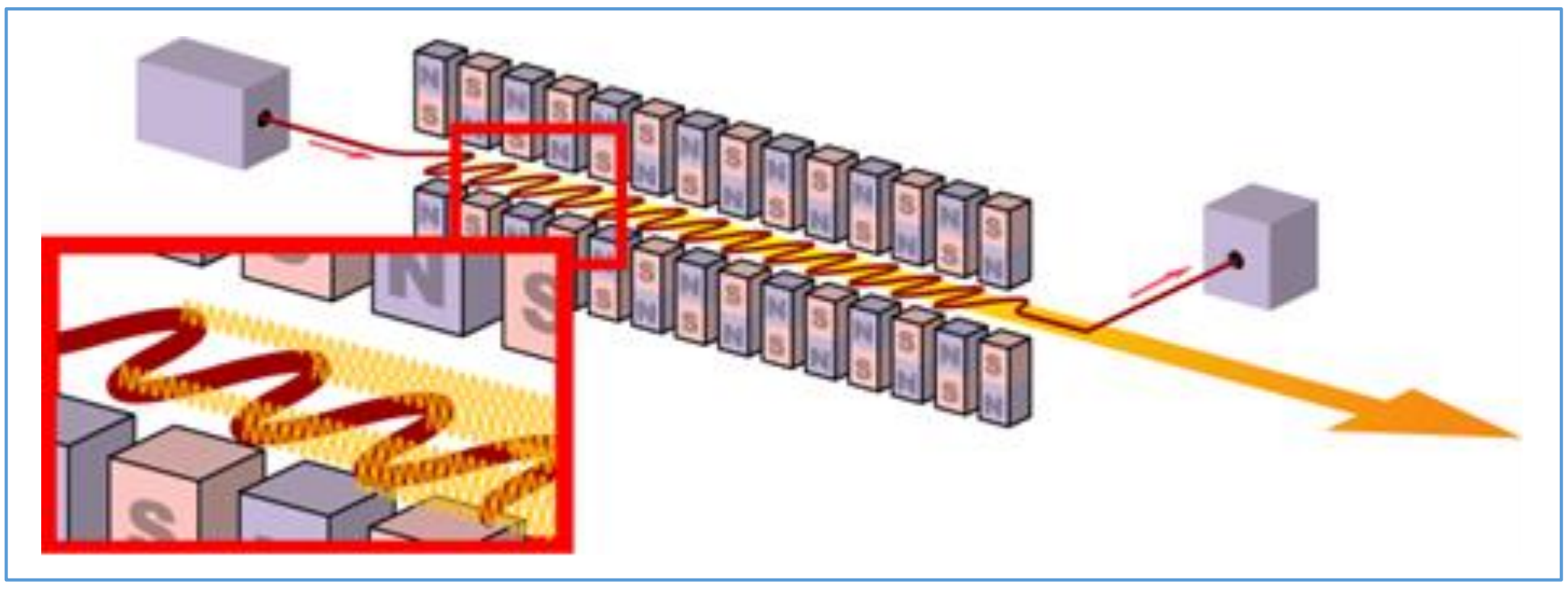

*Figure 60. Free Electron Laser functioning principle. By Horst Frank, CC BY-SA 3.0,* *https://commons.wikimedia.org/w/index.php?curid=3977203*

The necessary beam energy range from a few 10s of MeV to generate IR (~1 µm wavelength) to several GeV to reach soft X-rays (~1 nm wavelength) to 20 GeV for hard X-rays (~1 Å wavelength). With linacs; very short bunches (< ps) are available (several orders of magnitude better than in storage rings), which allows studying fast dynamic processes such as in-situ biologic reactions (think of the consequences of e.g. fully understand photosynthesis and being able to emulate it…)

Light sources provide users access to many different kinds of spectroscopies, enabling molecular structure and composition exploration. Users can be material scientists, biologists, agronomists, archeologists, industrial users… In 2020, they were more than 70 dedicated large scale Synchrotron radiation light sources worldwide [48]

XFEL (~800 cavities, Sweden) and LCLS II (~300 cavities, USA) are two examples of linear light sources based on the same technology as ILC

# 5 Conclusion

Nowadays, one can notice that most advanced technologies (materials for fusion, nuclear power, fuel cells, solar cells , batteries…and accelerators) are limited by material issues. Even if the technology is well mastered at the lab level, one will be facing reproducibility issue to switch to production level, the need of reducing costs, and finding solutions to aging problems.

It is particularly true when a technology is pushed to its ultimate limits as in the case of accelerators.

Most of the time the choice of the material results from a compromise between multiples constraints: mechanical, thermal, stability, superconductivity, costs… and interdisciplinary work is mandatory: multiple and complementary expertises are necessary

Meeting experts outside the accelerators community is often a way to prevent costly and long R&D steps, when the answer already exists elsewhere (do not re-invent the wheel!)

Let's also be prepared to meet diagnostics issues: where does the breakdown come from when many conjugate factors occurs together? A large part of the work then consists into diagnostics developments.

Also allow yourself to break traditions (but not too often because it costs money…)

# 6 Annex: Magnetic characterization

Magnetic measurement are one of the most current characterization method for superconductors. We present a short description of the most common techniques.

## 6.1 DC magnetization

Classical DC Magnetometry is an inductive method. The sample immersed in a uniform field, generated by a coil. It acquires a magnetic moment **M**. The sample is displaced in the field (e.g. oscillation “Vibrating Sample Magnetometer”-VSM is the most frequent). The variation of magnetic flux induced a change of voltage **V** in the measurement coil (**V α M**). Then a sensitive element is needed to translate the voltage into an electrical signal, e.g. Squid, Hall probe, sensing coil…

For superconducting samples, Squid VSM or DC scan are the most common, but many variants exist.

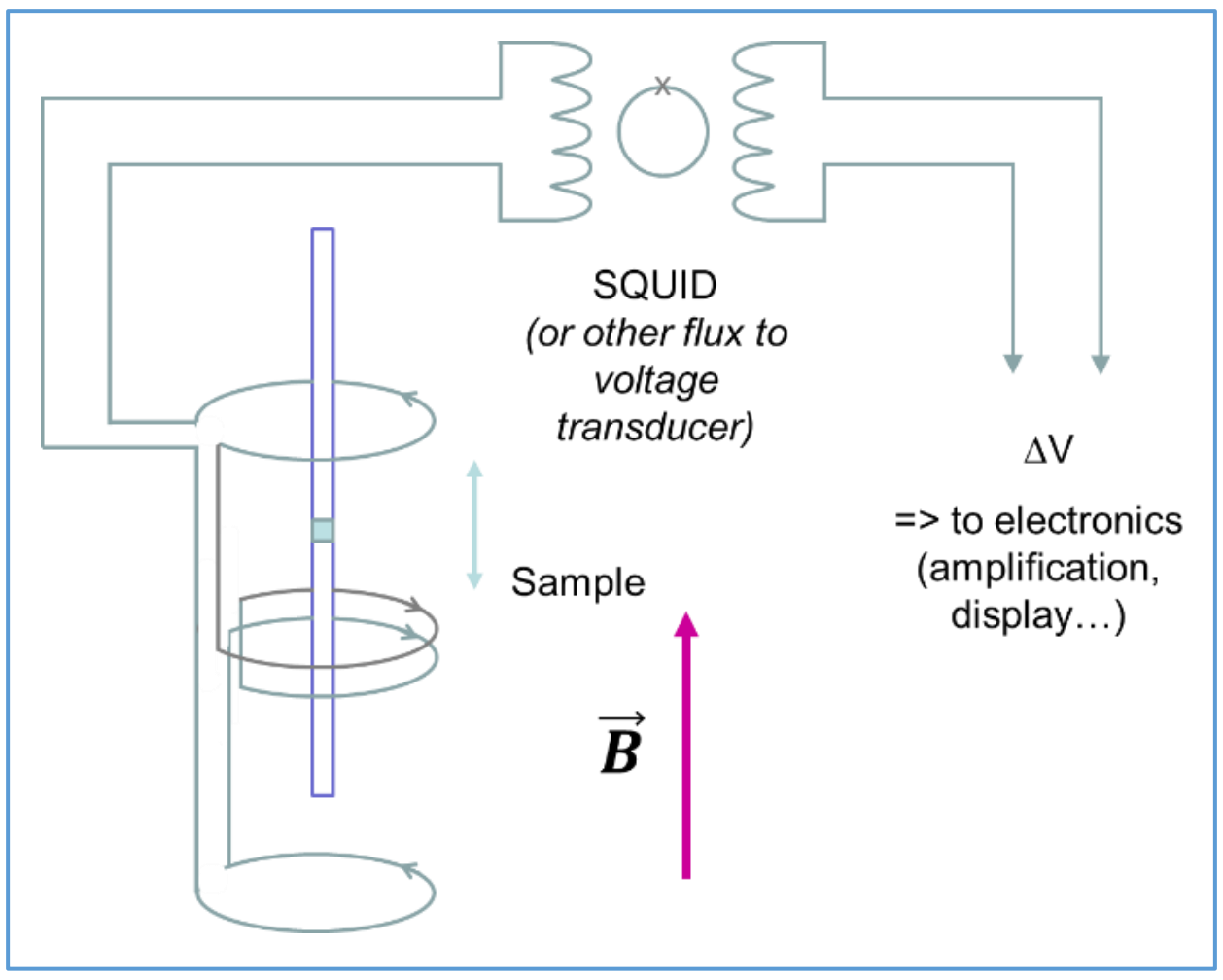

*Figure 61. Principle of magnetic flux-to-voltage measurement*

Figure 62 shows the typical aspect of magnetization hysteresis at T = constant, and summarizes the main interpretation of the results (after [49])

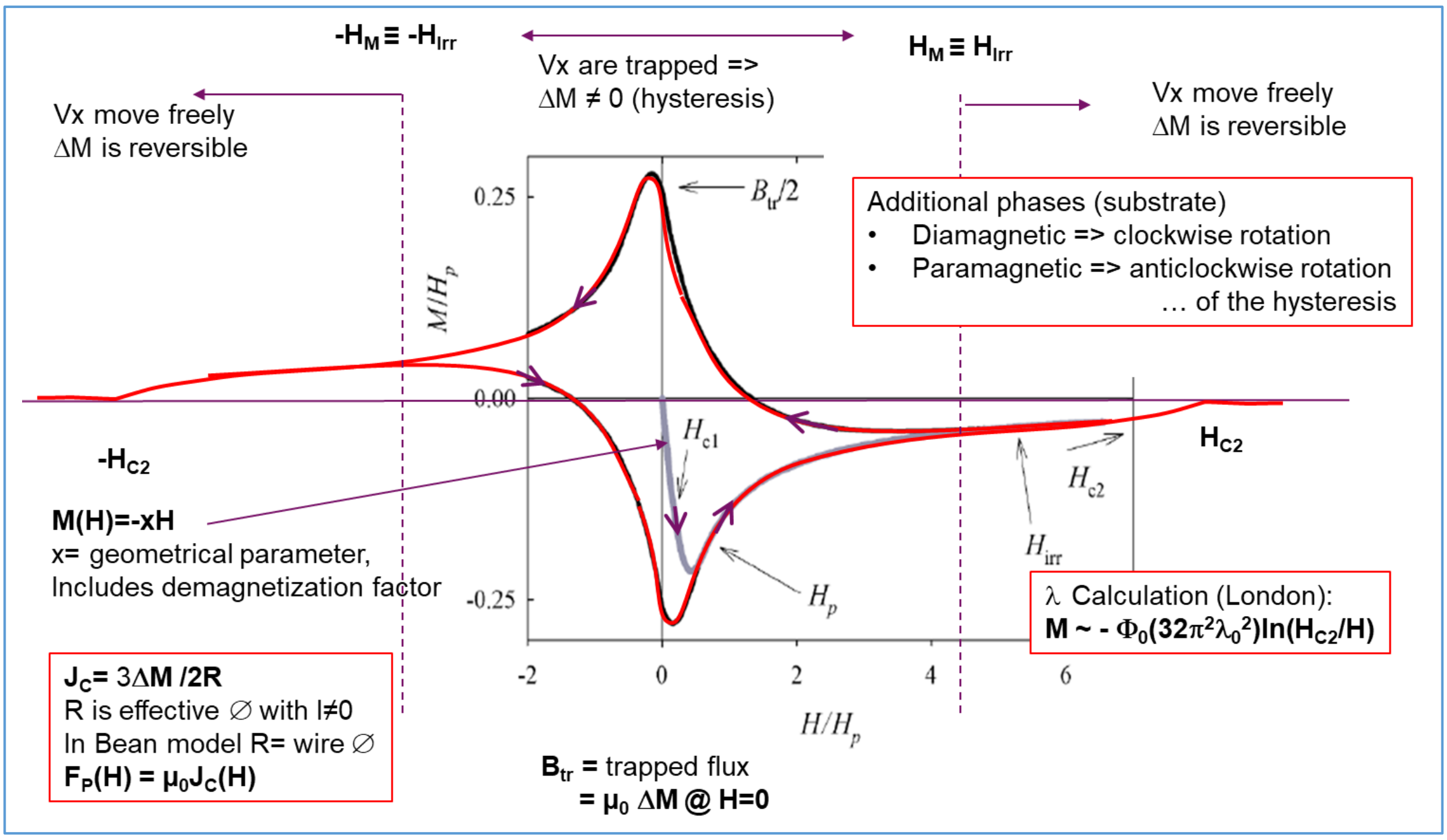

*Figure 62. Typical hysteresis shape and interpretable data gathered by DC magnetization*

The shape of the sample has an important effect on the slope and the shape of the initial magnetization curve. It is an important source of sample-to sample variation, especially since the roughness of the edges also play a role. For more details see [50, 51].

Another type of measurement is measuring the magnetic moment at constant Field in function of temperature. When the magnetic moment differs from 0, the transition temperature for this given field is reached (decreasing field).

## 6.2 AC techniques

AC Techniques are widely used to determine SC properties of cable material (surface susceptibility, surface impedance or surface resistivity) [23, 52]. The sample is immersed in a DC magnetic field, and an AC coil provides alternating field. Samples exhibit dissipative mechanisms in presence time varying field. The complex susceptibility $\chi$ is measured:

$$\chi = \chi' + i\chi'' \quad (24)$$

$\chi'$ is the component in-phase with the drive field ("reactive" or "dispersive" component). At low frequencies, the $\chi'$ term follows the applied field in phase (~instantaneous magnetization) and the dissipation keeps low. At high-frequency, the moments are not able to respond fast enough to the applied field and $\chi'$ (as well as $\chi''$) tends to 0.

$\chi''$ is the dissipative or absorptive component. Dissipation can be quantified graphically by the area enclosed by the hysteresis loop traced out over one AC cycle in the M(H) plane.

At Intermediate-frequency the moments are lagging behind the applied field, so the in-phase term $\chi'$ decreases and dissipative term $\chi''$ increases. There is a maximum where the oscillating field is most effectively dissipating heat in the sample.

One can also use the coefficients of the Fourier expansion of $\chi$ to describe components of the differential magnetic susceptibility. The n = 1 terms of the expansion correspond to a linear response of the material's magnetization to the applied field. Higher-order harmonics (n > 1) are due to nonlinear magnetic response.

Susceptibility curves look a lot like the one shown in Figure 43 (§3.4.4) vs frequency or temperature. Behaviors of materials upon increasing $\omega$ or T are similar since both actions provide energy to unpin vortices

## 6.3 3rd harmonic techniques:

In this case the sample is a relatively large slab and the magnetic field is provided by a coil with dimension much smaller than the sample one. Since the field decays quickly around the coil, at the first order, the sample can be considered like an infinite plane. With this technique, one can observe the first penetration of the field without contribution of the shape and the edge quality of the sample. The sample must in principle be thick enough to prevent full penetration, but corrections are available for thinner films [53]. Many variant exist, in particular in the excitation and signal measurement techniques.

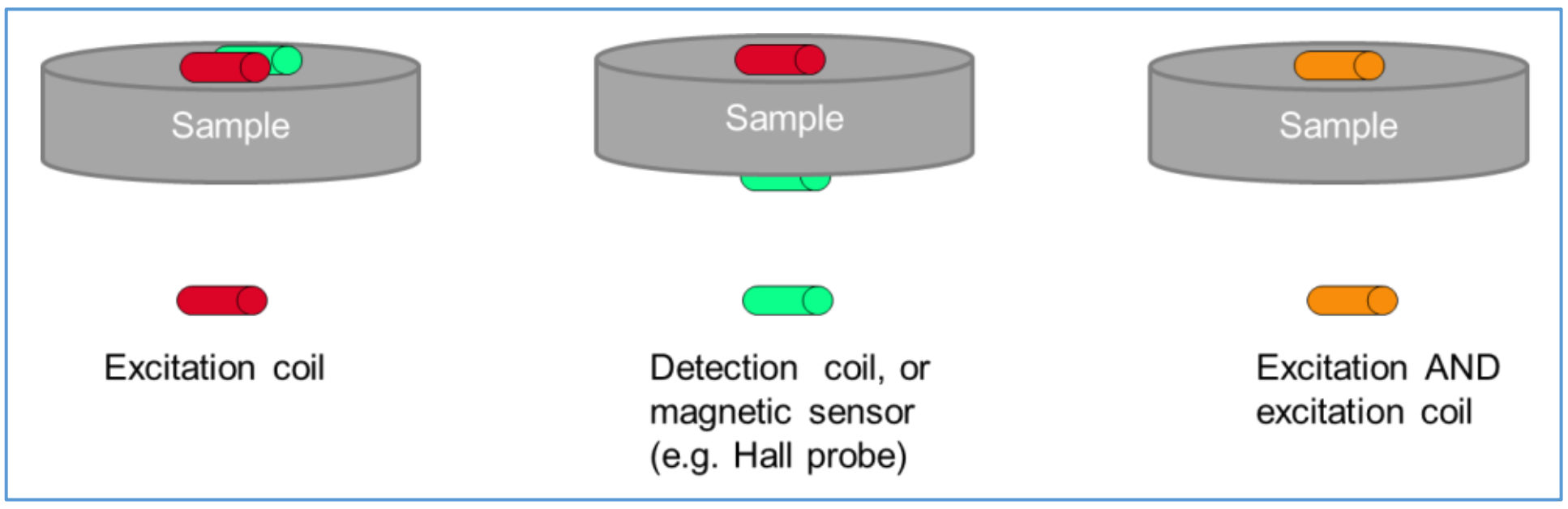

*Figure 63. Various measurement configuration for third harmonic techniques.*

The sample is 0-field cooled (ZFC) into the Meissner state. In this case, the samples behave like a perfect magnetic mirror. Then, either the field is increased at fixed temperature or the temperature is increased at fixed $b_0$ (an AC field is sent in the coil: $b_0 cos(\omega t+\varphi)$). When vortices start to enter the material, they are trapped and affect the reference signal. Higher harmonics start to appear. In general, the 3$^{rd}$ harmonic is the most intense and is monitored. The third harmonic signal provided information on the first penetration field, and on the pinning behaviour. Indeed the signal disappears once the vortices are unpinned and start to move “freely” in the material. For more details, see chapter 5 in [23].

# 7 Documentation

## 7.1 General documentation

More can be found e.g. in CERN accelerator schools on superconductivity:

- CAS-Superconductivity and Cryogenics for Accelerators and Detectors, 01 October 2002 - 10 October 2002, Erice, Italy
  https://cas.web.cern.ch/schools/erice-2002 .
  Proceeding CERN-2004-008: http://cdsweb.cern.ch/record/503603/files/CERN-2004-008.pdf
- CAS- Superconductivity, 24 April 2013 - 05 May 2013, Erice, Italy.
  https://cas.web.cern.ch/schools/erice-2013.
  Proceedings CERN-2014-005: https://cds.cern.ch/record/1507630